\documentclass[12pt,draftcls,onecolumn,twoside]{IEEEtran}
\pdfoutput=1 
\usepackage{amssymb}
\usepackage{amsmath}
\usepackage{amsopn}
\usepackage{graphicx,color}                   
\usepackage{rotating}
\usepackage{bm}
\usepackage{cite}              

\allowdisplaybreaks

\newcommand{\comment}[1]{}  
\definecolor{old}{gray}{0.6}

\newcommand{\eqn}[1]{\eqref{#1}}

\renewcommand{\vec}[1]{\mathbf{#1}}
\newcommand{\mat}[1]{\mathbf{#1}}

\newcommand{\equaldist}{\stackrel{\scriptscriptstyle d}{=}}
\newcommand{\equalasympt}{\stackrel{\scriptscriptstyle a}{\sim}}
\newcommand{\leqasympt}{\stackrel{\scriptscriptstyle }{\preceq}}
\newcommand{\gasympt}{\stackrel{\scriptscriptstyle }{\succ}}

\newcommand{\E}{{\mathbb E}}
\renewcommand{\P}{\text{P}}

\newcommand{\Rnum}{{\mathbb R}}
\newcommand{\Cnum}{{\mathbb C}}

\newcommand{\Nnum}{{\mathbb N}}

\newcommand{\garg}{{x}}

\newcommand{\tinfty}{{\widetilde{\infty}}}
\newcommand{\ltilde}{l^{\widetilde{\infty}}}

\newcommand{\linf}{l^{\infty}}
\newcommand{\ltwo}{l^{2}}

\newcommand{\rtwo}{C_2}
\newcommand{\rinf}{C_\infty}
\newcommand{\rinfsigma}{\kappa_\infty}
\newcommand{\rtwosigma}{\kappa_2}
\newcommand{\rtilde}{C_\tinfty}

\newcommand{\Nzeros}{{\widehat{m}}}

\newcommand{\rratio}{\rho_C}

\newcommand{\Qb}{\mat{Q}}

\newcommand{\nb}{{\mat{n}}}

\newcommand{\rgen}{C}

\newcommand{\deltab}{{\mathbf b}}
\newcommand{\deltanb}{b}

\newcommand{\CN}{{\cal C}{\cal N}}
\newcommand{\N}{{\cal N}}

\newcommand{\NN}{S}

\newcommand{\zb}{{\vec{z}}}

\newcommand{\layerk}{k}

\newcommand{\tdim}{M}
\newcommand{\rdim}{N}

\newcommand{\Hb}{\mat{H}}

\newcommand{\db}{\vec{d}}

\newcommand{\rb}{\vec{r}}
\newcommand{\wb}{\vec{w}}

\renewcommand{\d}{d}
\newcommand{\wvar}{{\sigma^2}}

\newcommand{\argument}{x}

\newcommand{\UBsymdiff}{{B}}
\newcommand{\hamdist}{{\xi}}
\newcommand{\tind}{a}

\newcommand{\Rb}{\mat{R}}
\newcommand{\RbOk}{\Big[ \hspace{-0.1cm}  \begin{array}{c}\Rb_\layerk \\[-0.3cm] \mat{0} \vspace{-0.05cm} \end{array}\hspace{-0.1cm} \Big]}
\newcommand{\RbO}{\Big[ \hspace{-0.1cm}  \begin{array}{c}\Rb \\[-0.3cm] \mat{0} \vspace{-0.05cm} \end{array}\hspace{-0.1cm} \Big]}
\newcommand{\nbknbL}{\Big[ \hspace{-0.1cm} \begin{array}{c}\nb_\layerk \\[-0.3cm] \vspace{0.1cm} \nb_L \vspace{-0.05cm} \end{array}\hspace{-0.1cm} \Big]}

\newcommand{\layer}{m}

\newcommand{\symbolset}{{\mathcal A}}

\newcommand{\yb}{\vec{y}}
\newcommand{\y}{y}

\newcommand{\real}{{\text{R}}}

\newcommand{\imag}{{\text{I}}}

\newcommand{\SNR}{\rho}

\title{Infinity-Norm Sphere-Decoding}

\author{{\large\it Dominik Seethaler and Helmut B\"olcskei}\\[3mm]
\normalsize Communication Technology Laboratory\\[-1.5mm]
ETH Zurich\\[-1.5mm]
8092 Zurich, Switzerland\\[-1.5mm]
\{seethal, boelcskei\}@nari.ee.ethz.ch\thanks{This work was supported in part by the STREP project No. IST-026905 (MASCOT) within the Sixth Framework Programme of the European Commission. Part of this work was performed while D.\ Seethaler was with the Institute of Communications and Radio-Frequency Engineering, Vienna University of Technology.
This paper was presented in part at IEEE ISIT 2008, Toronto, ON, Canada,  July 2008.}\vspace*{-2mm}}

\begin{document}

\maketitle

\begin{abstract}
The most promising approaches for efficient detection in multiple-input multiple-output (MIMO) wireless systems are based on 
sphere-decoding (SD). The conventional (and optimum) norm that is used to conduct the tree traversal step in SD is the $\ltwo$-norm. 
It was, however, recently observed that  using the $\linf$-norm instead reduces the hardware complexity of SD considerably at only a marginal performance loss. 
These savings result from a reduction in the length of the critical path in the circuit and the silicon area required for metric computation, but are also, as observed previously through simulation results, a consequence of a reduction in the 
 {\em computational (i.e., algorithmic) complexity}.  The aim of this paper is an analytical performance and computational complexity analysis of $\linf$-norm SD.  For i.i.d.\ Rayleigh fading MIMO channels, we 
 show that $\linf$-norm SD achieves full diversity order with an asymptotic SNR gap, compared to $\ltwo$-norm SD, that increases at most linearly in the number of receive antennas. Moreover, 
 we provide a closed-form expression for the  computational complexity of $\linf$-norm SD based on which we establish that its complexity scales exponentially in the system size. Finally, we characterize the tree pruning behavior of $\linf$-norm SD and show that it behaves fundamentally different from that of $\ltwo$-norm SD. 
 \end{abstract}


\section{Introduction}\label{sec.introduction}
Multiple-input multiple-output (MIMO) wireless systems offer considerable gains over single-antenna systems, in terms of throughput and link 
reliability, see, e.g., \cite{rohit03}. These gains come, however, at a significant increase in receiver complexity.  
In particular, one of the most challenging problems in MIMO receiver design is the development of hardware-efficient data detection algorithms achieving (close-to) optimum performance \cite{burg_phd}.
Among the most promising approaches to the solution of this problem is the so-called sphere-decoding (SD) algorithm  \cite{fincke_phost85,viterbo93,agrell_it02,damen03,burg05_vlsi,studer08}, which performs  
optimum, i.e., maximum-likelihood (ML), detection through a weighted tree search. SD exhibits (often significantly) smaller computational complexity than exhaustive search 
ML detection \cite{burg_phd,hass_sp03_part_i}.

\subsection{Hardware Implementation Aspects of SD}\label{sec.intro.hardware}

Hardware implementations of several variants of the SD algorithm are described in \cite{burg_phd,burg05_vlsi}.  
It is argued in 
 \cite{burg05_vlsi} that the overall hardware complexity of a SD  is essentially determined 
by (i) the {\em computational (i.e., algorithmic) complexity} in terms of the number of nodes visited in the tree search and  (ii) the {\em circuit complexity} in terms of the length of the critical path in the circuit and the required silicon area for metric computation. The length of the critical path limits the clock frequency of the circuit \cite{kaeslin08}. One of the main findings 
of \cite{burg05_vlsi} is that replacing the $\ltwo$-norm in the ML detector by the $\linf$-norm and hence traversing the search tree based on the $\linf$-metric incurs only a small performance loss while significantly reducing the overall hardware complexity of SD by virtue of a reduction of both the computational  and the circuit complexity. 

\begin{figure}[t]
\begin{center}
\includegraphics[height=48mm]{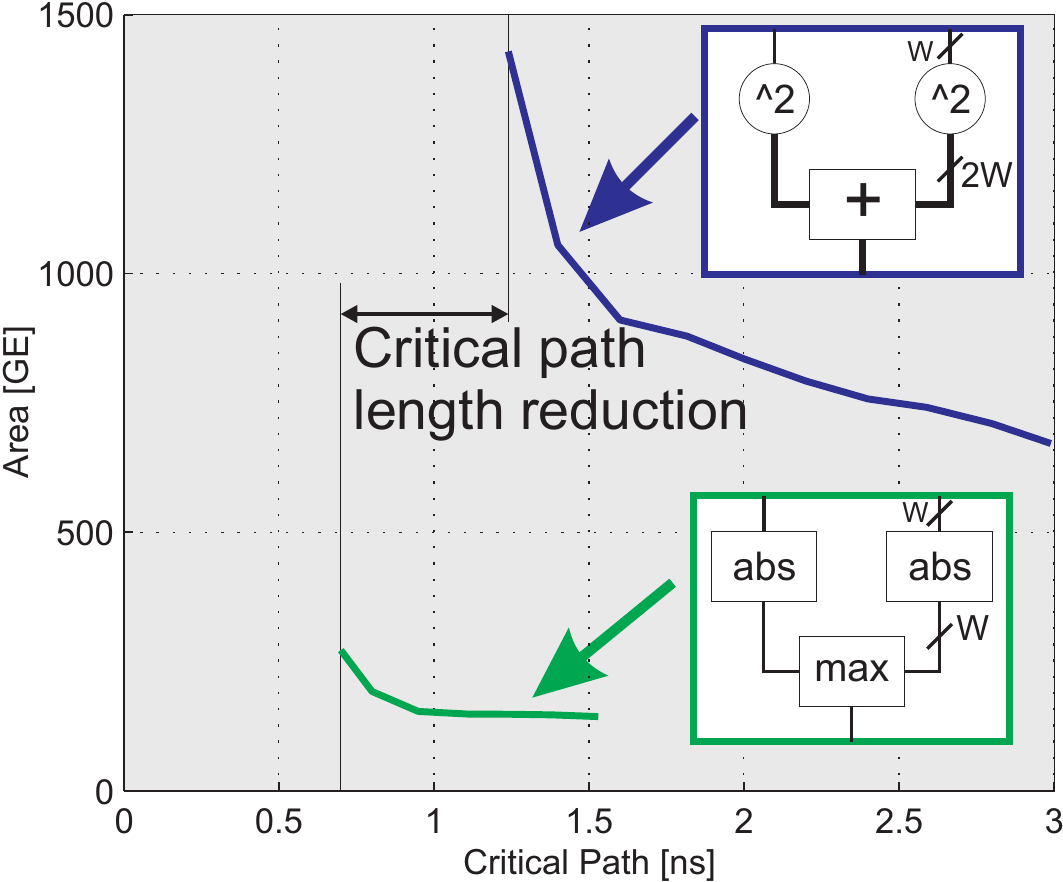}
\vspace{-0.25cm}
\caption{Circuit area and critical path length tradeoff curves corresponding to the computation of  $x_{1}^2+x_{2}^2$ (squared $\ltwo$-norm) and 
$\text{max}\{|x_{1}|,|x_{2}|\}$ ($\linf$-norm) for $x_{1}, x_{2} \in \Rnum$. The area is given in gate-equivalents (GE) and the length of the critical path is given in nano seconds (ns). 
 {\sf W} denotes the word length.}
\vspace{-0.6cm}
\label{fig.area-delay}
\end{center}
\end{figure}

To understand where the reduction in  circuit complexity  comes from, we refer to Fig.\  \ref{fig.area-delay}  (cf., \cite[Fig.\ 2]{burg05_vlsi})
showing tradeoff curves between circuit area and the length of the critical path corresponding to the computation of the metrics $x_{1}^2+x_{2}^2$  
(squared $\ltwo$-norm)  and $\text{max}\{|x_{1}|,|x_{2}|\}$ ($\linf$-norm) for $x_{1}, x_{2} \in \Rnum$.  
These tradeoffs can be achieved by choosing different hardware implementations of the corresponding metric computation circuit.  
From Fig.\  \ref{fig.area-delay} it can be seen that the computation of  $\text{max}\{|x_{1}|,|x_{2}|\}$ can be implemented much more efficiently in hardware than the computation of $x_{1}^2+x_{2}^2$. The main reason for this is that evaluating 
$\text{max}\{|x_{1}|,|x_{2}|\}$, in contrast to $x_{1}^2+x_{2}^2$, does not require squaring operations.  
Replacing the $\ltwo$- by the $\linf$-norm also has an impact on the computational complexity of SD. In particular,  it was observed in \cite{burg05_vlsi}, through simulation results, that SD based on the $\linf$-norm (referred to as SD-$\linf$) exhibits lower computational complexity than SD based on the $\ltwo$-norm (referred to as SD-$\ltwo$). Furthermore, the results in  \cite{burg05_vlsi} indicate  that the overall complexity (determined by both the circuit and the computational complexity)
of SD-$\linf$ is up to a factor of 5 lower than the overall complexity of SD-$\ltwo$. SD-$\linf$ therefore appears to be a promising approach to near-optimum MIMO detection at low hardware complexity.

\subsection{Contributions}
The aim of this paper is to deepen the understanding of SD-$\linf$ through an analytical performance and computational complexity analysis for i.i.d.\ Rayleigh fading MIMO channels. 
Our main contributions can be summarized as follows:  
\begin{itemize}
\item We show that SD-$\linf$ achieves the same (i.e., full) diversity order as SD-$\ltwo$.  
\item We show that the gap in signal-to-noise ratio (SNR) incurred by SD-$\linf$, compared to SD-$\ltwo$, increases at most linearly in the number of receive antennas.   
\item We derive a closed-form expression for the complexity of SD-$\linf$. Here and in the remainder of the paper, 
	complexity is defined as the average number of nodes visited in the tree search, where averaging is performed with respect to 
	 the (random) channel, noise, and transmit signal. Corresponding results for SD-$\ltwo$ can be found in \cite{hass_sp03_part_i,hass_sp03_part_ii,murugan06,Gowaikar_07}. 
\item We prove that the complexity of SD-$\linf$ scales 
	 exponentially in the number of transmit antennas. Our proof technique directly extends to SD-$\ltwo$ and thus yields an alternative (vis-$\grave{\text{a}}$-vis \cite{jalden_tsp05}) proof of the 
	 exponential complexity scaling behavior of SD-$\ltwo$. 
\item Finally, we provide insights into the tree pruning behavior of SD-$\linf$ relative to that of SD-$\ltwo$. In particular, based on an asymptotic (in SNR) analysis of 
	our closed-form complexity expressions, we show that SD-$\linf$ tends to prune more aggressively than SD-$\ltwo$ at tree levels closer to the root of the search tree, whereas this behavior is reversed at tree levels closer to the leaves. 
\end{itemize}

\subsection{Outline}

The paper is organized as follows. After introducing the system model and briefly reviewing relevant aspects of SD-$\ltwo$ and SD-$\linf$ in the remainder of Section \ref{sec.introduction}, we analyze the error probability behavior of SD-$\linf$ in terms of diversity order and SNR gap in Section  \ref{sec.error.performance}. In Section \ref{sec.expected.complexity}, we derive a closed-form expression for the complexity  of SD-$\linf$. This result is then used to establish the exponential complexity scaling 
behavior (in the number of transmit antennas)  of SD-$\linf$  and to analyze, in Section \ref{sec.pruning.behavior}, 
 the tree pruning behavior of SD-$\linf$ by means of an asymptotic  (in SNR) analysis.  In Section \ref{sec.ltilde.SD}, we report modifications of the results presented in Section  \ref{sec.error.performance} and Section  
 \ref{sec.expected.complexity}  to account for the slightly modified metric used in the hardware implementations reported in  \cite{burg05_vlsi}. Numerical results are provided in Section \ref{sec.numerical.results}. Section \ref{sec.conclusions} concludes the paper.  

\subsection{Notation}\label{sec.notation}
We write $A_{i,j}$ for the entry in the $i$th row and $j$th column of the matrix $\vec{A}$ and $x_i$ for the $i$th entry of the vector $\vec{x}$. For unitary $\vec{A}$, we have $\vec{A}^{\!H}  \vec{A} = \vec{A}  \vec{A}^{\!H} = \vec{I}$, where $^H$ denotes conjugate transposition, i.e., transposition $^T$ followed by element-wise complex conjugation $^*$, and $\vec{I}$ is the identity matrix. 
The $\ltwo$- and the $\linf$-norm of a vector $\vec{x} = (x_1 \cdots x_{\tdim})^T \in {\Cnum}^M$ are defined as 
$\|\vec{x}\|_2 =  \sqrt{|x_1|^2+\cdots+|x_M|^2}$ and $\|\vec{x}\|_\infty  = \text{max}\big\{|x_1|,\dots,|x_M|\big\}$, respectively. 
We will also need the $\ltilde$-norm $\|\vec{x}\|_{\tinfty} = \text{max}\big\{|x_{\text{R},1}|, |x_{\text{I},1}|, \dots, |x_{\text{I},M}|\big\}$, 
where $x_{\text{R}}$ and $x_{\text{I}}$ denote the real and imaginary parts, respectively, of $x \in \Cnum$. 
We note that the $\ltwo$-norm is invariant with respect to (w.r.t.) unitary transformations, 
i.e., $\|\vec{x}\|_2 = \|\vec{A}\vec{x}\|_2$ if $\vec{A}$ is unitary.
$\E\{\cdot\}$ stands for the expectation operator and $\Phi_{x}(s) = \E\!\left\{e^{s x}\right\}$ refers to the moment generating function (MGF) of the random variable (RV) $x$. We write $x \sim \chi_a$ if the RV $x$ is $\chi$-distributed with $a \geq 0$ degrees of freedom and  normalized such that 
$\E\!\left\{x^{2}\right\} = a$. The probability density function (pdf) of the RV $x \sim \chi_a$ is then  given by \cite{papoulis91}
\begin{equation}\label{eqn.chi}
f_{x}(t) = \frac{2^{1-a/2} }{\Gamma(a/2)} t^{a-1} e^{-\frac{t^2}{2}}, \quad t \geq 0 
\end{equation}
and $f_{x}(t)=0$, $t <0$, where $\Gamma(a) = \int_0^{\infty} y^{a-1} e^{-y} dy$ refers to the Gamma function. For the corresponding cumulative distribution function (cdf) we have
$P\big[x \leq t\big] = \gamma_{a/2} (t^2/2)$. Here, 
$\gamma_{a}(t)$ 
denotes the (regularized) lower incomplete Gamma function; some important properties of $\gamma_{a}(t)$ are stated in Appendix \ref{app.incomplete.gamma.function}. 
We denote a circularly symmetric complex Gaussian RV with variance $\sigma_{x}^2$ as $x \sim  \CN(0,\sigma_{x}^2)$;  
 $x \sim  \N(\mu_{x},\sigma_{x}^2)$ refers to a real-valued Gaussian distributed RV $x$ with mean $\mu_{x}$ and variance $\sigma_{x}^2$.
 For independently and identically distributed (i.i.d.) RVs $x_{i} \sim \N(0,1)$, $i=1, \dots, a$, we have  $z = \sqrt{x_{1}^2+\cdots+ x_{a}^2} \sim \chi_{a}$. Furthermore, if the RV $x$ is $\chi_{a}$-distributed, $x^2$ is $\chi^2_{a}$-distributed. We write $y \sim \chi^2_{a}$ if the RV $y$ is 
$\chi^2_{a}$-distributed with $\E\!\left\{y\right\} = a$.  In particular, the MGF of the RV $y \sim \chi^2_{a}$ is given by
\begin{equation}\label{eqn.chi2.mgf}
	\Phi_{y}(s) = (1\!-\!2s)^{-a/2}
\end{equation}
for any $s < 1/2$.  The Q-function is defined as $Q(x) = \big(1/\sqrt{2\pi}\,\big) \!\int_{x}^{\infty} e^{-y^2/2} dy$, for $x \geq 0$. 
For equality in distribution we write $\equaldist$. 
Furthermore, the ``Big O'' notation $g(x) = {\cal O}(f(x))$, $x \rightarrow x_{0}$, denotes that $|g(x)/f(x)|$ remains bounded as $x \rightarrow x_{0}$ \cite{Knuth_76}. 
The ``little o'' notation $g(x) = o(f(x))$, $x \rightarrow x_{0}$, stands for  $\text{lim}_{x \rightarrow x_{0}} \,g(x)/f(x) = 0$, and  $g(x) \equalasympt f(x)$, $x \rightarrow x_{0}$,  
means  that $\text{lim}_{x \rightarrow x_{0}} \,g(x)/f(x) = 1$. By $g(x) \leqasympt f(x)$, $x \rightarrow x_{0}$, and $g(x) \gasympt f(x)$, $x \rightarrow x_{0}$,
for positive functions $g(x)$ and $f(x)$, we denote 
$\text{lim}_{x \rightarrow x_{0}} \,g(x)/f(x) \leq 1$ and $\text{lim}_{x \rightarrow x_{0}} \,g(x)/f(x) > 1$, respectively.  The  Dirac delta function is referred to as $\delta(x)$, convolution is denoted as $\ast$, and the natural logarithm to the base $e$ is referred to as $\text{log}(\cdot)$. 
The summations in $\sum_{\mathbf x}$ and  $\sum_{{\mathbf x }\neq {\mathbf x'}}$ are over all possible values of $\mathbf x$ and over all possible values of $\mathbf x$ except for ${\mathbf x'}$, respectively.
Finally, $f^{(n)}(x)$ refers to the $n$th derivative of the function $f(x)$ and $f'(x) = f^{(1)}(x)$. 

\subsection{System Model}\label{sec.system.model}
We consider an $\rdim \times \tdim$ MIMO system with $\tdim$ transmit antennas and $\rdim \geq \tdim$
receive antennas. The corresponding complex-baseband input-output relation is given by 
\[
    \rb \,=\, \Hb\db' + \wb 
\]
where $\db' =  (\d_1' \cdots\, \d_{\tdim}')^T$ denotes the transmitted data vector, $\Hb$ is the 
$\rdim \times \tdim$ channel matrix,  $\rb = (r_1 \cdots\, r_{\rdim})^T$ is the received vector, and $\wb = (w_1 \cdots\, w_\rdim)^T\!\!$ denotes the additive noise vector. 
The symbols $\d_\layer'$, drawn from a finite alphabet $\symbolset$, have  
zero-mean and unit variance. 
Furthermore, we assume that the $H_{n,m}$ are i.i.d.\ 
$\CN(0,1/\tdim)$ and the $w_n$ are i.i.d.\  $\CN(0,\wvar)$.
 The SNR (per receive antenna)  is therefore given by  $\SNR = 1/\wvar$.
 
\subsection{Sphere-Decoding}  
We now briefly review SD based on the $\ltwo$-norm \cite{fincke_phost85,viterbo93,agrell_it02,damen03,studer08} and (suboptimum) SD based on 
the $\linf$-norm \cite{burg05_vlsi}. 

\subsubsection{SD based on the $\ltwo$-norm}
SD-$\ltwo$ performs ML detection by finding   
\begin{equation}\label{eqn.ml}
\widehat{\db}_\text{ML} \,=\, \underset{\db \in \symbolset^\tdim} { \mbox{arg min}}\,
\|\rb-\Hb\db\|^2_2
\end{equation}
 through a  tree search subject to a {\em sphere constraint} (SC), which amounts to  considering only those data vectors $\db$ that satisfy $\|\rb - \Hb\db\|^2_2 \leq \rtwo^2$ (known as  
the Fincke-Pohst \cite{fincke_phost85}  strategy).  
 Here, the radius $\rtwo$ has to be chosen sufficiently large for the corresponding search sphere to contain at 
 least one data vector.  Note, however, that if $\rtwo$ is chosen too large, too many points will satisfy the SC and the complexity of SD-$\ltwo$ will be high 
 (for guidelines on how to choose $\rtwo$ see  \cite{hass_sp03_part_i,hochbrink03} and Section \ref{eqn.choice.of.radii}). 
 The SC is then cast into a weighted tree search problem by first performing a QR-decomposition of $\Hb$ resulting in
\[
	\Hb = \Qb \RbO
\]
where $\Qb$ is an $\rdim\! \times \!\rdim$ unitary matrix, $\Rb$ is an $\tdim\! \times\! \tdim$ upper triangular matrix, and $\mat{0}$ denotes an all-zeros matrix 
of size $(\rdim \! -\! \tdim)\! \times\! \tdim$. 
Then, the SC can equivalently be written as
\begin{equation}\label{eqn.SC.qr}
    \|\zb(\db)\hspace{-0.04cm}\|^2_2 \leq \rtwo^2 
\end{equation}
where 
\begin{equation}\label{eqn.metric.2}
\zb(\db) = \yb -   \RbO \db \quad \text{with}\quad 
\yb =  \Qb^H \rb =   
 \RbO  \db' +  \nb. 
 \end{equation}
Here,  the unitarity of $\Qb$ implies that $\nb = \Qb^H \wb$ is again i.i.d.\ $\CN(0,\wvar)$.
The data subvectors $\db_\layerk \in \symbolset^\layerk$ of length $\layerk$ 
\[
\db_\layerk =  (\d_{\tdim-\layerk+1} \cdots\, \d_{\tdim})^T, \quad \layerk = 1, \dots, \tdim,
\] 
can be arranged in a tree with root above level $\layerk = 1$ and corresponding 
leaves at level $\layerk = \tdim$; a specific $\db_\layerk$ is associated with a node in this tree at level $\layerk$. 
Let us define
\[
\zb_{\layerk}(\db_{\layerk}) =  \yb_{\layerk} -   \RbOk \db_{\layerk}
\]
as the vector containing the bottom $\layerk+L$ with $L = \rdim-\tdim$ elements of $\zb(\db)$ in \eqn{eqn.metric.2}. Here, \sloppy
$\Rb_\layerk$ denotes the  $\layerk \times \layerk$ upper triangular submatrix of $\Rb$ associated with $\db_\layerk$ and 
$\yb_\layerk =  (\y_{\tdim-\layerk+1} \cdots\, \y_{\tdim}\,\, \y_{\tdim+1} \cdots\, \y_{\rdim})^T$.  
 The metric $\|\zb(\db)\hspace{-0.04cm}\|^2_2 = \|\zb_{\tdim}(\db_{\tdim})\hspace{-0.04cm}\|^2_2$ can then be computed recursively according to 
 \begin{equation}\label{eqn.recursive.metric.computation}
\|\zb_{\layerk}(\db_{\layerk})\hspace{-0.04cm}\|^2_2 = \|\zb_{\layerk-1}(\db_{\layerk-1})\hspace{-0.04cm}\|^2_2 + \big|[\zb(\db)]_{M-\layerk+1}\big|^2, \quad 
\layerk = 1,\dots,\tdim, 
\end{equation}
where
\begin{equation}\label{eqn.metric.inc} 
 \big|[\zb(\db)]_{M-\layerk+1}\big|^2 = \left| y_{\tdim-\layerk+1} -\!\!\! \sum_{i=\tdim-\layerk+1}^\tdim \!\!\!R_{\tdim-\layerk+1,i} \,d_i \right|^2.
\end{equation}
Thus, with \eqref{eqn.recursive.metric.computation}, a necessary condition for $\db$ to satisfy the SC \eqref{eqn.SC.qr} is that any associated $\db_\layerk$ satisfies the {\em partial SC} (PSC) 
\begin{equation}\label{eqn.PSC}
\|\zb_\layerk(\db_\layerk)\hspace{-0.04cm}\|^2_2 \leq \rtwo^2.  
\end{equation}
Consequently, we can find all data vectors inside the search sphere, i.e., all data vectors satisfying the SC \eqref{eqn.SC.qr}, through a weighted tree search. 
The tree is traversed starting at level $\layerk = 1$.  
If the PSC is violated by a given $\db_\layerk$, the node associated with that $\db_\layerk$ along with 
all its children is pruned from the tree. 
The ML solution  \eqref{eqn.ml} is found by choosing, among all surviving leaf nodes $\db = \db_\tdim$, the one with minimum $\|\zb(\db)\hspace{-0.04cm}\|_{2}$.
\subsubsection{SD based on the $\linf$-norm}\label{sec.SD.linf}
We define SD-$\linf$ as the algorithm obtained by replacing the 
SC  \eqref{eqn.SC.qr}  by the {\em box constraint} (BC) $\|\zb(\db)\hspace{-0.04cm}\|_\infty \leq \rinf$.  
The metric  $\|\zb(\db)\hspace{-0.04cm}\|_\infty$ can be computed recursively according to 
$\|\zb_{\layerk}(\db_{\layerk})\hspace{-0.04cm}\|_\infty = \text{max}\big\{\|\zb_{\layerk-1}(\db_{\layerk-1})\hspace{-0.04cm}\|_{\infty},\,  \big|[\zb(\db)]_{M-\layerk+1}\big|\big\}$. 
Consequently, the PSC is replaced by the {\em partial box constraint} (PBC) 
\begin{equation}\label{eqn.PBC}
    \|\zb_\layerk(\db_\layerk)\hspace{-0.04cm}\|_\infty \leq \rinf.
\end{equation}
If the PBC is violated by a given $\db_\layerk$, the node associated with that $\db_\layerk$ along with 
all its children is pruned from the tree. The $\linf$-optimal solution is obtained by choosing, among all surviving leaf nodes $\db = \db_\tdim$, the one with minimum $\|\zb(\db)\hspace{-0.04cm}\|_{\infty}$, i.e., 
\begin{equation}\label{eqn.ml.inf}
\widehat{\db}_\infty =\underset{\db \in \symbolset^\tdim} { \mbox{arg\, min}}\,
\|\zb(\db)\hspace{-0.04cm}\|_\infty.
\end{equation} 
Slightly abusing terminology, we call the side length $\rinf$ of the search box the ``{\em radius}'' associated with SD-$\linf$. 
Like in the SD-$\ltwo$ case with $\rtwo$, here the radius $\rinf$ has to be chosen large enough to ensure that at least one data vector is found by the algorithm. 
Again, however, choosing $\rinf$ too large will in general result in a high complexity of SD-$\linf$ (for guidelines on how to choose $\rinf$ we refer to Section \ref{eqn.choice.of.radii}). 

We emphasize the following aspects of SD-$\linf$:  

\begin{itemize}
\item The SD-$\linf$ hardware implementation reported in  \cite{burg05_vlsi} is actually based on the $\ltilde$-norm 
$\|\vec{x}\|_{\tinfty} = \text{max}\big\{|x_{\text{R},1}|, |x_{\text{I},1}|, \dots, |x_{\text{I},M}|\big\}$ rather than the $\linf$-norm 
 $\|\vec{x}\|_\infty  = \text{max}\big\{|x_1|,\dots,|x_M|\big\}$, ${\vec x} \in \Cnum^{\tdim}$. 
Here, the essential aspect is that the computation of the $\ltilde$-norm, as opposed to the $\linf$- and the $\ltwo$-norm, does not require squaring operations, which, as already noted in 
Section \ref{sec.intro.hardware}, results in significantly smaller circuit complexity. 
 Nevertheless, in the following, for the sake of simplicity of exposition, we first analyze SD-$\linf$, i.e., SD based on the conventional $\linf$-norm, thereby revealing the
 fundamental aspects (w.r.t.\ performance and complexity) of SD using the $\ltilde$-norm (referred to as SD-$\ltilde$). 
 The modifications of the results on SD-$\linf$ needed to account for the use of the $\ltilde$-norm are described in Section \ref{sec.ltilde.SD}. 

\item The tree search strategy underlying SD-$\linf$ is identical to that of {\em Kannan's strategy} (see, e.g., \cite{agrell_it02,banih_98}), which also finds all data vectors inside a hypercube. 
The difference between SD-$\linf$ and Kannan's algorithm lies in calculating the final detection result. SD-$\linf$ implements \eqref{eqn.ml.inf} while Kannan's approach is optimum as it implements \eqref{eqn.ml}. 
Optimality of Kannan's algorithm is achieved through (i) guaranteeing that the solution of \eqref{eqn.ml} is contained inside the search hypercube  (which, in general, necessitates choosing the search radius to be larger than the corresponding radius for SD-$\linf$ and hence incurs a higher complexity) and (ii) in the last step comparing all found data vectors with respect to their $\ltwo$-distance $\|\rb-\Hb\db\|_{2}$ 
(which, in contrast to SD-$\ltilde$, necessitates squaring operations). 

\item Finally, 
we emphasize that SD-$\linf$ as defined above does {\em not} correspond to $\linf$-norm decoding on the ``full'' channel matrix $\Hb$ 
according to 
\begin{equation}\label{eqn.ml.inf.full}
\widehat{\db}_{\infty,\text{full}} \,=\, \underset{\db \in \symbolset^\tdim} { \mbox{arg\, min}}\,
\|\rb-\Hb\db\|_{\infty}
\end{equation}
since  $\|\rb\!-\!\Hb\db\|_{\infty} \neq  \|\zb(\db)\hspace{-0.04cm}\|_{\infty}$, in general. This statement also holds true for SD based on the $\ltilde$-norm. In the $\ltwo$-norm case detection on the full channel matrix $\Hb$ is equivalent to detection on the upper triangular matrix $\Rb$. 

\end{itemize}
\section{Error Probability of SD-$\linf$}\label{sec.error.performance}
In this section, we show that SD-$\linf$ achieves the same diversity order as ML (i.e., SD-$\ltwo$) detection and we quantify the SNR loss incurred by SD-$\linf$. 

\subsection{Distance Properties}\label{sec.quality.of.SDlinf.solution}
We start by investigating distance properties of the SD-$\linf$ solution $\widehat{\db}_\infty$.  
Using the bounds 
\begin{equation}\label{eqn.linf.ltwo.bounds}
	\frac{1}{N} \| \vec{x} \|_2^2 \leq \| \vec{x} \|_\infty^2 \leq \| \vec{x} \|_2^2, 
\end{equation}
valid for any vector $\vec{x} \in {\Cnum}^N$,  we obtain
\begin{align}
	\big\|\rb-\Hb\, \widehat{\db}_\infty\big\|^2_2 & =  \Big\| \yb\! - \! \RbO \widehat{\db}_\infty \Big\|_2^2 
	\leq  \rdim \Big\| \yb\! -\! \RbO\widehat{\db}_\infty \Big\|_\infty^2 \nonumber \\
 	& \leq  \rdim \Big\| \yb\! - \!\RbO\widehat{\db}_\text{ML} \Big\|_\infty^2 
	 \leq   \rdim \Big\| \yb\! -\! \RbO\widehat{\db}_\text{ML} \Big\|_2^2  \nonumber
	\\ & =  \rdim \big\|\rb - \Hb \, \widehat{\db}_\text{ML}\big\|^2_2. \label{eqn.distance.property}
\end{align}
We are therefore guaranteed that $\big\|\rb-\Hb\, \widehat{\db}_\infty\big\|_2$ lies within a factor of $\sqrt{N}$ of the minimum distance $\big\|\rb-\Hb\, \widehat{\db}_\text{ML}\big\|_2$ realized by 
the ML detector \eqref{eqn.ml}. 
Trivially, SD-$\linf$ is optimum for $\rdim =1$ (simply because the $\linf$-norm equals the $\ltwo$-norm in this case). For increasing $\rdim$,   \eqref{eqn.distance.property} suggests 
an increasing performance loss incurred by 
SD-$\linf$ when compared to the ML detector (i.e., SD-$\ltwo$). In the next section, we quantify this performance loss in terms of diversity order and SNR gap. 
We note that for Babai's nearest plane algorithm  \cite{babai86} (which can be interpreted as a decision-feedback detector in combination with LLL lattice reduction, see, e.g., \cite{agrell_it02}), we 
get a result that is structurally similar to \eqref{eqn.distance.property} when $\widehat{\db}_\infty$ is replaced by Babai's detection result and the factor $N$ is replaced by $2^{(N-1)}$. Consequently, the performance loss incurred by Babai's nearest plane algorithm can be expected to be significantly larger than that incurred by SD-$\linf$. 

\subsection{Diversity Order and SNR Gap}\label{sec.PEP}
We denote the error probability as a function of SNR $\SNR$ as $\P(\SNR)$. In the following, we will only encounter error probabilities of the form 
$\P(\SNR)= (K \SNR)^{-\delta} + o(\SNR^{-\delta})$, $\SNR \rightarrow \infty$, with some constant $K >0$ not depending on $\rho$. 
We can define the corresponding 
SNR exponent $\delta$ as \cite{tarokh_it98,zheng02}
\begin{equation}\label{eqn.def.snr.exponent}
\delta = - \underset{\SNR \rightarrow\, \infty} {\lim}\, \frac{\text{log}\, \P(\SNR)}{\text{log}\,\SNR}. 
\end{equation}
Furthermore, if $\P_{{1}}(\SNR) $ and $\P_{{2}}(\SNR)$  have the same SNR exponent, we can define an asymptotic SNR gap $\alpha$ via $\P_{{1}}(\SNR) \equalasympt  \P_{{2}}(\alpha\, \SNR)$, $\SNR \rightarrow \infty$.
For example, if $\P_{{1}}(\SNR)= (K_{1}\,\SNR)^{-\delta}+ o(\SNR^{-\delta})$ and  $\P_{{2}}(\SNR) = (K_{2}\,\SNR)^{-\delta}+ o(\SNR^{-\delta})$,  we have $\alpha = K_{1}/K_{2}$. 
Our analysis corresponds to multiplexing gain $r=0$ in the framework of \cite{zheng02}. The corresponding results bear practical significance as it can be shown, for example, that 
 even for $r=0$ conventional suboptimum detection schemes like linear equalization-based or V-BLAST  detectors are unable to achieve the full diversity order of $\rdim$ \cite{loyka04,zheng02,rohit03}.
In the following, we first focus on the 
behavior of the pairwise error probability (PEP) and then analyze the total error probability.

\subsubsection{Pairwise Error Probability} \label{sec.PEP}
Assume that the data vector $\db'$ was transmitted. The probability of erroneously deciding in 
favor of another data vector $\db \neq \db'$  is denoted as $\P_{\db' \rightarrow  \db,\text{ML}}(\SNR)$ in the  SD-$\ltwo$ case and $\P_{\db' \rightarrow  \db,\infty}(\SNR)$ in the SD-$\linf$ case.  
To derive (an upper bound on) $\P_{\db' \rightarrow  \db,\infty}(\SNR)$, we first present a somewhat unconventional approach for upper-bounding 
$\P_{\db' \rightarrow  \db,\text{ML}}(\SNR)$, which lends itself nicely to an extension to the $\linf$-case.  
We start from
\begin{align*}
 \P_{\db' \rightarrow  \db, \text{ML}}(\SNR)
 & = \text{P}\Big[ \| \rb - \Hb \db \|_2  \leq \| \rb - \Hb \db' \|_2  \Big] \nonumber \\
& =  \text{P}\Big[ \| \Hb \deltab  + \wb \|_2 \leq  \|\wb\|_2 \Big]  \label{eqn.PEP.ml}
\end{align*}
with the error (difference) vector  $\deltab =  \db'-\db$.  Applying the inverse triangle inequality 
according to $\| \Hb \deltab  + \wb \|_2 \geq \big|\| \Hb \deltab \|_2   -  \| \wb \|_2 \big|$,  we further obtain
\begin{equation}\label{eqn.PEP.ml.UB1}
\P_{\db' \rightarrow  \db, \text{ML}}(\SNR) \leq \text{P}\bigg[ \|\wb\|_2 \geq  \frac{1}{2}  \| \Hb\deltab\|_2  \bigg]
\end{equation}
noting that $|x| \geq x$, for all $x \in \Rnum$. 
With $ \frac{\sqrt{2}} {\sigma} \|\wb\|_2 \sim \chi_{2 \rdim}$, conditioning on $\Hb$, and applying the Chernoff upper bound 
 yields
\[
\text{P}\bigg[ \|\wb\|_2 \geq  \frac{1}{2}  \| \Hb \deltab\|_2\,\Big|\, \Hb \bigg] \leq   \Phi_{\chi_{2\rdim}^2}\! (s) \,e^{-s\SNR \frac{\| \Hb \deltab\|_2^2}{2} }
\]
for $s\in [0, 1/2)$. Here, $\Phi_{\chi_{2\rdim}^2}\! (s)$ denotes the MGF of a $\chi^2_{2\rdim}$-distributed RV (see \eqref{eqn.chi2.mgf}).  
Averaging over $\Hb$ then results in 
\begin{align}
\P_{\db' \rightarrow  \db,\text{ML}}(\SNR) &  \leq  \Phi_{\chi_{2\rdim}^2}\! (s) \,   \E_\Hb \bigg\{ \!e^{-s\SNR \frac{\| \Hb\deltab\|_2^2}{2} } \bigg\}\nonumber \\ 
 & =   \Phi_{\chi_{2\rdim}^2}\! (s)  \bigg(1+s \SNR \frac{\|\deltab\|_{2}^2}{2 \tdim}\bigg)^{\!-\rdim} \label{eqn.PEP.ml.final}
\end{align}
because $2 \tdim \| \Hb\deltab\|_2^2/ \|\deltab\|_2^2  \sim \chi^2_{2\rdim}$ for a given $\deltab$. For high SNR,  the right hand side (RHS) of \eqref{eqn.PEP.ml.final} is minimized for $s = 1/4$, which gives 
\begin{equation}\label{eqn.upper.bound.PEP.min.s}
\P_{\db' \rightarrow  \db,\text{ML}}(\SNR) \leq 2^\rdim \bigg(1+\SNR \frac{\|\deltab\|_{2}^2}{8 \tdim}\bigg)^{\!-\rdim}. 
\end{equation}
Since $\rdim$ is the maximum diversity order that can be achieved over an $\rdim \times \tdim$ MIMO channel with the transmission setup considered in this paper (i.e., spatial multiplexing) \cite{zheng02},   
we can  immediately conclude that the SNR exponent of $\P_{\db' \rightarrow  \db,\text{ML}}(\SNR) $ equals $\rdim$ for any non-zero $\deltab$ (see also the lower bound 
\eqref{eqn.lower.bound.PEP.ML} on $\P_{\db' \rightarrow  \db,\text{ML}}(\SNR)$ having an SNR exponent of $\rdim$ as well).  

For SD-$\linf$ we can follow a similar approach.  Starting with \eqref{eqn.ml.inf}, we get  
\begin{align}
\P_{\db' \rightarrow  \db,\infty}(\SNR) & \leq  \text{P}\Big[  \|\zb(\db)\hspace{-0.04cm}\|_\infty \leq  \|\zb(\db')\hspace{-0.04cm}\|_\infty \Big] \nonumber \\ 
& =  \text{P}\bigg[  \Big\|   \RbO\deltab +  \nb \Big\|_\infty \!\! \leq \|\nb\|_\infty  \bigg]. \label{eqn.PEP.linf.intermediate}
\end{align}
Note that for SD-$\linf$, unlike for SD-$\ltwo$, the event $\|\zb(\db)\|_\infty =  \|\zb(\db')\|_\infty$ can, in general, occur with non-zero probability. 
Declaring an error in this case certainly yields an upper bound on $\P_{\db' \rightarrow  \db,\infty}(\SNR)$.
Next, we apply the upper and lower bounds in \eqref{eqn.linf.ltwo.bounds} and exploit the invariance of the $\ltwo$-norm to unitary transformations to get
\begin{align}
\P_{\db' \rightarrow  \db,\infty}(\SNR) & \leq 
\text{P}\bigg[ \frac{1}{\sqrt{\rdim}} 
\Big\| \RbO \deltab +  \nb \Big\|_2 \!\! \leq \|\nb\|_2  \bigg] \nonumber \\
& =  \text{P}\bigg[  \frac{1}{\sqrt{\rdim}} \| \Hb\deltab  + \wb \|_2  \leq  \|\wb\|_2 \bigg]. \label{eqn.PEP.SDlinf.UB0}
\end{align}
Finally, applying the inverse triangle inequality according to $\| \Hb \deltab  + \wb \|_2 \geq \big|\| \Hb \deltab \|_2   -  \| \wb \|_2\big|$, we have
\begin{equation}\label{eqn.PEP.SDlinf.UB1} 
\P_{\db' \rightarrow  \db,\, \infty}(\SNR)  \leq \text{P}\bigg[ \|\wb\|_2 \geq  \frac{1}{\sqrt{\rdim}\!+\!1}  \| \Hb \deltab\|_2  \bigg].
\end{equation}
Note the structural similarity of \eqref{eqn.PEP.SDlinf.UB1} and \eqref{eqn.PEP.ml.UB1}. Employing the same arguments as in the SD-$\ltwo$ case,  
we get 
\begin{equation*}
\text{P}\bigg[ \|\wb\|_2 \geq   \frac{1}{\sqrt{\rdim}\!+\!1}   \| \Hb \deltab\|_2\,\Big|\, \Hb \bigg] \leq   \Phi_{\chi_{2\rdim}^2}\! (s) \,e^{-s\SNR \frac{2\| \Hb \deltab\|_2^2}{(\sqrt{\rdim}\!+\!1)^2} }
\end{equation*}
for $s\in [0, 1/2)$.
Averaging over $\Hb$ then results in
\begin{equation}\label{eqn.upper.bound.PEP.SDlinf}
 \P_{\db' \rightarrow  \db, \infty}(\SNR) \leq 2^\rdim\! \bigg(\!1+\SNR\, \frac{\|\deltab\|_{2}^2}{2\big(\sqrt{\rdim}\!+\!1\big)^2 \tdim}\!\bigg)^{\!\!-\rdim} 
\!\! =   \text{UB}_{\infty}(\SNR)
\end{equation}
where we used the fact that $s = 1/4$ minimizes the upper bound for high SNR.  As in the SD-$\ltwo$ case for $\P_{\db' \rightarrow  \db,\text{ML}}(\rho)$, we can immediately conclude that the SNR exponent of $\P_{\db' \rightarrow  \db,\infty}(\SNR) $ equals $\rdim$ for any non-zero $\deltab$. There is, however, an SNR gap between  $\P_{\db' \rightarrow  \db, \infty}(\rho)$ and $\P_{\db' \rightarrow  \db,\text{ML}}(\rho)$, which 
can be quantified as follows. We start by evaluating \cite[Eq.\ (20)]{lu_wang_kumar_chugg03} for the case at hand to get
\begin{equation}\label{eqn.lower.bound.PEP.ML}
 \P_{\db' \rightarrow  \db,\text{ML}}(\SNR)  \geq    \frac{1}{2} \frac{1}{4^\rdim} {2 \rdim  \choose \rdim} \! \bigg(\!1 + \SNR\, \frac{\|\deltab\|_{2}^2}{4 \tdim} \bigg)^{\!\!-\rdim} 
 \! =  \text{LB}_{\text{ML}}(\SNR). 
\end{equation}
The asymptotic SNR gap \sloppy between $\text{UB}_{\infty}(\SNR)$ and $\text{LB}_{\text{ML}}(\SNR)$, denoted as $\beta$, i.e., $\text{UB}_{\infty}(\SNR)  \equalasympt   \text{LB}_{\text{ML}}(\SNR/\beta)$, 
$\SNR \rightarrow \infty$,
is directly obtained as 
\begin{equation}\label{eqn.snr.gap.beta}
\beta = 4\, \big(\!\sqrt{\rdim}\!+\!1\big)^2 \, \bigg[\frac{1}{2} {2 \rdim  \choose \rdim} \bigg]^{-\frac{1}{\rdim}}.
\end{equation}
We can thus conclude that the asymptotic SNR gap between the PEP for SD-$\linf$ and the PEP for SD-$\ltwo$ is upper-bounded by $\beta$, or, equivalently, we have 
\begin{equation}\label{eqn.PEP.SNR.gap}
	\P_{\db' \rightarrow  \db,\infty}(\SNR) \leqasympt \P_{\db' \rightarrow  \db,\text{ML}}(\SNR/\beta),  \quad \SNR \rightarrow \infty.
\end{equation}

\subsubsection{Total Error Probability}  \label{sec.overall.error.probability}
In the following, we consider the total error probability 
$\P_{\cal E}(\SNR) =  \P\big[\hspace{0.2mm} \db' \neq \widehat{\db}  \big]$ assuming equally likely transmitted data vectors $\db'$. If not specified, $\P_{\cal E}(\SNR)$ stands for the total error probability 
$\P_{{\cal E}_{\infty}}(\SNR)$ of SD-$\linf$ and  $\P_{{\cal E}_{\text{ML}}}(\SNR)$ of SD-$\ltwo$. 
We start by noting that 
\begin{equation}\label{eqn.total.prob}
\P_{\cal E}(\SNR) = |\symbolset|^{-\tdim} \sum_{ \db'} \P_{{\cal E}|\db'}(\rho). 
\end{equation}
 Here, $\P_{{\cal E}|\db'}(\SNR)$ refers to the total error probability conditioned on $\db'$ being transmitted, which 
 can be bounded as 
\begin{equation}\label{eqn.PE.2.PEP}
	 \P_{\db' \rightarrow  \text{any\,}\db}(\SNR) \leq \P_{{\cal E}|\db'}(\SNR) \leq \sum_{ \db \neq \db'}  \P_{\db' \rightarrow  \db}(\SNR).
\end{equation}
It follows that  
\begin{equation}\label{eqn.PE.2.PEP.overall.UB}
	 \P_{\cal E}(\SNR) \leq |\symbolset|^{-\tdim} \sum_{ \db'} \sum_{ \db \neq \db'}  \P_{\db' \rightarrow  \db}(\SNR)
\end{equation}
and 
\begin{equation}\label{eqn.PE.2.PEP.overall.LB}
	 \P_{\cal E}(\SNR) \geq |\symbolset|^{-\tdim} \sum_{ \db'}  \P_{\db' \rightarrow  \text{any\,}\db}(\SNR).
\end{equation}
As the SNR exponent of $\P_{\db' \rightarrow  \db,\infty}(\SNR)$ equals $\rdim$ for all $\deltab = \db-\db' \neq \vec{0}$ (cf.\ \eqref{eqn.upper.bound.PEP.SDlinf}), we can conclude that 
SD-$\linf$ {\em achieves full diversity order} $\rdim$ and hence the same diversity order as ML detection. 
The corresponding asymptotic SNR gap is obtained as follows. With \eqref{eqn.PEP.SNR.gap}-\eqref{eqn.PE.2.PEP.overall.UB}, 
we get  
\begin{align}
\P_{{\cal E}_{\infty}}(\SNR) & \leqasympt |\symbolset|^{-\tdim} \sum_{ \db'} \sum_{ \db \neq \db'} \P_{\db' \rightarrow  \db,\infty}(\SNR),  \quad \SNR \rightarrow \infty \nonumber \\
						&  \leqasympt |\symbolset|^{-\tdim} \sum_{ \db'} \sum_{ \db \neq \db'} \P_{\db' \rightarrow  \db,\text{ML}}(\SNR/\beta), \quad \SNR \rightarrow \infty \nonumber \\
						&   \leqasympt  |\symbolset|^{-\tdim} \sum_{ \db'} \sum_{ \db \neq \db'}  \P_{ {\cal E}_{\text{ML}}|\db'}(\SNR/\beta), \quad \SNR \rightarrow \infty \nonumber \\ 
						&   \leqasympt |\symbolset|^{\tdim} \P_{{\cal E}_{\text{ML}}}(\SNR/\beta), \quad \SNR \rightarrow \infty. \label{eqn.asympt.inequality.total.error.prob}
\end{align}
From \eqref{eqn.PE.2.PEP.overall.UB} together with  \eqref{eqn.upper.bound.PEP.min.s} and \eqref{eqn.PE.2.PEP.overall.LB} together with \eqref{eqn.lower.bound.PEP.ML}, we can conclude that 
$\P_{{\cal E}_{\text{ML}}}\!(\SNR)$ has SNR exponent $\rdim$ and can be written as $\P_{{\cal E}_{\text{ML}}}\!(\SNR) =  (K_{\text{ML}}\, \SNR)^{-\rdim}+ o(\SNR^{-\rdim})$, $\rho \rightarrow \infty$, with some constant $K_{\text{ML}} > 0$ that does not depend on $\SNR$. 
With  $\P_{{\cal E}_{\infty}}(\SNR)  \leqasympt  |\symbolset|^{\tdim} \P_{{\cal E}_{\text{ML}}}(\SNR/\beta)$ from \eqref{eqn.asympt.inequality.total.error.prob} and $\rdim \geq \tdim$, this yields 
$\P_{{\cal E}_{\infty}}(\SNR) \leqasympt  \P_{{\cal E}_{\text{ML}}}\!\big(\SNR/(|\symbolset|\beta )\big)$, which establishes that the asymptotic SNR gap incurred by SD-$\linf$ is upper-bounded by $|\symbolset| \beta$ with 
$\beta$ specified in \eqref{eqn.snr.gap.beta}. Furthermore, using ${m  \choose l} \geq \left(\frac{m}{l}\right)^{\!l}$, we have  ${2 \rdim  \choose \rdim} \geq 2^\rdim$, which, when employed in \eqref{eqn.snr.gap.beta}, shows that 
$\beta \leq 4\, \big(\!\sqrt{\rdim}\!+\!1\big)^2  \leq 16\rdim$. Thus, the asymptotic SNR gap between the total error probabilities $\P_{{\cal E}_{\text{ML}}}\!(\SNR)$ and  
$\P_{{\cal E}_{\infty}}(\SNR)$
is upper-bounded by $16|\symbolset|\rdim$. We can therefore conclude that the asymptotic SNR gap 
incurred by SD-$\linf$ scales at most linearly in the number of receive antennas. 
Simulation results (see Section \ref{sec.simu.error.probability}) reveal that the actual SNR gap is much smaller than $16|\symbolset|\rdim$. We finally note that applying \cite[Prop. 1]{jalden_icassp08i} shows that the statement on 
SD-$\linf$ achieving full diversity order for the i.i.d.\ Rayleigh fading case directly extends to more general fading statistics such as spatially correlated Rayleigh or Ricean fading.   

\subsubsection{$\linf$-Norm Decoding on Full Channel Matrix}
As pointed out in Section \ref{sec.SD.linf}, SD-$\linf$ does not correspond to $\linf$-norm decoding on the full channel matrix $\Hb$ according to  \eqref{eqn.ml.inf.full}. 
However, as the PEP of $\linf$-norm decoding on the full channel matrix satisfies 
\begin{equation*}
P_{\db' \rightarrow  \db, \infty, \Hb}(\SNR) 
=  \text{P}\Big[ \| \Hb \deltab  + \wb \|_\infty \leq  \|\wb\|_\infty \Big],
\end{equation*} 
we can apply the upper and lower bounds in \eqref{eqn.linf.ltwo.bounds} to arrive at 
\[
\P_{\db' \rightarrow  \db,\infty,\Hb}(\SNR)  \leq   \text{P}\bigg[  \frac{1}{\sqrt{\rdim}} \| \Hb\deltab  + \wb \|_2  \leq  \|\wb\|_2 \bigg]
\]
which is exactly the same upper bound as that obtained for SD-$\linf$ in \eqref{eqn.PEP.SDlinf.UB0}. We can therefore conclude that $\linf$-norm decoding on the full channel matrix $\Hb$ also achieves full diversity order with an 
asymptotic SNR gap, vis-$\grave{\text{a}}$-vis SD-$\ltwo$, that increases at most linearly in the number of receive antennas. 

\section{Complexity  of SD-$\linf$}\label{sec.expected.complexity}
In this section, we analyze the complexity  of SD-$\linf$ by deriving an analytic expression for the average number of nodes visited in the tree search when pruning according to the PBC \eqref{eqn.PBC} is performed. 
A node $\db_{\layerk}$ is visited if and only if its corresponding PBC \eqref{eqn.PBC} is satisfied.  We consider a fixed choice of $\rinf$
and average w.r.t.\ channel, noise, and transmit signal. 
Based on the analytic complexity expression for SD-$\linf$, it is then shown that the complexity  of SD-$\linf$ scales exponentially in the number of transmit antennas 
$\tdim$.  

\subsection{Basic Approach}
Our methodology  is similar to that adopted in \cite{hass_sp03_part_i} for SD-$\ltwo$. The key difference to the approach in \cite{hass_sp03_part_i} lies in the computation of the partial metric cdfs as 
detailed in Section \ref{sec.partial.metric.distributions}. 

For a given $\rinf$, a simple counting argument yields the number of nodes $\NN_{\infty,\layerk}$ visited at tree level $\layerk$, $\layerk = 1,\dots,\tdim$, as 
\begin{equation}\label{eqn.N.realization}
	\NN_{\infty,\layerk} =  \sum_{\db_\layerk} I(\zb_\layerk(\db_\layerk))
\end{equation}
where 
\[
	 I(\zb_\layerk(\db_\layerk)) = 
	 \begin{cases} 
	 1, & \text{if}\,\,\, \|\zb_\layerk(\db_\layerk)\hspace{-0.04cm}\|_{\infty} \leq \rgen_{\infty} \\[0.0cm] 
	 0, & \text{otherwise}.
	 \end{cases}
\]
We trivially have $\NN_{\infty,\layerk} \leq |\symbolset|^{\layerk}$.  
First, we note that $\E\{  I(\zb_\layerk(\db_\layerk))\} = \P\big[\|\zb_\layerk(\db_\layerk)\hspace{-0.04cm}\|_{\infty} \leq \rinf \big]$, where 
the expectation is w.r.t.\ the channel $\Rb$, noise $\nb$, and data vector $\db'$. 
Consequently, we have   
\begin{equation}\label{eqn.NN.intermediate}
	\E\{\NN_{\infty,\layerk}\} =  \sum_{\db_\layerk} \P\big[\|\zb_\layerk(\db_\layerk)\hspace{-0.04cm}\|_\infty \leq \rinf \big]
\end{equation}
with the total complexity   
\begin{equation}\label{eqn.total.complexity}
\E\{\NN_{\infty}\} = \sum_{\layerk = 1}^{\tdim}\E\{\NN_{\infty,\layerk}\}.
\end{equation}
Next, we condition on the data subvector $\db_\layerk' \in \symbolset^{\layerk}$ and write 
$\P\big[\|\zb_\layerk(\db_\layerk)\hspace{-0.04cm}\|_{\infty} \leq \rinf\,|\, \db_\layerk'\big] = \P\big[ \|\zb_\layerk(\deltab_\layerk)\hspace{-0.04cm}\|_{\infty} \!\leq \!\rinf\big]$ with 
\begin{equation}\label{eqn.part.metric.vec}
	\zb_\layerk(\deltab_\layerk) = \RbOk \deltab_\layerk + \nbknbL
\end{equation}
where $\deltab_\layerk =  \db_\layerk'\!-\!\db_\layerk$ is a pairwise error subvector,   
$\nb_\layerk =  (n_{\tdim-\layerk+1} \cdots\, n_{\tdim})^T$, and $\nb_L =  (n_{\tdim+1} \cdots\, n_{\rdim})^T$. 
We set $\zb(\deltab) =  \zb_{\tdim}(\deltab_{\tdim})$ and note that $\zb_\layerk(\deltab_\layerk) = \big([\zb(\deltab)]_{\tdim-\layerk+1} \cdots\, [\zb(\deltab)]_{\rdim}\big)^T$.
Formally, for a given $\db_{\layerk}'$, 
we will often speak of ``a node'' $\deltab_\layerk$, which, in a one-to-one fashion, refers to the node $\db_{\layerk} = \db_{\layerk}'-\deltab_\layerk$ in the search tree. For example, the node 
$\db_\layerk = \db_\layerk'$ corresponding to the transmitted data subvector $\db_{\layerk}' $ is equivalent to node 
$\deltab_\layerk =\vec{0}$.  If we speak of a node $\deltab_\layerk$ without specifying $\db_{\layerk}'$, the corresponding statements hold for all pairs $\db_{\layerk}', \db_{\layerk} \in {\symbolset}^{\layerk}$ satisfying 
 $\deltab_{\layerk} = \db_{\layerk}'-\db_\layerk$.
It follows that  
\eqref{eqn.NN.intermediate} can be written as 
\begin{align}
	\E\{\NN_{\infty,\layerk}\} & =  \frac{1}{|\symbolset|^\layerk} \sum_{\db_\layerk}  \sum_{\db_\layerk'}  \P\big[\|\zb_\layerk(\db_\layerk)\hspace{-0.04cm}\|_{\infty} \leq \rinf\,|\, \db_\layerk'\big] \nonumber \\
	& = \frac{1}{|\symbolset|^\layerk} \sum_{ \deltab_\layerk} \P\big[ \|\zb_\layerk(\deltab_\layerk)\hspace{-0.04cm}\|_{\infty} \leq \rinf\big]
	\label{eqn.NN.advancedk}
\end{align}
where we assumed equally likely transmitted data subvectors $\db_{\layerk}'$ for all tree levels $\layerk = 1,\dots, \tdim$; this  
holds, e.g., for statistically independent (across the transmit antennas) and equally likely data symbols. The sum in \eqref{eqn.NN.advancedk} is taken over all  
possible combinations of pairwise error subvectors  $\deltab_\layerk$. 

Equivalently, the complexity at the $\layerk$th tree level $\E\{\NN_{2,\layerk}\}$ for SD-$\ltwo$ is given by \eqref{eqn.NN.advancedk} with the $\linf$-norm replaced by the $\ltwo$-norm and $\rinf$ replaced by $\rtwo$, i.e., 
\begin{align}
	\E\{\NN_{2,\layerk}\} =   \frac{1}{|\symbolset|^\layerk} \sum_{ \deltab_\layerk} \P\big[ \|\zb_\layerk(\deltab_\layerk)\hspace{-0.04cm}\|_{2} \leq \rtwo\big]. 
	 \label{eqn.NN2.final.prob}
\end{align}
We finally note that $\P\big[ \|\zb_\layerk(\deltab_\layerk)\hspace{-0.04cm}\|_{\infty} \leq \rinf\big]$ in \eqref{eqn.NN.advancedk} and 
$\P\big[ \|\zb_\layerk(\deltab_\layerk)\hspace{-0.04cm}\|_{2} \leq \rtwo\big]$ in \eqref{eqn.NN2.final.prob} express the probability that node $\deltab_\layerk$ is visited by SD-$\linf$ and SD-$\ltwo$, respectively, 
and are equivalent to the average (w.r.t.\ the channel $\Rb$ and noise $\nb$) number of visits of node $\deltab_\layerk$ by SD-$\linf$ and SD-$\ltwo$, respectively. 
\subsection{Computation of the Partial Metric Cdfs}\label{sec.partial.metric.distributions}
From  \eqn{eqn.NN.advancedk}  and  \eqref{eqn.NN2.final.prob} we can see that the computation of $\E\{\NN_{\infty,\layerk}\}$ and $\E\{\NN_{2,\layerk}\}$ 
requires knowledge of the cdfs of the partial metrics $\|\zb_\layerk(\deltab_\layerk)\hspace{-0.04cm}\|_{\infty}$ and $\|\zb_\layerk(\deltab_\layerk)\hspace{-0.04cm}\|_{2}$, respectively (recall that 
the radii $\rinf$ and $\rtwo$ are assumed to be fixed and independent of the channel, noise, and data realizations). 
An analytic expression for $\P\big[\|\zb_\layerk(\deltab_\layerk)\hspace{-0.04cm}\|_2\leq \rtwo\big]$  was provided in \cite{hass_sp03_part_i,hass_sp03_part_ii}.  More specifically, it is shown in 
 \cite[Lemma 1]{hass_sp03_part_i} that thanks to the invariance of the $\ltwo$-norm w.r.t.\ unitary transformations 
\[
\|\zb_\layerk(\deltab_\layerk)\hspace{-0.04cm}\|_2 \equaldist \Big\|\Hb_{\layerk}\deltab_{\layerk} + \nbknbL \hspace{-0.04cm}\Big\|_2
\]
where the $(\layerk+L)\times \layerk$ matrix $\Hb_{\layerk}$ with $L =  \rdim -\tdim$  has i.i.d.\ $\CN(0,1/\tdim)$ entries. Conditioned on $\deltab_{\layerk}$, the RV  $\Big\|\Hb_{\layerk}\deltab_{\layerk} + \nbknbL \hspace{-0.04cm}\Big\|_2$ is then easily found to be 
$\chi_{\layerk+L}$-distributed, which leads to an expression for  $\P\big[\|\zb_\layerk(\deltab_\layerk)\hspace{-0.04cm}\|_2\leq \rtwo\big]$ in terms of the lower incomplete Gamma function (see also Section
\ref{sec.distribution.l2.metric}). As the $\linf$-norm is not invariant w.r.t.\ unitary transformations, 
 this approach does not carry over to the $\linf$-case considered here. 
Instead, we follow a direct approach as detailed below. 

\subsubsection{Cdf of $\|\zb_\layerk(\deltab_\layerk)\hspace{-0.04cm}\|_\infty$}\label{sec.partial.metric.distributions.overall}
Since the nonzero entries in $\Rb$ are statistically independent \cite[Lemma 2.1]{tulino04}, 
the elements of $\zb_\layerk(\deltab_\layerk)$  conditioned on $\deltab_\layerk$ are statistically independent as well. We thus have 
\begin{equation}\label{eqn.prod.stat.independent}
\P\big[\|\zb_\layerk(\deltab_\layerk)\hspace{-0.04cm}\|_\infty \leq \rinf\big] = \prod_{i=1}^{\layerk+L} \P\Big[ \big|[\zb_\layerk(\deltab_\layerk)]_i \big| \leq \rinf\Big]. 
\end{equation}
The bottom $L$ elements of $\zb_\layerk(\deltab_\layerk)$  are given by the i.i.d.\  $\CN(0,\sigma^2)$ vector $\nb_L$ (see \eqn{eqn.part.metric.vec}) so that
\[
\P\Big[ \big|[\zb_\layerk(\deltab_\layerk)]_i \big| \leq \rinf\Big] = \gamma_{1}\!\!\left(\!\frac{\rinf^2}{\sigma^2}\!\right), \quad i = \layerk+1,\dots,\layerk+L,
\]
which yields
\begin{align}
\P\big[\|\zb_\layerk(\deltab_\layerk)\hspace{-0.04cm}\|_\infty \leq \rinf\big] = & \nonumber \\ 
& \hspace{-2.5cm} \left[\gamma_{1}\!\!\left(\!\frac{\rinf^2}{\sigma^2}\!\right)\right]^{\!L}  \prod_{\layer=1}^{\layerk} \P\Big[ \big|[\zb(\deltab)]_{\tdim-\layer+1} \big| \leq \rinf\Big].\label{eqn.order.stat}
\end{align} 

\subsubsection{Cdf of $\big|[\zb(\deltab)]_{M-\layer+1}\big|$}\label{sec.dist.comp}
An analytic expression for $\text{P}\Big[\big|[\zb(\deltab)]_{M-\layer+1}\big| \leq C_\infty \Big]$ can be obtained via direct integration using the fact that the nonzero entries of $\Rb$ are statistically independent with 
$\sqrt{2\tdim} R_{i,i}  \sim  \chi_{2(\rdim-i+1)}$ and 
$R_{i,j}  \sim  \CN(0,1/\tdim)$,  for $i =1,\dots,\tdim$, $j > i$ \cite[Lemma 2.1]{tulino04}. 
In  Appendix \ref{app.distribution}  it is shown
that $\text{P}\Big[\big|[\zb(\deltab)]_{M-\layer+1}\big| \leq C_\infty \Big]$ is a binomial mixture of $\chi$-distributions with degrees of freedom reaching from $2$ up to $2(\layer+L)$.  More specifically, for 
$\layer = 1,\dots, \tdim$, we have 
\begin{align}
\nonumber  \text{P}\Big[\big|[\zb(\deltab)]_{M-\layer+1}\big| \leq C_\infty \Big]  = &  \\ \label{eqn.dist}
& \hspace{-3.0cm} 
\sum_{l=0}^{\layer + L-1} B_l(\deltab_\layer)\, \gamma_{\layer+L-l}\bigg(  
\frac{C_\infty^2}{\|\deltab_\layer \|_2^2/\tdim + \sigma^{2}}\bigg)
\end{align}
with the coefficients $B_l(\deltab_\layer)$ given by the binomial probabilities 
\begin{equation}\label{eqn.binomial.prob}
B_l(\deltab_\layer) =   {\layer\!+\!L\!-\!1 \choose l}\, (p(\deltab_\layer))^l\, (1\!-\!p(\deltab_\layer))^{\layer+L-1-l} 
\end{equation}
with parameter 
\begin{equation}\label{eqn.parameter}
 p(\deltab_\layer)  =  \frac{\|\deltab_{\layer-1}\|_2^2+\tdim \sigma^2}{\|\deltab_{\layer}\|_2^2+\tdim \sigma^2}
\end{equation}
and $\|\deltab_{0}\|_2^2 =  0$. 
In \cite{Gowaikar_07} the pdf of the RV $\big|[\zb(\deltab)]_{M-\layer+1}\big|^2$ associated with the distribution \eqref{eqn.dist} was obtained in a different form (i.e., not in terms of a binomial mixture of $\chi$-distributions) 
using an alternative derivation. More specifically, the derivation in \cite{Gowaikar_07} 
exploits the property 
$\|\zb_\layer(\deltab_\layer)\hspace{-0.04cm}\|_2^2 = \|\zb_{\layer-1}(\deltab_{\layer-1})\hspace{-0.04cm}\|_2^2 + \big|[\zb(\deltab)]_{M-\layer+1}\big|^2$
with $\|\zb_{\layer-1}(\deltab_{\layer-1})\hspace{-0.04cm}\|_2^2$  and $\big|[\zb(\deltab)]_{M-\layer+1}\big|^2$ being  statistically independent and 
 the MGFs of $\|\zb_\layer(\deltab_\layer)\hspace{-0.04cm}\|_2^2$ and $\|\zb_{\layer-1}(\deltab_{\layer-1})\hspace{-0.04cm}\|_2^2$ being known from \cite{hass_sp03_part_i}. 
This allows to compute the MGF of $\big|[\zb(\deltab)]_{M-\layer+1}\big|^2$ and, via the inverse Fourier transform, the corresponding pdf, which can then be used to establish \eqref{eqn.dist}.
Finally, we note that the direct integration approach used in this paper to obtain \eqn{eqn.dist} can, in contrast to the approach employed in \cite{Gowaikar_07}, be
applied to derive the distributions of $|[\zb(\deltab)]_{\text{R},M-\layer+1}|$ and $|[\zb(\deltab)]_{\text{I},M-\layer+1}|$, which are needed to compute (bounds on) the complexity of SD-$\ltilde$ (see Section \ref{sec.ltilde.SD.complexity} for more details). 
  
\subsubsection{Sum Representation and Moment Generating Function} 
The binomial mixture representation \eqn{eqn.dist} allows for an interesting alternative representation of the RV $\big|[\zb(\deltab)]_{M-\layer+1}\big|^2$ as the sum of independent RVs. In particular, 
using results from  \cite{behboodian72}, it is shown in Appendix \ref{app.behboodian} that 
\begin{equation}\label{eqn.equaldist.tm} 
	\big|[\zb(\deltab)]_{M-\layer+1}\big|^2 \equaldist t_\layer^2
\end{equation}
where 
\begin{equation}\label{eqn.sum.represent}
	t_\layer^2 =  \frac{\|\deltab_\layer\|_2^2/\tdim+\sigma^2}{2}\Bigg(\gamma^2 +\!\! \sum_{i=1}^{\layer+L-1}\! \lambda_i^2 \Bigg)
\end{equation}
with the independent RVs $\gamma^2$ and $\lambda_i^2$, $i=1,\dots, \layer+L-1$. 
Here, $\gamma^2 \sim \chi^2_2$ with pdf $f_{\!\chi^2_2}(\argument)$
and  the $\lambda_i^2$ have the mixture pdf 
\begin{equation}\label{eqn.mixture}
	f_{\lambda_i^2}(\argument) = (1\!-\!p(\deltab_\layer))f_{\!\chi^2_2}(\argument) + p(\deltab_\layer) \delta(\argument) 
\end{equation}
or, equivalently, with probability $p(\deltab_\layer)$ the $\lambda_i^2$ come from a population having pdf $\delta(\argument)$ (i.e., they are zero with probability $p(\deltab_\layer)$) and with probability $1-p(\deltab_\layer)$ they come from a population having a $\chi_2^2$ distribution. 

Besides being interesting in its own right, the representation \eqref{eqn.sum.represent} allows to compute the MGF of $\big|[\zb(\deltab)]_{M-\layer+1}\big|^2$ or, equivalently, of  $t_\layer^2$ 
in a straightforward manner, by using 
 \eqn{eqn.parameter} and \eqn{eqn.mixture}, as 
 (cf.\ \eqref{eqn.chi2.mgf})
\begin{align}
	\Phi_{t_\layer^2}\!(s) =  \E\big\{e^{s t_\layerk^2}\big\} & =  \frac{1}{1\!-\!(\|\deltab_\layer\|_2^2/\tdim+\sigma^2)s} 
	\left[\frac{1\!-\!p(\deltab_\layer)}{1\!-\!(\|\deltab_\layer\|_2^2/\tdim+\sigma^2)s} + p(\deltab_\layer) \right]^{\layer+L-1} \nonumber \\[0.5cm]\label{eqn.mgf.tm}
	& =  \frac{[1\!-\!(\|\deltab_{\layer-1}\|_2^2/\tdim+\sigma^2)s]^{\layer+L-1}}{[1\!-\!(\|\deltab_\layer\|_2^2/\tdim+\sigma^2)s]^{\layer+L}}.
\end{align}

 \subsection{Cdf of $\|\zb_\layerk(\deltab_\layerk)\hspace{-0.04cm}\|_2$}\label{sec.distribution.l2.metric}
Using the results of Section \ref{sec.partial.metric.distributions}, we can directly recover the cdf of $ \|\zb_\layerk(\deltab_\layerk)\hspace{-0.04cm}\|_{2}$ obtained in  \cite{hass_sp03_part_i}. 
The derivation in \cite{hass_sp03_part_i} is explicitly based on the rotational invariance of the $\ltwo$-norm (see \cite[Lemma 1]{hass_sp03_part_i}). Here, we follow an alternative approach and start by using  \eqn{eqn.part.metric.vec} to obtain
\[
\|\zb_\layerk(\deltab_\layerk)\hspace{-0.04cm}\|_2^2 = \|\nb_L\|^2_2 + \sum_{\layer=1}^{\layerk} \big|[\zb(\deltab)]_{M-\layer+1}\big|^2. 
\]
Thanks to \eqn{eqn.equaldist.tm} we then have 
\[
\|\zb_\layerk(\deltab_\layerk)\hspace{-0.04cm}\|_2^2 \equaldist \|\nb_L\|^2_2 +  \sum_{\layer = 1}^\layerk t_\layer^2.	
\] 
Since the RVs $\|\nb_L\|^2_2$ and $t_\layer^2$, $\layer = 1,\dots, \layerk$, are mutually statistically independent, 
the MGF of $\|\zb_\layerk(\deltab_\layerk)\hspace{-0.04cm}\|_2^2$ can be written as 
\begin{equation}\label{eqn.mgf.1}
	\Phi_{\|\zb_\layerk(\deltab_\layerk)\hspace{-0.02cm}\|_2^2}(s) 
		 =  \Phi_{\|\nb_L\|^2_2}(s) \prod_{\layer=1}^{\layerk} \Phi_{t_\layer^2}\!(s)
\end{equation}
where $\Phi_{\|\nb_L\|^2_2}(s)$ denotes the MGF of $\|\nb_L\|^2_2$ given by  (cf.\ \eqn{eqn.chi2.mgf}) 
\begin{equation}\label{eqn.mgf.nl}
	\Phi_{\|\nb_L\|^2_2}(s) = \frac{1}{(1\!-\!\sigma^2 s)^{L}}. 
\end{equation}
Inserting \eqref{eqn.mgf.tm} and \eqref{eqn.mgf.nl} into \eqref{eqn.mgf.1}, then yields 
\begin{align*}
\Phi_{\|\zb_\layerk(\deltab_\layerk)\hspace{-0.02cm}\|_2^2}(s) & =  \frac{1}{(1\!-\!\sigma^2 s)^{L}}  \prod_{\layer=1}^{\layerk} \frac{[1\!-\!(\|\deltab_{\layer-1}\|_2^2/\tdim+\sigma^2)s]^{\layer+L-1}}{[1\!-\!(\|\deltab_\layer\|_2^2/\tdim+\sigma^2)s]^{\layer+L}} \\
& = \frac{1}{[1\!-\!(\|\deltab_\layerk\|_2^2/\tdim+\sigma^2)s]^{\layerk+L}}
\end{align*}
which is the MGF of a $\chi^2_{2(\layerk+L)}$-distributed RV. Consequently, the cdf
of $\|\zb_\layerk(\deltab_\layerk)\hspace{-0.04cm}\|_2$ is given by   
\begin{equation}\label{eqn.dist2}
	\text{P}\big[\|\zb_\layerk(\deltab_\layerk)\hspace{-0.04cm}\|_2  \leq \rtwo \big] 
	= \gamma_{\layerk+L}\bigg(\frac{\rtwo^2}{\|\deltab_\layerk \|_2^2/\tdim+\sigma^2}\bigg)
\end{equation}
which is what was found in \cite{hass_sp03_part_i,hass_sp03_part_ii}. 

\pagebreak
\subsection{Final Complexity Expressions}\label{sec.final.complexity.expressions}
We are now ready to assemble our results to get the final complexity expressions for SD-$\linf$ and SD-$\ltwo$.  Inserting \eqn{eqn.dist} into \eqn{eqn.order.stat} and using \eqn{eqn.NN.advancedk}, we obtain 
\begin{align}
\E\{\NN_{\infty,\layerk}\} =  \frac{1}{|\symbolset|^\layerk}\, \left[\gamma_{1}\!\!\left(\!\frac{\rinf^2}{\sigma^2}\!\right)\right]^{\!L}\,\sum_{ \deltab_\layerk}  \prod_{\layer=1}^{\layerk} \sum_{l=0}^{\layer + L-1}& \nonumber \\ 
& \hspace{-3.4cm} 
 B_l(\deltab_\layer)\, \gamma_{\layer+L-l}\bigg(  
\frac{C_\infty^2}{\|\deltab_\layer \|_2^2/\tdim+\sigma^2}\bigg).\label{eqn.NNk.final}
\end{align}
The corresponding total complexity follows from \eqref{eqn.total.complexity}. 
In comparison, for SD-$\ltwo$, using \eqref{eqn.NN2.final.prob} and \eqref{eqn.dist2} yields \cite{hass_sp03_part_i,hass_sp03_part_ii}
\begin{equation}
 \label{eqn.NN2.final}
\E\{\NN_{2,\layerk}\} = \frac{1}{|\symbolset|^\layerk}  \sum_{ \deltab_\layerk} 
\gamma_{\layerk+L}\bigg(\frac{\rtwo^2}{\|\deltab_\layerk \|_2^2/\tdim+\sigma^2}\bigg). 
\end{equation}
The total complexity for SD-$\ltwo$ is then obtained as 
\begin{equation}\label{eqn.total.complexity.2}
\E\{\NN_{2}\} = \sum_{\layerk = 1}^{\tdim}\E\{\NN_{2,\layerk}\}.
\end{equation}

\subsection{Choice of Radii}\label{eqn.choice.of.radii}
For a meaningful comparison of the complexity of SD-$\linf$ and SD-$\ltwo$, the radii $\rinf$ and $\rtwo$ have to be chosen carefully. In our analysis below, 
we use the approach proposed in \cite{viterbo93,hass_sp03_part_i,hochbrink03} for SD-$\ltwo$, where the choice of $\rtwo$ is based on the 
noise statistics such that the probability of finding the transmitted data vector inside the search hypersphere is sufficiently high. 
Recall that our complexity analysis assumes a fixed choice of the radii that does not depend on the channel, noise, and data realizations. 
We start by noting that  $\|\zb(\db')\hspace{-0.04cm}\|_{2} = \|\nb\|_{2}$, which is $\chi_{2\rdim}$-distributed.
Choosing the radius $\rtwo$ such that the transmitted data vector $\db'$ is found inside the search hypersphere with probability $1-\epsilon  \in [0, 1]$ is accomplished by setting 
\begin{equation} \label{eqn.noise.ctwo}
	\text{P}\big[\|\nb\|_2 \leq \rtwo \big] = \gamma_{\rdim}\!\bigg(\frac{\rtwo^2}{\sigma^2}\bigg) = 1-\epsilon.
\end{equation}
Solving  \eqref{eqn.noise.ctwo} for $\rtwo^2$ yields
\begin{equation}\label{eqn.ctwo.epsilon}
\rtwo^2  =\sigma^2\,\gamma_{\rdim}^{-1}\!\left(1-\epsilon\right).
\end{equation}
For the SD-$\linf$ case, we adopt an analogous approach arguing that we choose $\rinf$ such that $\db'$ is contained in the search hypercube with sufficiently high probability. 
Specifically, for the complexity comparisons in the remainder of the paper, we choose the radius $\rinf$  such that the probability of finding the 
 transmitted data vector $\db'$ through SD-$\linf$ equals that for SD-$\ltwo$. 
 This is accomplished by setting 
\begin{equation}
	\text{P}\big[\|\nb\|_\infty \leq \rinf \big]  =\left[\gamma_{1}\!\!\left(\!\frac{\rinf^2}{\sigma^2}\!\right)\right]^{\!\rdim}=  1-\epsilon \label{eqn.noise.cinf}  
\end{equation}
which results in 
\begin{equation}
	\rinf^2 = -\sigma^2\,\text{log}\!\left(1-\sqrt[\rdim]{1-\epsilon}\,\right) \label{eqn.cinf.epsilon}.
\end{equation}
For any $\epsilon > 0$, for both SD-$\ltwo$ and SD-$\linf$, there is a nonzero probability that no leaf node is 
found by the detector, i.e., $\|\zb(\db)\hspace{-0.04cm}\|_{2} > \rtwo$ or $\|\zb(\db)\hspace{-0.04cm}\|_{\infty} > \rinf$, respectively, for all $\db \in \symbolset^{\tdim}$.  Stopping the detection procedure and declaring an error in this case, it follows that the corresponding SD
does not implement exact SD-$\ltwo$ (i.e., ML) or exact SD-$\linf$ detection, respectively. 
To obtain exact ML or SD-$\linf$ performance, the corresponding SD algorithm has to be 
restarted using a schedule of increasing radii (or equivalently a schedule of decreasing values for  $\epsilon$) until a leaf node is found within the search hypersphere or 
hypercube, respectively  (see, e.g., \cite{hass_sp03_part_i,damen00} for SD-$\ltwo$). 

Let us investigate the case of SD without restarting as described above  and denote any of the corresponding SD algorithms, i.e., SD-$\ltwo$ or SD-$\linf$,  based on 
a fixed radius $C$ (according  to either \eqref{eqn.ctwo.epsilon} for SD-$\ltwo$  or 
\eqref{eqn.cinf.epsilon} for SD-$\linf$ with a fixed $\epsilon$) as SD-NoR.  
Denote the corresponding error probability as  
$\P_{{\cal E}, \text{SD-NoR}}(\SNR) =  \P\left[\hspace{0.2mm}\widehat{\db} \neq \db'; C \right]$ in contrast to  $\P_{{\cal E},\text{SD}}(\SNR) = \P\left[\hspace{0.2mm}\widehat{\db} \neq \db'\right]$, which denotes the error probability of exact SD-$\linf$ or SD-$\ltwo$, implemented, e.g., through restarting using a schedule of increasing radii 
as explained above. In the following, we show that the consequence of not restarting the SD is an error floor of $\epsilon$.  
In the remainder of this section, $\|\cdot\|$ stands for either the $\linf$- or $\ltwo$-norm. 
By the law of total probability and recalling \eqref{eqn.noise.cinf} and \eqref{eqn.noise.ctwo}, we have 
\begin{equation}\label{eqn.error.fixed.epsilon}
\P_{{\cal E}, \text{SD-NoR}}(\SNR)   =  \P\left[\hspace{0.2mm}\widehat{\db} \neq \db';  C \, \big|\, \|\nb\| \leq \rgen \right] (1-\epsilon)
	+ \P\left[\hspace{0.2mm}\widehat{\db} \neq \db';  C \, \big|\, \|\nb\| > \rgen \right] \epsilon
\end{equation}
and, similarly,  
\begin{equation}\label{eqn.error.restarted}
\P_{{\cal E},\text{SD}}(\SNR)  =  \P\left[\hspace{0.2mm}\widehat{\db} \neq \db'\big|\, \|\nb\| \leq \rgen \right] (1-\epsilon)
	+ \P\left[\hspace{0.2mm}\widehat{\db} \neq \db'\big|\, \|\nb\| > \rgen \right] \epsilon. 
\end{equation}
In the case $\|\nb\| \leq \rgen$, SD-NoR and exact SD yield identical results since at least one data vector (namely the transmitted data vector $\db'$) is found inside the search space with radius $C$. Hence,  $ \P\left[\hspace{0.2mm}\widehat{\db} \neq \db';  C \, \big|\, \|\nb\| \leq \rgen \right] = \P\left[\hspace{0.2mm}\widehat{\db} \neq \db'\big|\,\|\nb\| \leq \rgen \right]$, which, together with 
 $\P\left[\hspace{0.2mm}\widehat{\db} \neq \db'\big|\, \|\nb\| \leq \rgen \right] (1-\epsilon) \leq \P_{{\cal E},\text{SD}}(\SNR)$  by \eqn{eqn.error.restarted}, yields 
\begin{equation}\label{eqn.first.term.of.fixed.error}
 \P\left[\hspace{0.2mm}\widehat{\db} \neq \db';  C \, \big|\, \|\nb\| \leq \rgen \right]  (1-\epsilon) \leq  \P_{{\cal E},\text{SD}}(\SNR)
\end{equation}
for the first term on the \sloppy RHS of \eqref{eqn.error.fixed.epsilon}. 
In the case $\|\nb\| > \rgen$, the transmitted data vector $\db'$  is not found inside the search space with radius $C$. Thus, 
for SD-NoR, this case will certainly result in an error, i.e., 
\begin{equation}\label{eqn.error.dtx.not.found}
\P\left[\hspace{0.2mm}\widehat{\db} \neq \db';  C \, \big|\, \|\nb\| > \rgen \right] = 1
\end{equation}
 which together with \eqref{eqn.first.term.of.fixed.error} yields the following upper bound on the error probability of SD-NoR:  
\[
\P_{{\cal E},\text{SD-NoR}}(\SNR)  \leq  \P_{{\cal E},\text{SD}}(\SNR) + \epsilon. 
\]
Using \eqref{eqn.error.dtx.not.found} in \eqn{eqn.error.fixed.epsilon} immediately yields the lower bound 
\[
\P_{{\cal E},\text{SD-NoR}}(\SNR)  \geq  \epsilon. 
\]
Since 
${\lim}_{\SNR \rightarrow \infty}\, 
\P_{{\cal E},\text{SD}}(\SNR) =  0$ (see Section \ref{sec.overall.error.probability}), we have 
 ${\lim}_{\SNR \rightarrow \infty}\, \P_{{\cal E},\text{SD-NoR}}(\SNR) =  \epsilon$. We can finally conclude that if the system operates at a target error rate that is much higher than this error floor, a fixed radius
and the absence of restarting will have a negligible impact on the total error probability.

\subsection{Asymptotic Complexity Analysis}\label{sec.asympt.complexity.analysis}
In  \cite{jalden_tsp05} it is shown that the complexity of SD-$\ltwo$ scales exponentially in the number of transmit antennas $\tdim$. 
Motivated by this result, we will next show that the complexity scaling behavior of SD-$\linf$ is also exponential in $\tdim$. 
For simplicity of exposition, we set $\tdim = \rdim$ in the following.

\subsubsection{Impact of Choice of Radius} 

The asymptotic complexity scaling behavior of SD-$\ltwo$ is studied in detail in \cite{jalden_tsp05}, where it is shown that 
$\E\{\NN_{2}\} \geq e^{\gamma \tdim}$ for large $\tdim$ and some $\gamma > 0$. 
This result is derived under the assumption of $\rtwo^2$ increasing (at least) {\em linearly} in $\tdim$, which guarantees a nonvanishing probability 
of finding at least one leaf node inside the search hypersphere \cite[Theorem 1]{jalden_ISIT05}. 
It is furthermore shown in  \cite{jalden_ISIT05} that the exponential complexity scaling behavior of SD-$\ltwo$ extends to the case where the sphere radius is chosen optimally, i.e., 
when the radius is set to the minimum value still guaranteeing that at least one leaf node is found (this would, of course, correspond to a genie-aided choice of the sphere radius since it essentially necessitates the knowledge of the 
ML detection result). We finally note that $\rtwo^2$ chosen according to 
\eqn{eqn.ctwo.epsilon} results in linear scaling in $\tdim$ for large $\tdim$. For a proof of this statement the reader is referred to 
Appendix \ref{app.calc.linearM.SD2}. Linear scaling of $\rtwo^2$ in $\tdim$ is also obtained, for example, by setting 
 $\rtwo^2 \propto \E\big\{\|\nb\|_2^2 \big\} = \sigma^{2} \tdim$ as was done in \cite{jalden_tsp05,hass_sp03_part_i}.

For SD-$\linf$ it is shown in Appendix \ref{app.calc.linearM.SD2} that the radius $\rinf^2$ according to \eqref{eqn.cinf.epsilon} scales logarithmically in 
$\tdim$ for large $\tdim$. We will next show that this is also the case if 
$\rinf^2$ is chosen to be proportional to 
$\E\big\{\|\nb\|_\infty^2 \big\}$. Consider the $\tdim$ i.i.d.\ $\chi_{2}^2$-distributed RVs $2y_{i} \sim \chi_{2}^2$, $i = 1,\dots, \tdim$. 
Then, \cite[Eq.\ (2.5.5)]{david03} 
\begin{equation*}\label{eqn.david.result}
 \text{max}\{y_{1}, y_{2}, \dots, y_{\tdim}\} \equaldist \sum_{i=1}^{\tdim}  \frac{1}{i} z_{i}
\end{equation*}  
where the RVs $z_{i}$, $i = 1,\dots, \tdim$, are also i.i.d.\ with $2z_{i} \sim \chi_{2}^2$. Since we have  
$\|\nb\|_\infty^2 \equaldist \sigma^2\, \text{max}\{y_{1}, y_{2}, \dots, y_{\tdim}\}$ and $\E\{z_{i}\} = 1$, we obtain
\[
 \E\big\{\|\nb\|_\infty^2 \big\} = \sigma^2\,  {H}_{M}
\]
with ${H}_{\tdim} = \sum_{i=i}^{\tdim} 1/i $ denoting the $\tdim$th harmonic number. 
For large $\tdim$ and with $\beta \approx 0.5772$ denoting the Euler-Mascheroni constant, we have ${H}_{\tdim} = \beta + \text{ln}(\tdim)+ {\cal O}(\tdim^{-1})$, $\tdim \rightarrow \infty$  \cite{Finch_03}, which establishes the result. 
At first sight, the logarithmic scaling of $\rinf^2$ in $\tdim$ versus the linear scaling of $\rtwo^2$ suggests a difference in the asymptotic complexity 
behavior of SD-$\linf$ and SD-$\ltwo$. While the complexity and pruning (see Section \ref{sec.pruning.behavior}) behavior for finite $\tdim$ are indeed quite different in general, we will, however, next show that SD-$\linf$ also exhibits 
exponential complexity scaling in $\tdim$.

\subsubsection{Lower Bound on Complexity}\label{sec.lower.bound} 
Computing the asymptotics of the exact SD-$\linf$ complexity expression (\eqref{eqn.NNk.final} together with \eqref{eqn.total.complexity}) seems involved. 
We therefore tackle the problem by computing a lower bound on complexity and by showing that this lower bound scales exponentially in the problem size $\tdim$. Our technique can readily be extended to the SD-$\ltwo$ case resulting in an alternative, w.r.t.\  \cite{jalden_tsp05}, proof of the exponential complexity scaling behavior of SD-$\ltwo$. We note, however, that while our proof seems to be shorter and more direct, the result in \cite{jalden_tsp05} is more general in the sense that it applies to MIMO channels with very general fading statistics. Our approach, in contrast, explicitly hinges on the channel matrix $\Hb$ being i.i.d.\ Rayleigh fading. 
On a conceptual basis,  our proof is more closely related to the approach in \cite{Gowaikar_07}, where bounds on the complexity of SD-$\ltwo$ (and variants thereof) are studied.

We start by focusing on the expression for $\text{P}\big[\|\zb_\layerk(\deltab_\layerk)\|_\infty  \leq \rinf \big]$, $\layerk = 1,\dots, \tdim$, obtained by inserting \eqref{eqn.dist} into the RHS of 
\eqref{eqn.order.stat}. 
Considering only the summand with  
index $l = \layer-1$  in \eqref{eqn.dist}, we obtain (recall that $L = \rdim\! -\!\tdim = 0$)
\begin{equation}\label{eqn.lower.bound.first.step}
\text{P}\big[\|\zb_\layerk(\deltab_\layerk)\|_\infty  \leq \rinf \big] \geq  \prod_{\layer=1}^{\layerk}  \Bigg( \frac{\|\deltab_{\layer-1}\|_2^2/\tdim+\sigma^2}{\|\deltab_{\layer}\|_2^2/\tdim+\sigma^2} \Bigg)^{\!\! \layer-1}
 \!\gamma_{1}\bigg(  
\frac{C_\infty^2}{\|\deltab_\layer \|_2^2/\tdim+\sigma^2}\bigg). 
\end{equation}
Using \eqref{eqn.gamma.bound} in Appendix \ref{app.incomplete.gamma.function} according to 
\[
	\gamma_{1}\bigg(\frac{C_\infty^2}{\|\deltab_\layer \|_2^2/\tdim+\sigma^2}\bigg) \geq \gamma_{1}\bigg(\frac{C_\infty^2}{\sigma^2}\bigg)
	\frac{\sigma^{2}}{\|\deltab_\layer \|_2^2/\tdim +\sigma^{2}}, 
\]
we get
\begin{align}
\text{P}\big[\|\zb_\layerk(\deltab_\layerk)\|_\infty  \leq \rinf \big] & \geq  
\left[\gamma_{1}\!\!\left(\!\frac{\rinf^2}{\sigma^2}\!\right)\right]^{\!\layerk}
\prod_{\layer=1}^{\layerk}  \sigma^2  \frac{(\|\deltab_{\layer-1}\|_2^2/\tdim+\sigma^2)^{\layer-1} }{(\|\deltab_{\layer}\|_2^2/\tdim+\sigma^2)^{ \layer}} \nonumber \\
 & = \left[\gamma_{1}\!\!\left(\!\frac{\rinf^2}{\sigma^2}\!\right)\right]^{\!\layerk}
   \bigg(1+\frac{\|\deltab_\layerk \|_2^2}{\tdim \sigma^2}\bigg)^{\!-\layerk}. \label{eqn.lower.bound.intermediate}
 \end{align}
Furthermore, by \eqref{eqn.UB.on.gamma} in Appendix \ref{app.incomplete.gamma.function} we have $\gamma_{a}(x) \leq 1$, $x \geq 0$, so that 
\begin{equation}\label{eqn.lower.bound.prefactor}
   \left[\gamma_{1}\!\!\left(\!\frac{\rinf^2}{\sigma^2}\!\right)\right]^{\!\layerk}
 \geq \left[\gamma_{1}\!\!\left(\!\frac{\rinf^2}{\sigma^2}\!\right)\right]^{\!\tdim}. 
\end{equation}
Inserting the specific choice of $\rinf^2$ according to \eqref{eqn.cinf.epsilon} into the RHS of \eqref{eqn.lower.bound.prefactor}, we obtain
\[
	\left[\gamma_{1}\!\!\left(\!\frac{\rinf^2}{\sigma^2}\!\right)\right]^{\!\layerk}
 \geq 1-\epsilon
\]
and hence  \eqref{eqn.lower.bound.intermediate} becomes
\begin{equation}\label{eqn.lower.bound.linf}
\text{P}\big[\|\zb_\layerk(\deltab_\layerk)\|_\infty  \leq \rinf \big] \geq   (1-\epsilon) \bigg(1+\frac{\|\deltab_\layerk \|_2^2}{\tdim \sigma^2}\bigg)^{\!-\layerk}.
\end{equation}
With \eqref{eqn.NN.advancedk} and \eqref{eqn.total.complexity} we then obtain
\begin{equation}\label{eqn.simple.lower.bound.linf}
	\E\{\NN_{\infty}\} \geq (1-\epsilon) \sum_{\layerk = 1}^{\tdim}  \frac{1}{|\symbolset|^\layerk}  \sum_{\deltab_\layerk} \bigg(1+\frac{\|\deltab_\layerk \|_2^2}{\tdim \sigma^2}\bigg)^{\!-\layerk}
\end{equation}
which can be further simplified using 
\[
	\|\deltab_\layerk \|_2^2 \leq \UBsymdiff^2\, \hamdist(\deltab_\layerk)
\]
where $\UBsymdiff^2 =  \underset{d, d' \in \symbolset}{\text{max}} |d'-d|^{2}$ is the maximum Euclidean distance in the scalar symbol constellation and 
$\hamdist(\deltab_\layerk) = \hamdist(\db'_\layerk,\db_\layerk)$ denotes the Hamming distance between $\db_{\layerk}$ and $\db'_{\layerk}$, i.e., the number of non-zero entries (symbol errors) in $\deltab_\layerk = \db_{\layerk}'-\db_{\layerk}$.  Note that every data vector $\db'_{\layerk}$ induces the same set of Hamming distances 
$\{ \hamdist(\db'_\layerk,\db_\layerk)$, $\db_{\layerk} \in \symbolset^{\layerk}\}$. From \eqref{eqn.simple.lower.bound.linf}, we therefore get 
\begin{align}
\E\{\NN_{\infty}\} & \geq 
(1-\epsilon) \sum_{\layerk = 1}^{\tdim}  \frac{1}{|\symbolset|^\layerk}  \sum_{\deltab_\layerk} \bigg(1+\frac{\UBsymdiff^2\, \hamdist(\deltab_\layerk)}{\tdim \sigma^2}\bigg)^{\!-\layerk}
\nonumber \\
& = (1-\epsilon) \sum_{\layerk = 1}^{\tdim} \sum_{\db_\layerk} \bigg(1+\frac{\UBsymdiff^2\, \hamdist(\db'_\layerk,\db_\layerk)}{\tdim \sigma^2}\bigg)^{\!-\layerk}, \quad\text{for any}\,\, \db_{\layerk}', 
\nonumber \\
& = (1-\epsilon) \sum_{\layerk = 1}^{\tdim}  \sum_{i = 0}^{k}\, W_{i}\, \bigg(1+\frac{\UBsymdiff^2\, i}{\tdim \sigma^2}\bigg)^{\!-\layerk}
\nonumber 
\end{align}
where, in the last step,  all terms having the same Hamming distance $\hamdist(\db'_\layerk,\db_\layerk) = i$ have been merged.
Here, $W_{i}  =  {\layerk \choose i } (|\symbolset|-1)^{i} \geq   {\layerk \choose i }$ 
denotes the number of data vectors $\db_{\layerk} \in {\symbolset}^{\layerk}$ that have Hamming distance $i$  from $\db'_{\layerk}$.
Furthermore, with 
${\layerk \choose i }  \geq \left(\frac{\layerk}{i}\right)^{i}$,  $i = 1, \dots, \layerk$, we get $W_{i} \geq \left(\frac{\layerk}{i}\right)^{i}$, so that 
\begin{equation}\label{eqn.simple.lower.bound.linf2}
	\E\{\NN_{\infty}\} \geq 
	(1-\epsilon) \sum_{\layerk = 1}^{\tdim}  \sum_{i = 1}^{k}  \left(\frac{\layerk}{i}\right)^{i} \bigg(1+\frac{\UBsymdiff^2\, i}{\tdim \sigma^2}\bigg)^{\!-\layerk}
\end{equation}
where the $i = 0$ ($W_{0} = 1$) term is omitted for all $\layerk$. 
\paragraph*{Lower Bound on Complexity for SD-$\ltwo$} As already mentioned, the technique used to derive \eqref{eqn.simple.lower.bound.linf2} can readily be extended to the SD-$\ltwo$ case. We start by considering 
$\text{P}\big[\|\zb_\layerk(\deltab_\layerk)\|_2  \leq \rtwo \big]$, $\layerk = 1,\dots, \tdim$, given in \eqref{eqn.dist2} and applying the lower bound \eqn{eqn.gamma.bound} in Appendix \ref{app.incomplete.gamma.function}, to obtain (recall that $L = \rdim -\tdim = 0$)
 \[
 \text{P}\big[\|\zb_\layerk(\deltab_\layerk)\|_2  \leq \rtwo \big] = 
	\gamma_{\layerk}\bigg(\frac{\rtwo^2}{\|\deltab_\layerk \|_2^2/\tdim+\sigma^2}\bigg) \geq   \gamma_{\layerk}\bigg(\frac{\rtwo^2}{\sigma^2}\bigg) \bigg(1+\frac{\|\deltab_\layerk \|_2^2}{\tdim \sigma^2}\bigg)^{\!-\layerk}. 
\] 
Employing \eqref{eqn.lower.bound.gamma.higher.orders} in Appendix \ref{app.incomplete.gamma.function} and using \eqn{eqn.ctwo.epsilon}, we get 
\[
   \gamma_{\layerk}\bigg(\frac{\rtwo^2}{\sigma^2}\bigg)  \geq \gamma_{\tdim}\bigg(\frac{\rtwo^2}{\sigma^2}\bigg) = 1-\epsilon 
\]
which yields
\[
 \text{P}\big[\|\zb_\layerk(\deltab_\layerk)\|_2  \leq \rtwo \big] \geq   (1-\epsilon) \bigg(1+\frac{\|\deltab_\layerk \|_2^2}{\tdim \sigma^2}\bigg)^{\!-\layerk}
\]
and finally, by \eqref{eqn.NN2.final} and  \eqref{eqn.total.complexity.2}, results in   
\begin{equation}\label{eqn.lower.bound.l2}
 	\E\{\NN_{2}\} \geq (1-\epsilon) \sum_{\layerk = 1}^{\tdim}  \frac{1}{|\symbolset|^\layerk}  \sum_{\deltab_\layerk} \bigg(1+\frac{\|\deltab_\layerk \|_2^2}{\tdim \sigma^2}\bigg)^{\!-\layerk}. 
\end{equation} 
The RHS of \eqref{eqn.lower.bound.l2} is precisely the lower bound \eqref{eqn.simple.lower.bound.linf} on $\E\{\NN_{\infty}\}$. Consequently, the simplified lower bound \eqref{eqn.simple.lower.bound.linf2} on  
$\E\{\NN_{\infty}\}$ is also a lower bound on $\E\{\NN_{2}\}$.

\subsubsection{Asymptotic Analysis of Lower Bound} \label{sec.asymptotic.analysis.of.lower.bound}
In the following, we show that the lower bound \eqn{eqn.simple.lower.bound.linf2} exhibits exponential scaling in the system size $\tdim = \rdim$, which, together with the trivial upper bound 
$\E\{\NN_{\infty}\} \leq |\symbolset|^{\tdim+1}$,  establishes exponential complexity scaling of SD-$\linf$ (and of SD-$\ltwo$ together with $\E\{\NN_{2}\} \leq |\symbolset|^{\tdim+1}$).  

We start by noting that a trivial lower bound on $\E\{\NN_{\infty}\}$ is obtained by considering only one term in the RHS of \eqref{eqn.simple.lower.bound.linf2}, resulting in 
\begin{equation}\label{eqn.asymptotic.analysis.fM}
 	\E\{\NN_{\infty}\} \geq (1-\epsilon) \left(\frac{\layerk}{i}\right)^{i} \bigg(1+\frac{\UBsymdiff^2\, i}{\tdim \sigma^2}\bigg)^{\!-\layerk} =  f(\tdim). 
\end{equation}
Evidently, establishing that 
\begin{equation}\label{eqn.def.exponentially.complex}
 	 \underset{\tdim \rightarrow\, \infty} {\lim}\, \frac{\text{log}\, f(\tdim) }{\tdim}  > 0
\end{equation}
is sufficient to prove that SD-$\linf$ (and SD-$\ltwo$) exhibits exponential complexity scaling. To this end, we set $\layerk = \lceil \alpha \tdim \rceil$ and $i = \lceil \beta \tdim \rceil$ with $\alpha  \in\,\, ]0, 1]$ and $\beta \in \,\, ] 0, \alpha]$. 
We then have 
\begin{align*}
\frac{\text{log}\, f(\tdim) }{\tdim} & =  \frac{\text{log}(1-\epsilon)}{\tdim}+\frac{ \lceil \beta \tdim \rceil}{\tdim}\, \text{log}\left(\frac{ \lceil \alpha \tdim \rceil}{ \lceil \beta \tdim \rceil}\right) - \frac{ \lceil \alpha \tdim \rceil}{\tdim}\, \text{log}\left(1+\frac{\UBsymdiff^2\,  \lceil \beta \tdim \rceil}{\tdim \sigma^2} \right). 
\end{align*}
Furthermore, writing $\lceil \alpha \tdim \rceil =  \alpha \tdim + \Delta_{\alpha}$ and $\lceil \beta \tdim \rceil = \beta \tdim + \Delta_{\beta}$ for some values $\Delta_{\alpha}$ and $\Delta_{\beta}$ satisfying $0 \leq \Delta_{\alpha}, \Delta_{\beta} < 1$ gives
\begin{align}
\frac{\text{log}\, f(\tdim) }{\tdim}  = \frac{\text{log}(1-\epsilon)}{\tdim}+ (\beta +\Delta_{\beta}/\tdim)\, \text{log}\!\left(\frac{\alpha + \Delta_{\alpha}/\tdim}{ \beta + \Delta_{\beta}/\tdim} \right) -  & \nonumber \\ 
& \hspace{-4.4cm} 
 ( \alpha + \Delta_{\alpha}/\tdim) \, \text{log}\left(1+\frac{\UBsymdiff^2 \beta}{\sigma^2}+ \frac{\UBsymdiff^2}{\sigma^2} \Delta_{\beta}/\tdim   \right) \nonumber
\end{align}
which results  in 
\begin{equation}
 	 \underset{\tdim \rightarrow\, \infty} {\lim}\, \frac{\text{log}\, f(\tdim) }{\tdim}  =  \beta\, \text{log}(\alpha/\beta) - \alpha\, \text{log}(1+\UBsymdiff^2\, \beta/\sigma^2)  
		=   \gamma(\alpha,\beta) \label{eqn.exponential.complexity.gamma}. 
\end{equation}
Indeed, for any SNR (i.e., for any $\sigma^2$), there exist values of $\alpha$ and $\beta$ for which $\gamma(\alpha,\beta) > 0$. 
For example, with $\beta = \alpha/2$, any $\alpha$ satisfying 
\begin{equation}\label{eqn.exponential.complexity.final}
	0 < \alpha <  \text{min}\left\{ \frac{1}{\UBsymdiff^2}\, 2\,\sigma^2 \,(\sqrt{2}-1),\, 1\right \}
\end{equation}
results in $\gamma(\alpha,\beta) > 0$, which establishes the desired result. 

\section{Tree Pruning Behavior}\label{sec.pruning.behavior}
In the previous section, we showed that both SD-$\linf$ and SD-$\ltwo$ exhibit exponential complexity scaling in $\tdim$. The analytic results for $\E\{\NN_{\infty}\}$ and $\E\{\NN_{2}\}$ 
in Section \ref{sec.final.complexity.expressions} indicate, however, that the finite-$\tdim$ complexity can be very different for SD-$\linf$ and SD-$\ltwo$.
While it seems difficult to draw general conclusions based on the analytic expressions for $\E\{\NN_{\infty}\}$ and $\E\{\NN_{2}\}$, interesting insights on the difference in the corresponding
 {\em tree pruning behavior} (TPB) can be obtained. 
Here, we predominantly focus on the {\em average} (w.r.t.\ channel, data, and noise) TPB; some comments on the {\em instantaneous}  (i.e., for a given channel, data, and noise realization) TPB will be made at the end of this section. Our analytic results will be corroborated by numerical results in Section \ref{sec.simu.pruning.behavior}. 
  
\subsection{Average TPB}\label{sec.TPB.asymptotic.analyis} 
The average TPB of SD-$\linf$ and SD-$\ltwo$ will be studied through a high-SNR analysis of the probabilities  $\P\big[\|\zb_\layerk(\deltab_\layerk)\hspace{-0.04cm}\|_\infty \leq \rinf\big]$ (see \eqref{eqn.order.stat} with 
 \eqref{eqn.dist}) and $\text{P}\big[\|\zb_\layerk(\deltab_\layerk)\|_2  \leq \rtwo \big] $ \eqref{eqn.dist2} of a certain node $\deltab_{\layerk}$ being visited by SD-$\linf$ and SD-$\ltwo$, respectively. 
Equivalently, $1-\P\big[\|\zb_\layerk(\deltab_\layerk)\hspace{-0.04cm}\|_\infty \leq \rinf\big]$ and $1-\text{P}\big[\|\zb_\layerk(\deltab_\layerk)\|_2  \leq \rtwo \big]$ refer to the probabilities of node $\deltab_{\layerk}$ being pruned by SD-$\linf$ and SD-$\ltwo$, respectively. While \eqref{eqn.dist2} shows that the probability of a node $\deltab_{\layerk}$ being visited by SD-$\ltwo$ 
depends only on $\|\deltab_{\layerk}\|_2$, i.e., on the Euclidean distance between $\db_{\layerk}$ and $\db_{\layerk}'$,  in the SD-$\linf$ case this dependence on $\deltab_{\layerk}$ seems in general rather involved.  However, the high-SNR analysis of 
 $\P\big[\|\zb_\layerk(\deltab_\layerk)\hspace{-0.04cm}\|_\infty \leq \rinf\big]$ reveals simple characteristics of $\deltab_{\layerk}$
 that determine the probability of node $\deltab_{\layerk}$ being visited by SD-$\linf$, which then enables us to characterize the fundamental differences in the TPB of SD-$\linf$ and SD-$\ltwo$. 
The corresponding results will be supported by simple geometric considerations.  
 Throughout this section, the radii $\rinf$ and $\rtwo$ are chosen according to 
\eqref{eqn.cinf.epsilon} and \eqref{eqn.ctwo.epsilon}, respectively, and we define
\begin{align}
	\rinfsigma & = \frac{\rinf^2}{\sigma^2} = -\text{log}\!\left(1-\sqrt[\rdim]{1-\epsilon}\,\right) \label{eqn.rinfsigma.def}\\
	\rtwosigma & = \frac{\rtwo^2}{\sigma^2} = \gamma_{\rdim}^{-1}\!\left(1-\epsilon\right).\nonumber 
\end{align}
 
 \subsubsection{High-SNR Analysis}\label{sec.TPB.asymptotic.analyis} 
Consider a node\footnote{\label{page.footnote.trivial.high.SNR.behavior}
For $\deltab_\layerk = \vec{0}$ the high-SNR behavior of $\P\big[\|\zb_\layerk(\deltab_\layerk)\hspace{-0.04cm}\|_\infty \leq \rinf\big]$ and $\text{P}\big[\|\zb_\layerk(\deltab_\layerk)\|_2  \leq \rtwo \big]$ is trivial since the expressions    
$\text{P}\big[\|\zb_\layerk(\vec{0})\|_\infty  \leq \rinf \big] =  [\gamma_{1}(\rinfsigma)]^{\layerk+L}$ and 
$\text{P}\big[\|\zb_\layerk(\vec{0})\|_2  \leq \rtwo \big] = \gamma_{\layerk+L}(\rtwosigma)$ 
do not depend on the SNR 
$\SNR$. 
}
 $\deltab_\layerk \neq \vec{0}$ at tree level $\layerk$ and denote the index of the corresponding first tree level exhibiting a symbol error by
  $\Nzeros(\deltab_\layerk) \in [1,2,\dots,\layerk]$. More precisely, we have  
  $[\deltab_\layerk]_{k-i+1} = 0$, for $i = 1, \dots, \Nzeros(\deltab_\layerk)-1$ and $[\deltab_\layerk]_{k-\Nzeros(\deltab_\layerk)+1} \neq 0$. 
   In Appendix \ref{app.asympt.prop.b.l.inf},  it is shown that 
\begin{equation}\label{eqn.asymptotic.behavior.Pbk.inf}
	\P\big[\|\zb_\layerk(\deltab_\layerk)\hspace{-0.04cm}\|_\infty \leq \rinf\big] \equalasympt 
	A\big(\Nzeros(\deltab_\layerk)\big)\rinfsigma^{\layerk+L}\big(\rho\,\|\deltab_{\layerk}\|_2^2/\tdim \big)^{-(\layerk+L)}, 
	\quad\rho \rightarrow \infty 
\end{equation}
where 
\begin{equation}\label{eqn.function.Am0}
	A\big(\Nzeros(\deltab_\layerk)\big) = \left[\gamma_{1}(\rinfsigma)\right]^{\Nzeros(\deltab_\layerk)+L-1} 
 	\sum_{l=0}^{\Nzeros(\deltab_\layerk)+L-1} \!\!\! {\Nzeros(\deltab_\layerk)\!+\!L\!-\!1 \choose l} \frac{1}{(\Nzeros(\deltab_\layerk)\! +\!L\!-\!l)!}
	\rinfsigma^{-l}.
\end{equation} 
Note that $A\big(\Nzeros(\deltab_\layerk)\big)$ does not depend on the SNR $\SNR$.
Furthermore, using \eqref{eqn.gamma.series1} in \eqref{eqn.dist2}, we directly obtain a corresponding result for SD-$\ltwo$ as  
\begin{equation}\label{eqn.asymptotic.behavior.Pbk.two}
	\text{P}\big[\|\zb_\layerk(\deltab_\layerk)\|_2  \leq \rtwo \big] \equalasympt  \frac{1}{(\layerk+L)!} \rtwosigma^{\layerk+L}\,
	\big(\rho\, \|\deltab_{\layerk}\|_2^2/\tdim\big)^{-(\layerk+L)}, 
	\quad\rho \rightarrow \infty.
\end{equation}
From \eqref{eqn.asymptotic.behavior.Pbk.inf} and \eqref{eqn.asymptotic.behavior.Pbk.two} we can infer that the only characteristics of $\deltab_{\layerk}$, which determine the 
high-SNR probability of node $\deltab_{\layerk}$ being visited, are $\|\deltab_{\layerk}\|_2^2$ in the case of SD-$\ltwo$  and 
 $\|\deltab_{\layerk}\|_2^2$ and  $\Nzeros(\deltab_\layerk)$ in the case of SD-$\linf$. Moreover, for SD-$\linf$ the dependence on $\Nzeros(\deltab_\layerk)$ is through the function  
$A\big(\Nzeros(\deltab_\layerk)\big)$  \eqref{eqn.function.Am0}, which has the following properties.  By inspection we get 
\begin{equation}\label{eqn.Abo.lim.LB}
\underset{\rinfsigma \rightarrow \infty} {\lim}\,\, 
A\big(\Nzeros(\deltab_\layerk)\big) = \frac{1}{(\Nzeros(\deltab_\layerk)+L)!}
\end{equation}
and, as shown in Appendix \ref{app.Am0.limit}, 
\begin{equation}\label{eqn.Abo.lim.UB}
\underset{\rinfsigma \rightarrow 0} {\lim}\,\, 
A\big(\Nzeros(\deltab_\layerk)\big) = 1. 
\end{equation}
Note that $\rinfsigma \rightarrow \infty$ for $\epsilon \rightarrow 0$ and $\rinfsigma \rightarrow 0$  for $\epsilon \rightarrow 1$. 
In Appendix \ref{app.Am0.decreasing.function} it is furthermore shown that $A\big(\Nzeros(\deltab_\layerk)\big)$ is a nonincreasing function of $\rinfsigma$, which, together 
 with \eqn{eqn.Abo.lim.LB} and \eqn{eqn.Abo.lim.UB},  yields
\begin{equation}\label{eqn.Ab0.UB.LB}
 \frac{1}{(\Nzeros(\deltab_\layerk)+L)!}  \leq A\big(\Nzeros(\deltab_\layerk)\big) \leq 1. 
\end{equation}
The lower bound in \eqref{eqn.Ab0.UB.LB} allows us to conclude that, in the best case, the high-SNR probability of SD-$\linf$ visiting node $\deltab_\layerk$ decreases as $1/((\Nzeros(\deltab_\layerk)+L)!)$ for increasing $\Nzeros(\deltab_\layerk)$. This  
suggests that nodes corresponding to a first symbol error at high tree levels, i.e, nodes with large $\Nzeros(\deltab_\layerk)$, are in general pruned with higher probability than those
corresponding to a first symbol error at low tree levels, i.e., nodes with small $\Nzeros(\deltab_\layerk)$ (provided, of course, that $\|\deltab_{\layerk}\|_{2}$ is constant in this comparison). 

\subsubsection{Average TPB Comparison}\label{sec.average.TPB.comparison}
Let us next compare the high-SNR TPB of SD-$\linf$ to that of SD-$\ltwo$. We start by defining 
\begin{equation}\label{eqn.radii.ratio.def}
\rratio =  \frac{\rtwo^2}{\rinf^2}.
\end{equation} 
For $\deltab_\layerk \neq \vec{0}$, the results in \eqref{eqn.asymptotic.behavior.Pbk.inf} and \eqref{eqn.asymptotic.behavior.Pbk.two} imply
\begin{equation}\label{eqn.asympt.comparison.SDl2linf}
	\P\big[\|\zb_\layerk(\deltab_\layerk)\hspace{-0.04cm}\|_\infty \leq \rinf\big] \leqasympt  \text{P}\big[\|\zb_\layerk(\deltab_\layerk)\|_2  \leq \rtwo \big],  \quad\SNR \rightarrow \infty
\end{equation}
if 
\begin{equation}\label{eqn.cond.more.efficient.TPB}
A(\Nzeros(\deltab_\layerk)) \leq \frac{1}{(\layerk+L)!} \,\rratio^{\layerk+L}
\end{equation}
and vice-versa, i.e.,  
\begin{equation}\label{eqn.asympt.comparison.SDl2linf2}
	\P\big[\|\zb_\layerk(\deltab_\layerk)\hspace{-0.04cm}\|_\infty \leq \rinf\big] \gasympt  \text{P}\big[\|\zb_\layerk(\deltab_\layerk)\|_2  \leq \rtwo \big],  \quad\SNR \rightarrow \infty
\end{equation}
if 
\begin{equation}\label{eqn.cond.more.efficient.TPB2}
A(\Nzeros(\deltab_\layerk)) > \frac{1}{(\layerk+L)!} \,\rratio^{\layerk+L}. 
\end{equation}
Hence, the high-SNR average pruning probability of a node $\deltab_{\layerk}\neq \vec{0}$ for SD-$\linf$ as compared to SD-$\ltwo$ is entirely described by the two functions 
$A(\Nzeros(\deltab_\layerk))$ and $1/(\layerk+L)! \,\rratio^{\layerk+L}$, $\layerk = 1,\dots, \tdim$. 
Since $A(\Nzeros(\deltab_\layerk)) \leq 1$, the condition in \eqref{eqn.cond.more.efficient.TPB} is certainly satisfied for {\em all} nodes $\deltab_\layerk \neq \vec{0}$ and tree levels  $\layerk = 1, \dots, \bar{\layerk}$, with $\bar{\layerk}$ being the largest integer satisfying  
\begin{equation}\label{eqn.cond.more.efficient.TPB.all.nodes}
	\sqrt[\layerk+L]{(\layerk+L)!} \leq   \rratio. 
\end{equation}
We set $\bar{\layerk} = 0$ if no integer satisfies  \eqref{eqn.cond.more.efficient.TPB.all.nodes}. 
Using \eqref{eqn.asympt.comparison.SDl2linf} in the expressions for $\E\{\NN_{\infty,\layerk}\}$ \eqref{eqn.NN.advancedk}  and $\E\{\NN_{2,\layerk}\}$ \eqref{eqn.NN2.final.prob}  for the terms 
with $\deltab_\layerk \neq \vec{0}$ and\footnote{
As already noted in footnote 2, we have $\P\big[\|\zb_\layerk(\vec{0})\hspace{-0.04cm}\|_\infty \leq \rinf\big] =  [\gamma_{1}(\rinfsigma)]^{\layerk+L}$ and 
$\text{P}\big[\|\zb_\layerk(\vec{0})\|_2  \leq \rtwo \big] = \gamma_{\layerk+L}(\rtwosigma)$, where $\left[\gamma_{1}(\rinfsigma)\right]^\rdim =  \gamma_{\rdim}(\rtwosigma ) = 1-\epsilon$ due to \eqref{eqn.ctwo.epsilon}
and \eqref{eqn.noise.ctwo}. 
It follows that the condition $\P\big[\|\zb_\layerk(\vec{0})\hspace{-0.04cm}\|_\infty \leq \rinf\big] \leq \text{P}\big[\|\zb_\layerk(\vec{0})\|_2  \leq \rtwo \big]$ is equivalent to $[\gamma_{1}(\rinfsigma)]^{\layerk+L} \leq \gamma_{\layerk+L}(\rtwosigma)$. Furthermore, using $\gamma_{1}(\rinfsigma) = (1-\epsilon)^{1/\rdim} = [\gamma_{\rdim}(\rtwosigma)]^{1/\rdim}$, this condition can be written as  
$[\gamma_{\rdim}(\rtwosigma)]^{1/\rdim} \leq [\gamma_{\layerk+L}(\rtwosigma)]^{1/(\layerk+L)}$,  which according to \eqref{eqn.UB.root.lower.DOFs} holds for all $\layerk =1,\dots,\tdim$. 
}  $\P\big[\|\zb_\layerk(\deltab_\layerk)\hspace{-0.04cm}\|_\infty \leq \rinf\big] \leq \text{P}\big[\|\zb_\layerk(\deltab_\layerk)\|_2  \leq \rtwo \big]$
for the terms with $\deltab_{\layerk} = \vec{0}$,  we can now infer that 
\begin{equation}\label{eqn.asympt.TPB.lower.compl}
 \E\{\NN_{\infty,\layerk}\}  \leqasympt  \E\{\NN_{2,\layerk}\}, \quad\SNR \rightarrow \infty
 \end{equation}
for $\layerk = 1, \dots, \bar{\layerk}$, or equivalently, in the high-SNR regime, the average number of nodes 
visited by SD-$\linf$ up to tree level $\bar{\layerk}$ (corresponding to tree levels closer to the root) is smaller than that for SD-$\ltwo$. 
Furthermore, if $\bar{\layerk}=\tdim$
\begin{equation}\label{eqn.asympt.TPB.lower.compl.total}
 \E\{\NN_{\infty}\}  \leqasympt  \E\{\NN_{2}\}, \quad\SNR \rightarrow \infty 
 \end{equation} 
 since \eqref{eqn.asympt.TPB.lower.compl} then holds for all tree levels $\layerk=1,\dots, \tdim$.  We will next show that for the radii chosen according to
 \eqref{eqn.cinf.epsilon} and \eqref{eqn.ctwo.epsilon}, we can, indeed,  have $\bar{\layerk}=\tdim$. In the following, we write $\rratio(\epsilon)$ to emphasize the dependence of the radii ratio on the parameter $\epsilon \in [0,1]$. 
 In Appendix \ref{app.radii.ratio} it is shown that $\rratio(\epsilon)$ is a nondecreasing function of $\epsilon$ and furthermore 
\begin{equation}\label{eqn.limit.ratii.ratio.1}
\underset{\epsilon \rightarrow 0} {\lim}\,\, \rratio(\epsilon) = 1
 \quad \text{and}\quad \underset{\epsilon \rightarrow 1} {\lim}\,\, \rratio(\epsilon) = \sqrt[\rdim]{\rdim!}
\end{equation}
which implies 
\begin{equation}\label{eqn.bound.radii.ratio}      
 1 \leq \rratio(\epsilon) \leq \!\sqrt[\rdim]{\rdim!}\,. 
\end{equation} 
We can therefore conclude that $\bar{\layerk}$ is a nondecreasing function of $\epsilon$ taking on any value in $[0,M]$ (achieved by varying the parameter $\epsilon$) with the following two extreme cases:
\begin{itemize}
\item For $\epsilon \rightarrow 1$, we get $\bar{\layerk} \rightarrow \tdim$ so that  \eqref{eqn.asympt.TPB.lower.compl.total} holds. This indicates that in the high-SNR regime 
SD-$\linf$ will have a smaller total complexity than SD-$\ltwo$ if $\epsilon$  is sufficiently close to $1$. 
\item For $\epsilon \rightarrow 0$,  we have $\bar{\layerk} \rightarrow 1$ for $L=0$ and $\bar{\layerk} \rightarrow 0$  for $L >0$. 
Equivalently, if $\epsilon$ is sufficiently close to $0$, \eqref{eqn.asympt.TPB.lower.compl} holds for the first tree level if $L =0$ and holds for none of the tree levels if $L>0$.  
In particular, for $\epsilon \rightarrow 0$, we have $\rratio(\epsilon) \rightarrow 1$ and hence $\rinf \equalasympt \rtwo$, $\epsilon \rightarrow 0$. This implies that the hypercube of radius $\rinf$ contains the hypersphere of radius $\rtwo$ and the total complexity of SD-$\linf$ will trivially be higher than that of 
SD-$\ltwo$. In general, SD-$\linf$ will have a higher total complexity than SD-$\ltwo$ if $\epsilon$ is small. 
\end{itemize}
In summary, varying the parameter $\epsilon$ has a significant impact on the total complexity of SD-$\linf$ relative to that of SD-$\ltwo$. In particular, the total complexity of SD-$\linf$ can be higher or lower 
than that of SD-$\ltwo$. 
 
Let us next study the average TPB of SD-$\linf$ as compared to SD-$\ltwo$ for the tree levels $\layerk =  \bar{\layerk}+1, \dots, \tdim$.
Here, we have $1/((\layerk+L)!) \rratio^{\layerk+L} \leq 1$ so that condition \eqref{eqn.cond.more.efficient.TPB}  will not necessarily be satisfied for all 
nodes $\deltab_\layerk \neq \vec{0}$. This means that for tree levels $\layerk \geq  \bar{\layerk}+1$, in the high-SNR regime, SD-$\ltwo$ may prune certain nodes with higher probability than SD-$\linf$.  Since $A(\Nzeros(\deltab_\layerk)) \geq 1/(\Nzeros(\deltab_\layerk)+L)!$, the 
condition in \eqref{eqn.cond.more.efficient.TPB2} is certainly satisfied for all nodes \sloppy
$\deltab_\layerk$, $\layerk =  \bar{\layerk}+1, \dots, \tdim$, with $\Nzeros(\deltab_\layerk) = 1,\dots, \overline{\layer}(\layerk)$, where $\overline{\layer}(\layerk)$ is the largest integer 
 $\layer \in  [1,\layerk]$ satisfying
\begin{equation}\label{eqn.bar.m}
\frac{1}{(\layer+L)!} > \frac{1}{(\layerk+L)!} \rratio^{\layerk+L}. 
\end{equation}
A high-SNR statement dual to \eqref{eqn.asympt.TPB.lower.compl} based on \eqref{eqn.bar.m} can, in general, not be given for tree levels $\layerk \geq \bar{\layerk}+1$ as \eqref{eqn.bar.m} applies only to 
a certain subset of nodes at a specific tree level $\layerk \geq \bar{\layerk}+1$. 
Some insight can, however, be gained by studying the cardinalities of these subsets at high tree levels. Let us denote  the cardinality of the set of nodes $\deltab_{\layerk}$  
satisfying \eqref{eqn.asympt.comparison.SDl2linf2}  for some given transmitted data subvector $\db_{\layerk}'$ (recall that $\deltab_{\layerk} = \db_{\layerk}-\db_{\layerk}'$ with $\db_{\layerk}, \db_{\layerk}' \in \symbolset^{\layerk}$)  by 
$\overline{\NN_{\layerk}}$.
From the previous paragraph we know that this set includes for sure all nodes with $\Nzeros(\deltab_\layerk) = 1,\dots, \overline{\layer}(\layerk)$, $\layerk =  \bar{\layerk}+1, \dots, \tdim$.  
Hence,  we have 
\[
\overline{\NN_{\layerk}} \geq \sum_{\layer=1}^{\overline{\layer}(\layerk)} (|\symbolset|-1) |\symbolset|^{\layerk-\layer} =  |\symbolset|^{\layerk} (1- |\symbolset|^{-\overline{\layer}(\layerk)}), \quad  
\layerk =  \bar{\layerk}+1, \dots, \tdim.
\]
For $\overline{\layer}(\layerk) \geq 1$ (which  is always the case for $L=0$ and all tree levels $\layerk \geq  \bar{\layerk}+1$) and $|\symbolset| > 2$, 
we see that more nodes at tree level $\layerk \geq  \bar{\layerk}+1$ will be pruned with a 
higher probability by SD-$\ltwo$ than by SD-$\linf$. 
Even more, since the RHS of \eqref{eqn.bar.m} is a decreasing function\footnote{
This can be proved by showing that 
$\layerk + 1+L > \rratio$, for $\layerk \geq  \bar{\layerk}+1$. Applying the definition of $\bar{\layerk}$ in \eqref{eqn.cond.more.efficient.TPB.all.nodes},  
we get $\sqrt[\bar{\layerk}+1+L]{(\bar{\layerk}+1+L)!} >   \rratio$, which together with
$\sqrt[n]{n!} \leq n$, for $n \in \Nnum$, establishes the desired result. 
}
 of $\layerk$ for all $\layerk \geq  \bar{\layerk}+1$, $\overline{\layer}(\layerk)$ 
is a nondecreasing function of $\layerk$ that becomes large if $\layerk$ is large. In this case, most out of the $|\symbolset|^{\layerk}$ nodes at tree level $\layerk$ 
will be pruned with a higher probability by SD-$\ltwo$ than by SD-$\linf$.
Note, as shown above, that this behavior is reversed at tree levels close to the root (i.e., up to tree level $\bar{\layerk}$), where {\em all} 
 nodes are pruned with higher probability by SD-$\linf$ than by SD-$\ltwo$. 

\subsubsection{Relation to Geometric Properties}\label{sec.average.pruning.geometric}
The results on the average high-SNR TPB are nicely supported by simple geometric considerations. 
We now assume $L = \rdim - \tdim =0$ and argue that the average number of visited nodes at tree level $\layerk$ is roughly determined by the volume of the involved search space of dimension $\layerk$ (see, e.g., \cite{hass_sp03_part_i}). 
In the SD-$\ltwo$ case the search spaces are {\em hyperspheres}, whereas in the SD-$\linf$ case they are  {\em hypercubes}. We will next see that 
analyzing the volume behavior of the hyperspheres and hypercubes associated to SD-$\ltwo$ and SD-$\linf$, respectively, as a function of the dimension, or equivalently, as a function of the tree level   $\layerk$, recovers many of the insights obtained in the previous section. 
 
For SD-$\ltwo$, the search space at tree level $\layerk$ is a hypersphere of radius $\rtwo$ in $2\layerk$ real-valued dimensions with volume 
(e.g., \cite{ball_97})
\begin{equation}\label{eqn.volume.sphere}
	V_{2,\layerk} = \frac{\pi^{\layerk} (\rtwo^2)^{\layerk}}{\layerk !}. 
\end{equation}
For SD-$\linf$, the search space consists of the set of all $\layerk$ pairs $x_{i,1}, x_{i,2} \in \Rnum$, $i = 1,\dots, \layerk$, that satisfy $x_{i,1}^{2} + x_{i,2}^{2} \leq \rinf^2$,  $\forall i$, with the corresponding volume 
\begin{equation}\label{eqn.volume.cube}
	V_{\infty,\layerk} = \pi^{\layerk} (\rinf^2)^{\layerk}.
\end{equation}

From \eqref{eqn.volume.sphere} and \eqref{eqn.volume.cube}  it follows that $V_{\infty,\layerk} \leq V_{2,\layerk}$ for all tree levels $\layerk = 1, \dots, \bar{\layerk}$ with $\bar{\layerk}$ being the largest integer
satisfying 
 \begin{equation}\label{eqn.kbar.radii.ratio}
	\sqrt[\layerk]{\layerk!} \leq   \rratio
\end{equation}
and vice-versa, i.e., $V_{\infty,\layerk} > V_{2,\layerk}$ for $\layerk = \bar{\layerk}+1,\dots, \tdim$.  This indicates that SD-$\linf$ prunes more nodes than SD-$\ltwo$ at tree levels closer to the root, whereas this behavior is reversed at tree levels closer to the leaves. 
Even more, the threshold tree level $\bar{\layerk}$ defined by \eqref{eqn.kbar.radii.ratio} (found through analyzing the volume behavior of the search spaces) equals the threshold tree level 
\eqref{eqn.cond.more.efficient.TPB.all.nodes} found through a high-SNR analysis of the pruning probabilities. 

\subsection{Instantaneous TPB} \label{sec.pruning.behavior.close.root}
The insights and results on the average TPB found in the previous section extend, to a certain degree, to 
 the instantaneous TPB (i.e., the TPB for a given channel, data, and noise realization).  Recall that a node $\deltab_{\layerk}$ is pruned by SD-$\linf$ if $\|\zb_\layerk(\deltab_\layerk)\hspace{-0.04cm}\|_{\infty}^2 > \rinf^2$ and by SD-$\ltwo$ if
  $\|\zb_\layerk(\deltab_\layerk)\hspace{-0.04cm}\|_{2}^2 > \rtwo^2$. 
Noting that $\zb_\layerk(\deltab_\layerk)$ is a length $\layerk\!+\!L$ vector and applying \eqref{eqn.linf.ltwo.bounds} yields 
\[
(\layerk\!+\!L) \|\zb_\layerk(\deltab_\layerk)\|^2_\infty \geq \|\zb_\layerk(\deltab_\layerk)\|^2_2. 
\]
A node pruned by SD-$\ltwo$ is therefore guaranteed to be pruned by SD-$\linf$ as well if 
\[
	   \frac{ \rtwo^2}{\layerk\!+\!L} \geq \rinf^2. 
\]
Consequently, we have  $\NN_{\infty,\layerk} \leq \NN_{2,\layerk}$ for $\layerk = 1,\dots, \bar{\layerk}_{\text{I}}$ with
\begin{equation}\label{eqn.inst.cross.over.tree.level}
 \bar{\layerk}_{\text{I}} = \text{max}\{\lfloor \rratio \rfloor-L, 0\}. 
\end{equation}
We can therefore conclude that SD-$\linf$ prunes (in an instantaneous sense) more nodes than SD-$\ltwo$ at tree levels close to the root, more specifically, for all tree levels up to level $\bar{\layerk}_{\text{I}}$
 (cf.\ Section \ref{sec.average.TPB.comparison} for the corresponding result in terms of average TPB).  We furthermore note that the radii ratio $\rratio$ not only determines the average TPB but also the instantaneous TPB.

Next, let us compare the instantaneous and the average high-SNR TPB results quantitatively. We have $\NN_{\infty,\layerk} \leq \NN_{2,\layerk}$, for $\layerk = 1,\dots, \bar{\layerk}_{\text{I}}$, with 
$\bar{\layerk}_{\text{I}}$ defined in \eqref{eqn.inst.cross.over.tree.level}, while in terms of the average TPB, we have $\E\{\NN_{\infty,\layerk}\}  \leqasympt  \E\{\NN_{2,\layerk}\}$, $\SNR \rightarrow \infty$, for 
$\layerk = 1, \dots, \bar{\layerk}$,  
 with $\bar{\layerk}$ defined in \eqref{eqn.cond.more.efficient.TPB.all.nodes}.
Due to $\sqrt[\layerk+L]{(\layerk+L)!} \leq \layerk+L$ (since, evidently, $(\layerk+L)! \leq (\layerk+L)^{\layerk+L}$), we obtain $\bar{\layerk}_{\text{I}} \leq \bar{\layerk}$, which shows that the instantaneous TPB 
result $\NN_{\infty,\layerk} \leq \NN_{2,\layerk}$ extends, in general, to fewer tree levels than the average TPB  result $\E\{\NN_{\infty,\layerk}\}  \leqasympt  \E\{\NN_{2,\layerk}\}$, $\SNR \rightarrow \infty$. 
This, of course, makes sense since $\NN_{\infty,\layerk} \leq \NN_{2,\layerk}$ 
implies $\E\{\NN_{\infty,\layerk}\}  \leqasympt  \E\{\NN_{2,\layerk}\}$, $\SNR \rightarrow \infty$, but not vice-versa. 

\section{The Truth and the Beautiful: $\ltilde$-Norm SD}   \label{sec.ltilde.SD}
As already mentioned, the SD-$\linf$ VLSI implementation in \cite{burg05_vlsi} is actually based on the $\ltilde$-norm rather than the $\linf$-norm; the corresponding tree search is conducted using 
the recursive metric computation rule 
$\|\zb_{\layerk}(\db_{\layerk})\hspace{-0.04cm}\|_\tinfty = \text{max}\big\{\|\zb_{\layerk-1}(\db_{\layerk-1})\hspace{-0.04cm}\|_{\tinfty},\,  \|[\zb(\db)]_{M-\layerk+1}\|_{\tinfty}\big\}$ together with 
the 
partial BC 
\begin{equation}\label{eqn.PBC.schlange}
    \|\zb_\layerk(\db_\layerk)\hspace{-0.04cm}\|_\tinfty \leq \rtilde
\end{equation}
where $\rtilde$ denotes the  ``radius'' associated with SD-$\ltilde$. Consequently, SD-$\ltilde$ finds all data vectors $\db$ satisfying 
$ \|\zb(\db)\hspace{-0.04cm}\|_\tinfty \leq \rtilde$ and chooses, within this set, the vector
\begin{equation}\label{eqn.ml.schlange}
\widehat{\db}_\tinfty =\underset{\db \in \symbolset^\tdim} { \mbox{arg\, min}}\,
\|\zb(\db)\hspace{-0.04cm}\|_\tinfty.
\end{equation} 
We next show how the error probability (see Section \ref{sec.error.performance}) and complexity  (see Section \ref{sec.expected.complexity}) 
results obtained for SD-$\linf$ carry over to SD-$\ltilde$. Most  results in this section are based on the simple inequalities   
\begin{equation}\label{eqn.lschlange.linf.bounds}
	\frac{1}{2} \| \vec{x} \|_\infty^2 \leq \| \vec{x} \|_\tinfty^2 \leq \| \vec{x} \|_\infty^2, \quad\vec{x} \in \Cnum^{\rdim}. 
\end{equation}

\subsection{Error Probability of SD-$\ltilde$}\label{eqref.error.prob.tilde}
\paragraph{Distance Properties} Combining \eqref{eqn.lschlange.linf.bounds} with \eqref{eqn.linf.ltwo.bounds} and following the steps in \eqref{eqn.distance.property} yields 
\[
	\big\|\rb-\Hb\, \widehat{\db}_\tinfty\big\|^2_2 \leq 2\rdim \big\|\rb - \Hb \, \widehat{\db}_\text{ML}\big\|^2_2
\]
which shows that, compared to SD-$\linf$, we  essentially incur at most a factor of $\sqrt{2}$ increase in terms of the distance $\big\|\rb-\Hb\, \widehat{\db}_\tinfty\big\|_{2}$ realized by SD-$\ltilde$. 

\paragraph{Diversity Order and SNR Gap} 
With \eqref{eqn.ml.schlange}, an upper bound on the PEP of SD-$\ltilde$ is given by 
\[
\P_{\db' \rightarrow  \db,\tinfty}(\SNR) \leq  \text{P}\Big[  \|\zb(\db)\hspace{-0.04cm}\|_\tinfty \leq  \|\zb(\db')\hspace{-0.04cm}\|_\tinfty \Big].
\]
Next, following the steps in \eqref{eqn.PEP.linf.intermediate} -- \eqref{eqn.PEP.SDlinf.UB1} for SD-$\linf$,  using the bounds \eqref{eqn.lschlange.linf.bounds} and  \eqref{eqn.linf.ltwo.bounds}, yields  
\begin{equation}\label{eqn.PEP.SDschlange.UB1}
\P_{\db' \rightarrow  \db,\, \tinfty}(\SNR)  \leq \text{P}\bigg[ \|\wb\|_2 \geq  \frac{1}{\sqrt{2\rdim}\!+\!1}  \| \Hb \deltab\|_2  \bigg].
\end{equation}
Employing the same arguments as in the SD-$\linf$ or in the SD-$\ltwo$ case in Section \ref{sec.PEP}, we can conclude 
that $\P_{\db' \rightarrow  \db,\, \tinfty}(\SNR)$ has the same SNR exponent as $\P_{\db' \rightarrow  \db,\, \infty}(\SNR)$ and $\P_{\db' \rightarrow  \db,\, \text{ML}}(\SNR)$. 
Furthermore, from \eqref{eqn.PEP.SDschlange.UB1} we obtain $\P_{\db' \rightarrow  \db,\, \tinfty}(\SNR)  \leq \text{UB}_{\tinfty}(\SNR)$, where $ \text{UB}_{\tinfty}(\SNR)$ is given by  
$\text{UB}_{\infty}(\SNR)$ in \eqref{eqn.upper.bound.PEP.SDlinf} 
with the factor $\sqrt{\rdim}$ replaced by $\sqrt{2\rdim}$.  Accordingly, the asymptotic SNR gap $\widetilde{\beta}$ between $\text{UB}_{\tinfty}(\SNR)$ and $\text{LB}_{\text{ML}}(\SNR)$, as defined in  
\eqref{eqn.lower.bound.PEP.ML}, i.e., $\text{UB}_{\tinfty}(\SNR)  \equalasympt   \text{LB}_{\text{ML}}(\SNR/\widetilde{\beta})$, $\SNR \rightarrow \infty$, is given by $\beta$ in \eqref{eqn.snr.gap.beta} with 
the factor $\sqrt{\rdim}$ replaced by $\sqrt{2\rdim}$. This corresponds to an increase of a factor of roughly two in the corresponding upper bound on the SNR gap as compared to that achieved by SD-$\linf$ 
\eqref{eqn.snr.gap.beta}. 
Finally, employing the arguments used in Section  \ref{sec.overall.error.probability},  these statements carry over to the total error probability in a straightforward fashion
 showing that SD-$\ltilde$ (like SD-$\linf$ and SD-$\ltwo$)  {\em achieves full diversity order} $\rdim$ with an asymptotic SNR gap to ML detection that increases at most linearly in $\rdim$. 

\subsection{Complexity of SD-$\ltilde$}\label{sec.ltilde.SD.complexity}

With \eqref{eqn.PBC.schlange} and following the steps \eqref{eqn.N.realization} -- \eqref{eqn.NN.advancedk}, we obtain the complexity $\E\{\NN_{\tinfty,\layerk}\}$ of SD-$\ltilde$ at tree level $\layerk$ 
as 
\begin{equation}\label{eqn.NN.advancedk.schlange}
	\E\{\NN_{\tinfty,\layerk}\} = \frac{1}{|\symbolset|^\layerk} \sum_{ \deltab_\layerk} \P\big[ \|\zb_\layerk(\deltab_\layerk)\hspace{-0.04cm}\|_{\tinfty} \leq \rtilde\big]
\end{equation}
with the total complexity given by $\E\{\NN_{\tinfty}\}  = \sum_{\layerk=1}^{\tdim} \E\{\NN_{\tinfty,\layerk}\}$. As in the case of SD-$\linf$,  invoking the fact that the 
 elements of $\zb_\layerk(\deltab_\layerk)$  (conditioned on $\deltab_\layerk$) are statistically independent 
(cf.\ \eqref{eqn.prod.stat.independent}), we get 
\begin{equation}\label{eqn.prod.stat.independent.tilde}
 \P\big[\|\zb_\layerk(\deltab_\layerk)\hspace{-0.04cm}\|_\tinfty \leq \rtilde\big] = \prod_{i=1}^{\layerk+L} \P\Big[ \big\|[\zb_\layerk(\deltab_\layerk)]_i \big\|_{\tinfty} \leq \rtilde\Big].
\end{equation}
The real and imaginary parts of the bottom $L$ elements of $\zb_\layerk(\deltab_\layerk)$ are i.i.d.\  $\N(0,\sigma^2/2)$ so that
\[
\P\Big[ \big\|[\zb_\layerk(\deltab_\layerk)]_i \big\|_{\tinfty} \leq \rtilde\Big] =
\left[
\gamma_{\frac{1}{2}}\!\!\left(\frac{\rtilde}{\sigma^2}\right)
\!\right]
^{2}, \,\,\,\, i = \layerk+1,\dots,\layerk+L,
\]
which, upon insertion into \eqref{eqn.prod.stat.independent.tilde}, yields 
\begin{align}
\P\big[\|\zb_\layerk(\deltab_\layerk)\hspace{-0.04cm}\|_\tinfty \leq \rtilde\big]  = & \nonumber \\ 
& \hspace{-2.5cm} 
\left[
\gamma_{\frac{1}{2}}\!\!\left(\frac{\rtilde}{\sigma^2}\right)
\!\right]
^{2L} \prod_{\layer=1}^{\layerk} \P\Big[ \big\|[\zb(\deltab)]_{\tdim-\layer+1} \big\|_{\tinfty} \leq \rtilde\Big], \label{eqn.order.stat.schlange}
\end{align} 
analogously to \eqref{eqn.order.stat}. 
An analytic expression for $\P\Big[ \big\|[\zb(\deltab)]_{\tdim-\layer+1} \big\|_{\tinfty} \leq \rtilde\Big]$ can be obtained if 
 $\deltanb_{\tdim-\layer+1}$ is  {\em purely real, purely imaginary,} or {\em equal to zero}. 
 For these cases the real- and imaginary parts of $[\zb(\deltab)]_{\tdim-\layer+1}$ are statistically independent, which gives (see Appendix  \ref{app.distribution.schlange.independent}) 
 \begin{equation}\label{eqn.dist.schlange}
\P\Big[ \big\|[\zb(\deltab)]_{\tdim-\layer+1} \big\|_{\tinfty} \leq \rtilde\Big] =  \gamma_{\frac{1}{2}}\!\bigg(\frac{C_\tinfty^2}{\sigma^2_\layer}\bigg) \sum_{s = 0}^{\infty} D_{s}(\deltab_\layer) \,
\gamma_{s+\frac{1}{2}}\! \bigg(\frac{C_\tinfty^2}{\sigma^2_\layer}\bigg)
\end{equation}
where 
$D_{s}(\deltab_\layer)$ is defined in \eqn{eqn.coefficient.schlange} and $\sigma^2_\layer$ is specified in \eqref{eqn.def.sigmam}. 
For the general case of $\deltanb_{\tdim-\layer+1}$ having a nonzero real and a nonzero imaginary part, i.e., $\deltanb_{\real, \tdim-\layer+1} \neq 0$ and $\deltanb_{\imag, \tdim-\layer+1} \neq 0$, 
the real- and imaginary parts of $[\zb(\deltab)]_{\tdim-\layer+1} $ are statistically dependent, which seems to make it difficult to find a closed-form expression for 
$\P\Big[ \big\|[\zb(\deltab)]_{\tdim-\layer+1} \big\|_{\tinfty} \leq \rtilde\Big]$. On can, however, resort to upper and lower bounds. 
In particular, it follows from \eqref{eqn.lschlange.linf.bounds} that 
\begin{align}
 \P\Big[ \big\|[\zb(\deltab)]_{\tdim-\layer+1} \big\|_{\tinfty} \leq \rtilde\Big] &\geq  \P\Big[ \big|[\zb(\deltab)]_{\tdim-\layer+1} \big| \leq \rtilde\Big] \label{eqn.Nk.tilde.LB} \\  
\P\Big[ \big\|[\zb(\deltab)]_{\tdim-\layer+1} \big\|_{\tinfty} \leq   \rtilde\Big] &\leq \P\Big[ \big|[\zb(\deltab)]_{\tdim-\layer+1} \big| \leq \sqrt{2}\,\rtilde\Big] \label{eqn.Nk.tilde.UB}.
\end{align}
The RHS expressions of \eqref{eqn.Nk.tilde.LB} and \eqref{eqn.Nk.tilde.UB} can now be expressed analytically using  \eqref{eqn.dist}.  Together with  \eqn{eqn.order.stat.schlange} and \eqref{eqn.dist.schlange}  this provides upper and lower bounds on $\P\big[\|\zb_\layerk(\deltab_\layerk)\hspace{-0.04cm}\|_\tinfty \leq \rtilde\big]$ and thus on $\E\{\NN_{\tinfty,\layerk}\}$, $\layerk=1,\dots, \tdim$, and $\E\{\NN_{\tinfty}\}$. We do not display the resulting final expressions as they are rather involved and do not contribute to deepening the understanding. 
Corresponding numerical results are provided in Section \ref{sec.simu.complexity.versus.tree.level}.
 
Following the choice of the radii for SD-$\linf$ and SD-$\ltwo$ in \eqref{eqn.noise.cinf} and \eqref{eqn.noise.ctwo}, respectively, $\rtilde$ is obtained by setting 
\begin{equation}\label{eqn.noise.ctilde}
	\text{P}\big[\|\nb\|_\tinfty \leq \rtilde  \big]  = 
	\bigg[
	\gamma_{\frac{1}{2}\!\!}\!\left(\frac{\rtilde^2}{\sigma^2}\right)
	\!\bigg]
	^{2\rdim} =  1-\epsilon 
\end{equation}
which results in (cf.\ \eqref{eqn.cinf.epsilon} and  \eqref{eqn.ctwo.epsilon}) 
\begin{equation}\label{eqn.ctilde.epsilon}
	\rtilde^2 =\sigma^2\,\gamma_{\frac{1}{2}\!\!}^{-1}\!\!\left(\!\sqrt[2\rdim]{1-\epsilon}\,\right).
\end{equation}

\subsection{Asymptotic Complexity Analysis}
We next show that SD-$\ltilde$ with $\rtilde$ chosen according to \eqref{eqn.ctilde.epsilon} exhibits  exponential complexity scaling in the problem size $\tdim$.  This will be accomplished by following the same approach as for SD-$\linf$ and SD-$\ltwo$ (see Sections 
\ref{sec.lower.bound} and  \ref{sec.asymptotic.analysis.of.lower.bound}, respectively), i.e., by developing an analytically tractable lower bound on $\E\{\NN_{\tinfty}\}$ and then establishing that this bound scales 
exponentially in $\tdim$. For the sake of simplicity of exposition, we set  $L = \rdim-\tdim = 0$ in the remainder of this section. 

The approach we take is to lower bound the complexity of SD-$\ltilde$ by the complexity of SD-$\linf$ with a suitably scaled radius. Once this is accomplished, exponential complexity scaling of SD-$\ltilde$ can be established  by straightforward modifications of the key steps in the corresponding proof for the SD-$\linf$ case. 
We start by applying \eqref{eqn.lschlange.linf.bounds} to get $\text{P}\big[\|\nb\|_\tinfty \leq \rtilde  \big] \leq \text{P}\big[\|\nb\|_\infty \leq \sqrt{2}\,\rtilde\big]$, which, together with 
\eqref{eqn.noise.ctilde}, 
 results in 
 \begin{equation}\label{eqn.asympt.SD.ltilde.zw1}
 \text{P}\big[\|\nb\|_\infty \leq \sqrt{2}\,\rtilde\big] \geq 1-\epsilon.
 \end{equation}
According to \eqref{eqn.noise.cinf}, we also have  $\text{P}\big[\|\nb\|_\infty \leq \rinf \big] = 1-\epsilon$, which by comparing with \eqref{eqn.asympt.SD.ltilde.zw1} results in 
$\rtilde \geq  \rinf/\sqrt{2}$  for any given $\epsilon$. This, together with 
 $ \|\zb_\layerk(\deltab_\layerk)\hspace{-0.04cm}\|_{\tinfty}  \leq \|\zb_\layerk(\deltab_\layerk)\hspace{-0.04cm}\|_{\infty}$, implies 
\[
 \P\big[ \|\zb_\layerk(\deltab_\layerk)\hspace{-0.04cm}\|_{\tinfty} \leq \rtilde\big] \geq  \P\Big[ \|\zb_\layerk(\deltab_\layerk)\hspace{-0.04cm}\|_{\infty} \leq \rinf/\sqrt{2} \Big].
\]
Hence, the complexity of SD-$\ltilde$ with radius $\rtilde$ is lower-bounded by the complexity of SD-$\linf$ with radius $\rinf/\sqrt{2}$, where the radii $\rtilde$ and $\rinf$ are related through the parameter 
$\epsilon$. It remains to follow the asymptotic complexity analysis of 
SD-$\linf$ performed in Section \ref{sec.asympt.complexity.analysis}, where $\rinf$ is now replaced by $\rinf/\sqrt{2}$. 
 Invoking the lower bounds 
\eqref{eqn.lower.bound.intermediate} and \eqref{eqn.lower.bound.prefactor} with $\rinf$ replaced by $\rinf/\sqrt{2}$ , we get 
\[
 \P\big[ \|\zb_\layerk(\deltab_\layerk)\hspace{-0.04cm}\|_{\tinfty} \leq \rtilde\big] \geq  
 \left[
 \gamma_{1}\!\!\left(\frac{C_\infty^2}{2\sigma^2}\right)
 \right]
 ^{\!\tdim}
    \bigg(1+\frac{\|\deltab_\layerk \|_2^2}{\tdim \sigma^2}\bigg)^{\!-\layerk}.
\]
Noting that  
\begin{equation}\label{eqn.tilde.epsilon}
	\left[
	\gamma_{1}\!\!\left(\frac{C_\infty^2}{2\sigma^2}\right)
	\right]
	^{\!\tdim} = \left[1-\sqrt{1-\sqrt[M]{1-\epsilon}} \,\right]^{\!\tdim}
\end{equation}
we furthermore obtain 
\[
 \P\big[ \|\zb_\layerk(\deltab_\layerk)\hspace{-0.04cm}\|_{\tinfty} \leq \rtilde\big] \geq  
 \left[1-\sqrt{1-\sqrt[M]{1-\epsilon}} \,\right]^{\!\tdim}
    \bigg(1+\frac{\|\deltab_\layerk \|_2^2}{\tdim \sigma^2}\bigg)^{\!-\layerk}.
\]
Comparing this result with \eqref{eqn.lower.bound.linf}, we can immediately conclude, following the steps \eqref{eqn.lower.bound.linf} -- \eqref{eqn.asymptotic.analysis.fM}, that 
$\E\{\NN_{\tinfty}\} \geq   \tilde{f}(\tdim)$ with 
\[ 
 \tilde{f}(\tdim) =  \left[1-\sqrt{1-\sqrt[M]{1-\epsilon}} \,\right]^{\!\tdim}  \left(\frac{\layerk}{i}\right)^{i} \bigg(1+\frac{\UBsymdiff^2\, i}{\tdim \sigma^2}\bigg)^{\!-\layerk}. 
 \]	
Evidently, we have   
\[
\underset{\tdim \rightarrow\, \infty} {\lim} \text{log}\!\left(1-\sqrt{1-\sqrt[M]{1-\epsilon}} \,\right) = 0 
\]
which implies that 
\[
 \underset{\tdim \rightarrow\, \infty} {\lim}\, \frac{\text{log}\,  \tilde{f}(\tdim) }{\tdim}  = \underset{\tdim \rightarrow\, \infty} {\lim}\, \frac{\text{log}\,  f(\tdim) }{\tdim}
\]
where $f(\tdim)$ was defined in \eqref{eqn.asymptotic.analysis.fM}. Finally, following the steps \eqref{eqn.def.exponentially.complex} -- \eqref{eqn.exponential.complexity.final} establishes
 that the complexity of  SD-$\ltilde$ scales exponentially in the problem size $\tdim$.

\begin{figure}[t]
\begin{center}
\resizebox{8.5cm}{!}{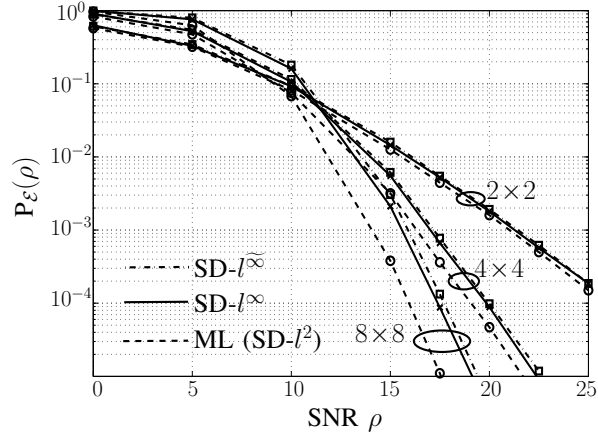}
\vspace{-0.45cm}
\caption{Uncoded total error probability $\P_{\cal E}(\SNR)$ as a function of SNR $\SNR$ for SD-$\ltilde$, SD-$\linf$, and SD-$\ltwo$ (ML) detection for a $2 \times 2$, $4 \times 4$, and $8 \times 8$ MIMO system, respectively, 
and a $4$-QAM symbol alphabet.
 } 
\vspace{-1.0cm}
\label{fig.error.rate.versus.SNR}
\end{center}
\end{figure}

\section{Numerical Results}\label{sec.numerical.results}
In this section, we provide numerical results quantifying some of our analytical findings. All the results in the remainder of this section are based on independently and equally likely transmitted data symbols. 

\subsection{Error Probability}\label{sec.simu.error.probability}

We compare the uncoded error-rate performance of SD-$\linf$ and SD-$\ltilde$ to that of SD-$\ltwo$ (ML) detection by means of Monte-Carlo simulations.   Fig.$\!$ \ref{fig.error.rate.versus.SNR} shows total error probabilities $\P_{\cal E}(\SNR)$ as functions of SNR $\SNR$ for a $2 \times 2$, $4 \times 4$, and $8 \times 8$ MIMO system, respectively, using $4$-QAM symbols in all three cases.  We can observe that both SD-$\linf$ and SD-$\ltilde$ achieve full diversity order and show near-ML performance.  
Indeed, SD-$\linf$ and SD-$\ltilde$ perform much better than suggested by the corresponding upper bounds on the SNR gap  (i.e., $|\symbolset| \beta$ with $\beta$ given by \eqref{eqn.snr.gap.beta} for 
SD-$\linf$ and $|\symbolset| \tilde{\beta}$ with $\tilde{\beta}$ given by \eqref{eqn.snr.gap.beta} with 
the factor $\sqrt{\rdim}$ replaced by $\sqrt{2\rdim}$ for SD-$\ltilde$).  
Consistent with the $\sqrt{2}$-difference in the upper bounds on the corresponding SNR gaps, we can observe that SD-$\ltilde$ performs slightly worse than SD-$\linf$. 
Finally, the results in Fig.\ \ref{fig.error.rate.versus.SNR} show that the performance loss incurred by SD-$\linf$ and SD-$\ltilde$ increases for 
increasing $\tdim = \rdim$.

\subsection{Complexity}\label{sec.simu.complexity}

Next we consider the complexity of SD-$\linf$ and SD-$\ltwo$  for the case of fixed radii $\rinf$ and $\rtwo$  chosen according to 
\eqref{eqn.cinf.epsilon} and  \eqref{eqn.ctwo.epsilon}, respectively, with the same value of $\epsilon$ in both cases (for numerical results on the complexity of SD-$\ltilde$, we refer to Section \ref{sec.simu.pruning.behavior}).   
 The total complexity $\E\{\NN\}$ as a function of $\epsilon$ for SD-$\linf$ 
(see \eqref{eqn.total.complexity} with \eqref{eqn.NNk.final}) 
and SD-$\ltwo$ (see \eqref{eqn.total.complexity.2} with \eqref{eqn.NN2.final}) 
is shown 
 in Fig.\ \ref{fig.C.versus.epsilon} for a $4 \times 4$, $6 \times 6$, and $8 \times 8$ MIMO system, respectively, operating at an SNR of $\rho = 15$dB. 
The following conclusions can be drawn from these results:

\begin{figure}[t]
\begin{center}
\resizebox{8.0cm}{!}{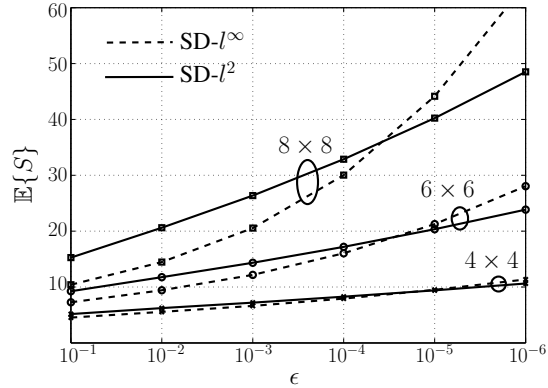}
\vspace{-0.45cm}
\caption{Total complexity  $\E\{\NN\}$ as a function of $\epsilon$
for SD-$\linf$ and SD-$\ltwo$ for a $4 \times 4$, $6 \times 6$, and $8 \times 8$ MIMO system, respectively, and a $4$-QAM symbol alphabet at an SNR of $\rho = 15$dB.}
\vspace{-1.0cm}
\label{fig.C.versus.epsilon}
\end{center}
\end{figure}

\begin{itemize}

\item For a given $\epsilon$, the complexity  of SD-$\linf$ can be higher or lower than that of SD-$\ltwo$.  

\item SD-$\linf$ exhibits a lower complexity than SD-$\ltwo$ for larger values of $\epsilon$, while for smaller values of $\epsilon$, SD-$\linf$ has a higher complexity than SD-$\ltwo$. 
This behavior was indicated by the high SNR-analysis of the TPB of SD-$\linf$ and SD-$\ltwo$ (in particular, see the discussion on the two extreme cases $\epsilon \rightarrow 0$ and $\epsilon \rightarrow 1$ in 
Section \ref{sec.average.TPB.comparison}).

\item The complexity savings of SD-$\linf$ over SD-$\ltwo$ for values of $\epsilon$ close to 1 are more pronounced for increasing $\tdim = \rdim$. 

\item In practice, $\epsilon$ is matched to the target error rate of the system (see the discussion in Section \ref{eqn.choice.of.radii}). In the present example, we operate at $15$dB SNR and the corresponding target error rates can be inferred from 
Fig.\ \ref{fig.error.rate.versus.SNR}, which results 
in 
$\epsilon$ values (target error rates) for which SD-$\linf$ has a lower complexity than SD-$\ltwo$ (cf.\ Fig.\ \ref{fig.C.versus.epsilon}). 
For the $8\times 8$ system, for example, our target error rate at $15$dB SNR, according to Fig.\ \ref{fig.error.rate.versus.SNR},  
is around $10^{-3}$. For this case, the complexity savings of SD-$\linf$ as compared to SD-$\ltwo$ are around $25\%$ according to Fig.\ \ref{fig.C.versus.epsilon},. 
\end{itemize}

\begin{figure}[t]
\begin{center}
\resizebox{8.5cm}{!}{\input Am0_vs_m0_6x6_xfig.pdftex_t}
\vspace{-0.45cm}
\caption{$A\big(\Nzeros(\deltab_\layerk)\big)$ as a function of $\Nzeros(\deltab_\layerk) = 1,\dots, \layerk$ (including the corresponding lower bound $1/(\Nzeros(\deltab_\layerk)!)$) and
the RHS of \eqref{eqn.cond.more.efficient.TPB}  given by $1/(\layerk!) \,\rratio^{\layerk}$ as a function of $\layerk = 1,\dots, \tdim$ for $\epsilon = 10^{-2}$ and  
$\epsilon = 10^{-5}$, respectively, for a $6\times 6$ MIMO system. 
 } 
\vspace{-1.0cm}
\label{fig.mbk}
\end{center}
\end{figure}

\begin{figure}[t]
\begin{center}
\unitlength1mm
\begin{picture}(170,65)
\put(3,5){\resizebox{8.0cm}{!}{\input N_versus_k_6x6_4QAM_001_xfig.pdftex_t}}
\put(84,5){\resizebox{8.0cm}{!}{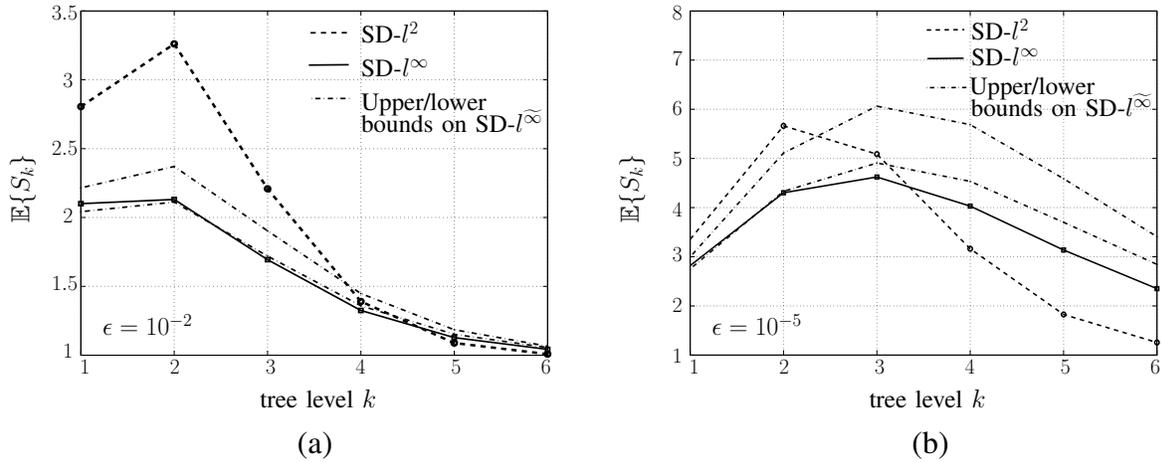}}
\put(42.2,0){(a)}
\put(123.9,0){(b)}
\vspace{-0.65cm}
\end{picture}
\vspace{-0.45cm}
\caption{Complexity $\E\{\NN_\layerk\}$ as a function of the tree level $\layerk$ for SD-$\ltilde$, SD-$\linf$, and SD-$\ltwo$ with (a) $\epsilon = 10^{-2}$ and
(b) $\epsilon = 10^{-5}$ for  
a $6\times6$ MIMO system at an SNR of $\SNR = 15$dB using $4$-QAM modulation. For SD-$\ltilde$ upper and lower bounds are shown (see Section \ref{sec.ltilde.SD.complexity}). Fig.\  \ref{fig.C.versus.epsilon} shows the corresponding complexity 
results for  SD-$\linf$ and SD-$\ltwo$.   }
\vspace{-1.0cm}
\label{fig.Nk_inf_Nk_two_versus_k}
\end{center}
\end{figure}

\subsection{Tree Pruning Behavior} \label{sec.simu.pruning.behavior}
Next, we quantify some of the results on the average TPB reported in Section \ref{sec.TPB.asymptotic.analyis}. Specifically, we consider a $6 \times 6$ MIMO system with the radii $\rtwo$, $\rinf$, and $\rtilde$  
chosen according to \eqref{eqn.ctwo.epsilon}, \eqref{eqn.cinf.epsilon}, and \eqref{eqn.ctilde.epsilon}, respectively, for $\epsilon = 10^{-2}$ and  $\epsilon = 10^{-5}$. 

\subsubsection{High-SNR Results} 
Fig.\ \ref{fig.mbk} shows $A\big(\Nzeros(\deltab_\layerk)\big)$ in \eqref{eqn.function.Am0} as a function of $\Nzeros(\deltab_\layerk)$ (including the corresponding 
lower bound $1/(\Nzeros(\deltab_\layerk)!)$); we also display  
 $1/(\layerk!) \,\rratio^{\layerk}$ as a function of $\layerk$. Recall that the high-SNR average pruning probability of a node $\deltab_{\layerk}\neq \vec{0}$ for 
 SD-$\linf$ as compared to SD-$\ltwo$ is entirely described by the two functions $A(\Nzeros(\deltab_\layerk))$, $\Nzeros(\deltab_\layerk) = 1,\dots, \layerk$, and 
 $1/(\layerk!) \,\rratio^{\layerk}$, $\layerk = 1,\dots, \tdim$ (see \eqref{eqn.asympt.comparison.SDl2linf} -- \eqref{eqn.cond.more.efficient.TPB2}). Hence, from 
 Fig.\ \ref{fig.mbk} one now can directly infer  the high-SNR average TPB of SD-$\linf$ as compared to that of SD-$\ltwo$ for {\em every} node $\deltab_{\layerk}$, $\layerk = 1,\dots, \tdim$. 
 Considering the case  $\epsilon = 10^{-5}$ in Fig.\ \ref{fig.mbk}, one can, for example, observe that, in the high-SNR regime,  at tree level $\layerk=4$   SD-$\ltwo$ prunes all nodes $\deltab_{\layerk}$ with  
 $\Nzeros(\deltab_\layerk) = 1, 2$ with higher probability than SD-$\linf$ (and vice-versa) or that SD-$\linf$ prunes all nodes $\deltab_{\layerk}$ up to tree level $\layerk=2$ with higher probability than SD-$\ltwo$.  
 Furthermore, the following general conclusions can be drawn: 
\begin{itemize}
\item For the two considered $\epsilon$-values, the function $A\big(\Nzeros(\deltab_\layerk)\big)$ is close to the lower bound $1/(\Nzeros(\deltab_\layerk)!)$.  
\item The function  $A\big(\Nzeros(\deltab_\layerk)\big)$ decreases in $\Nzeros(\deltab_\layerk)$. Therefore, SD-$\linf$ prunes nodes that 
	correspond to a first symbol error at high tree levels, i.e, nodes with large $\Nzeros(\deltab_\layerk)$, in general, with higher probability (in the high-SNR regime) than those that correspond to a first  symbol error    		at low tree levels, i.e., nodes with small $\Nzeros(\deltab_\layerk)$ (provided that $\|\deltab_{\layerk}\|_{2}$ is constant in this comparison). 
\item The function $A\big(\Nzeros(\deltab_\layerk)\big)$ increases by going from $\epsilon = 10^{-5}$ to $\epsilon = 10^{-2}$ for a given $\Nzeros(\deltab_\layerk)>1$
	(see also Appendix \ref{app.Am0.decreasing.function} showing that $A\big(\Nzeros(\deltab_\layerk)\big)$ is a nondecreasing function of $\epsilon$). 
\end{itemize}

\subsubsection{Complexity Versus Tree Level and Complexity Bounds for SD-$\ltilde$}\label{sec.simu.complexity.versus.tree.level}
The goal of this section is to quantify the level-wise complexities $\E\{\NN_{\layerk}\}$ for SD-$\linf$, SD-$\ltilde$, and SD-$\ltwo$, as well as to illustrate the quality of the
 upper and lower bounds on the complexity of SD-$\ltilde$ reported in Section \ref{sec.ltilde.SD.complexity}. Note that for the
 cases of SD-$\linf$ and SD-$\ltwo$ exact complexity expressions according to  \eqref{eqn.NNk.final} and  \eqref{eqn.NN2.final}, respectively, are available.
 Fig.\  \ref{fig.Nk_inf_Nk_two_versus_k} shows $\E\{\NN_{\layerk}\}$ as a function of the tree level $\layerk$ for SD-$\ltwo$ and for SD-$\linf$ including the corresponding upper and lower bounds on $\E\{\NN_{\layerk}\}$ for SD-$\ltilde$ at an SNR of $\SNR = 15 \text{dB}$  (Fig.\  \ref{fig.Nk_inf_Nk_two_versus_k}(a) for $\epsilon = 10^{-2}$ and Fig.\  \ref{fig.Nk_inf_Nk_two_versus_k}(b) for $\epsilon = 10^{-5}$).    The following conclusions can be drawn from these results:

\begin{itemize}

\item At tree levels close to the root (i.e., for small $\layerk$), SD-$\linf$ (SD-$\ltilde$) visits fewer nodes  than SD-$\ltwo$ on average; at tree levels close to the leaves this behavior is reversed.  
This observation is supported by the results on the average TPB reported in Section \ref{sec.pruning.behavior} (in particular, see \eqref{eqn.asympt.TPB.lower.compl} and the discussion in the last paragraph of 
Section \ref{sec.average.TPB.comparison}).  

\item The complexity savings of SD-$\linf$  (SD-$\ltilde$) over SD-$\ltwo$ close to the root extend to higher tree levels for the larger $\epsilon$ value of $10^{-2}$. 
This behavior is consistent with the average TPB analysis in Section \ref{sec.pruning.behavior} stating that 
$\E\{\NN_{\infty,\layerk}\}  \leqasympt  \E\{\NN_{2,\layerk}\}$, $\SNR \rightarrow \infty$, up to tree level $\bar{\layerk}$, where $\bar{\layerk}$ was shown to be a nondecreasing function of $\epsilon$ 
(see Section \ref{sec.average.TPB.comparison}). For example,  
we have $\bar{\layerk} = 3$ for  $\epsilon = 10^{-2}$, while $\bar{\layerk} = 2$ for $\epsilon = 10^{-5}$ (see also Fig.\ \ref{fig.mbk}). 

\item For $\epsilon = 10^{-2}$, the complexity savings of SD-$\linf$ at tree levels close to the root are dominant enough to result 
in a smaller total complexity of SD-$\linf$ as compared to the complexity of SD-$\ltwo$ (cf.\ Fig.\ \ref{fig.C.versus.epsilon}).  
For $\epsilon = 10^{-5}$, however, the increased complexity  of SD-$\linf$ at tree levels close to the leaves outweighs the savings close to the root resulting in higher total complexity of SD-$\linf$  
when compared to the complexity of SD-$\ltwo$  (cf.\ Fig.\ \ref{fig.C.versus.epsilon}). 

\item The upper and lower bounds on the complexity of SD-$\ltilde$ are sufficiently tight to capture the essential aspects of the level-wise complexity of SD-$\ltilde$ since they both show the same
behavior over the tree levels; as for SD-$\linf$,  we can again observe complexity savings of SD-$\ltilde$ over SD-$\ltwo$ close to the root, whereas this behavior is reversed at tree levels close to the leaves. Furthermore, for the examples  considered, the lower bounds on the complexity of SD-$\ltilde$ show that SD-$\ltilde$ has a higher total complexity than SD-$\linf$ (see also next Section). 
		
\end{itemize}

\subsection{Complexity of Sphere-Decoding with Restarting} \label{sec.complexity.restarted.SDs}
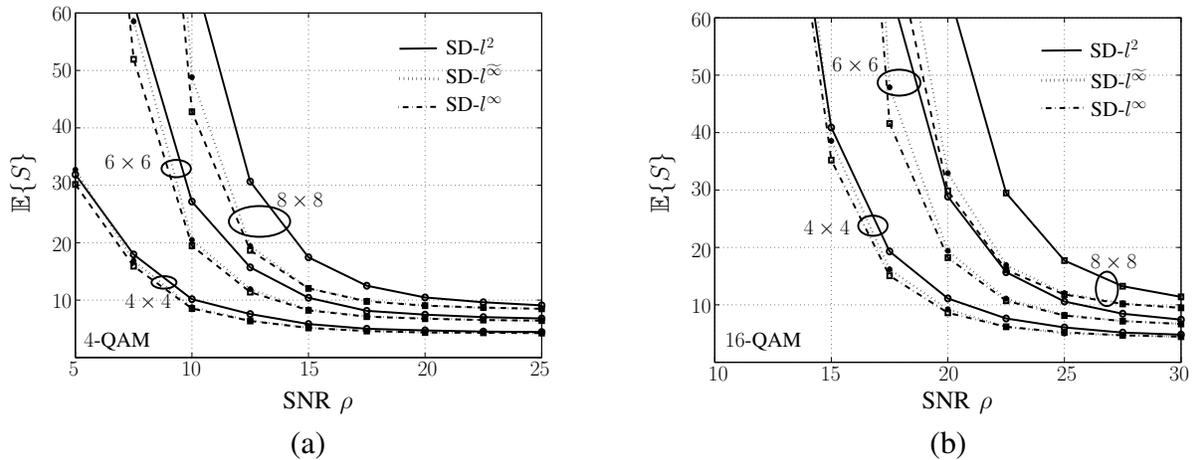
\begin{figure}[t]
\begin{center}
\unitlength1mm
\begin{picture}(170,65)
\put(2,5){\resizebox{8.0cm}{!}{\input N_versus_SNR_4QAM_xfig.pdftex_t}}
\put(87,5){\resizebox{8.0cm}{!}{\input N_versus_SNR_16QAM_xfig.pdftex_t}}
\put(41.0,0){(a)}
\put(126,0){(b)}
\end{picture}
\caption{Total complexity versus SNR $\SNR$ for SD-$\ltilde$, SD-$\linf$, and SD-$\ltwo$, all with restarting for a $4 \times 4$, $6\times 6$, and $8 \times 8$ MIMO system (for the $\epsilon$-schedule see text), using (a) 
$4$-QAM modulation and (b) 16-QAM modulation.}
\vspace{-1cm}
\label{fig.N.versus.SNR}
\end{center}
\end{figure}

As already mentioned in Section \ref{eqn.choice.of.radii},  to guarantee ML or exact SD-$\linf$ performance the corresponding SD algorithm has to be restarted with an increased radius in cases where the initial radius 
was chosen too small for the search sphere (or box) to contain a valid leaf node. The same is, of course, true for SD-$\ltilde$.  
To evaluate the overall (across potential multiple SD runs) complexity of SD-$\ltwo$, SD-$\linf$, and SD-$\ltilde$ we choose an increasing radii schedule obtained by setting 
$\epsilon = 0.1^{i}$, $i=1,2, \dots$, in the $i$th run of the SD.  Corresponding average (w.r.t.\ channel, noise, and data) complexity results for $4\times 4$, $6\times 6$, and $8 \times 8$ MIMO systems 
using $4$-QAM and $16$-QAM modulation obtained through Monte-Carlo simulations can be found in Fig.\  \ref{fig.N.versus.SNR}. We note that analytical expressions for the overall complexity of SD (with any norm considered here) with restarting are not available since 
the statistics of the corresponding required number of SD runs seem to be difficult to obtain. 
From Fig.\  \ref{fig.N.versus.SNR} we can observe that  in the relevant SNR regime (e.g., about $10$dB to $15$dB for the $4$-QAM case corresponding to error probabilities of about $10^{-1}$ to $10^{-3}$, 
cf.\ Fig.\  \ref{fig.error.rate.versus.SNR}) SD-$\linf$ and SD-$\ltilde$ exhibit lower complexity than SD-$\ltwo$. For example, at $12.5$dB, we can infer from Fig.\ \ref{fig.N.versus.SNR}(a)  that the corresponding complexity savings of SD-$\linf$ and SD-$\ltilde$ over SD-$\ltwo$ are about $30\%$. 
Furthermore, it can be observed that the complexity savings of SD-$\linf$ and SD-$\ltilde$ over SD-$\ltwo$ are more pronounced for increasing $\tdim = \rdim$. 
We finally emphasize that these computational (algorithmic) complexity savings of SD-$\ltilde$ over SD-$\ltwo$ go along with a significant reduction in the circuit complexity for metric computation  \cite{burg05_vlsi} (see the discussion in Section \ref{sec.intro.hardware}). Indeed, the overall (circuit and algorithmic) complexity of SD-$\ltilde$ is up to a factor of 5 lower than the overall complexity of SD-$\ltwo$.
\section{Conclusions}\label{sec.conclusions}   
We analyzed sphere-decoding (SD) based on the $\linf$-norm and provided theoretical underpinning for the observations reported in \cite{burg05_vlsi}. The significance of $\linf$-norm SD is supported by the fact  
that its overall implementation complexity in hardware is up to a factor of 5 lower than that for SD based on the $\ltwo$-norm (corresponding to optimum detection).  In particular, we found that using the $\linf$-norm instead of the $\ltwo$-norm does not result in a reduction of diversity order while leading to an SNR gap, compared to optimum performance, that increases at most linearly in the number of receive antennas. We furthermore showed that for many cases of practical interest $\linf$-norm SD, besides having a smaller circuit complexity for metric computation (thanks to the fact that it avoids squaring operations) also exhibits smaller computational (algorithmic) complexity (in terms of the number of nodes visited in the search tree) than $\ltwo$-norm SD.  The computational complexity of $\linf$-norm SD was found to scale exponentially in the number of transmit antennas as is also the case for $\ltwo$-norm SD. 

Besides the $\linf$-norm, VLSI implementations are often based on the $l^{1}$-norm (which does not require squaring operations either). The tools developed in this paper could turn out useful in analyzing the performance of SD based on the $l^{1}$-norm as well. From a computational complexity point-of-view, however, the results in  \cite{burg05_vlsi} suggest that $\linf$-norm SD is more attractive than 
$l^{1}$-norm SD. More generally,  it would be interesting to understand the impact of $l^{p}$-norm (sphere) decoding with general $p$ and to investigate  this impact for other channel models (such as ISI-channels, for example).
%
%
\section*{Acknowledgments}
The authors would like to thank G.\ Matz for suggesting the direct integration approach for deriving \eqref{eqn.NNk.final} and for pointing out reference \cite{behboodian72}, M.\ Borgmann for valuable discussions on the diversity order of SD-$\linf$,  A.\ Burg for helpful discussions on VLSI implementation aspects of SD-$\linf$, and S.\ Gerhold for pointing out the proof in Appendix \ref{app.radii.limit.eps0}. 

\appendices 
%
%
%
%
\section{Calculation of $\P\Big[\big|[\zb(\deltab)]_{M-\layer+1}\big| \leq C_\infty\Big]$}  \label{app.distribution}
In the following, we derive \eqn{eqn.dist}. 
 We start by introducing the RVs 
\begin{equation}\label{def.vm.um}
	v_\layer =  R_{\tdim-\layer+1,\tdim-\layer+1}|\deltanb_{\tdim-\layer+1}|, \quad u_\layer =  \sum_{i=\tdim-\layer+2}^{\tdim}\! R_{\tdim-\layer+1,i}\, \deltanb_i + n_{\tdim-\layer+1}.
\end{equation}
Since the nonzero entries in $\Rb$ and the entries in $\nb$ are all statistically independent, $v_\layer$ and $u_\layer$ are statistically independent as well. 
 Here, $v_\layer$ is a $\chi_{2(\layer+L)}$-distributed RV with pdf (cf.\  \eqref{eqn.chi})
\begin{equation}\label{eqn.dist.p}
g_{\layer}(v)
= \frac{2\,\tdim^{\layer+L}}{\Gamma(\layer\!+\!L) |\deltanb_{\tdim-\layer+1}|^{2(\layer+L)} }\,  v^{2(\layer+L)-1}e^{-\frac{v^2}{|\deltanb_{\tdim-\layer+1}|^2/\tdim}}. 
\end{equation}
The RV $u_\layer$ is $\CN(0, \sigma^2_\layer)$ distributed, where 
\begin{equation}\label{eqn.def.sigmam}
\sigma^2_\layer =  \|\deltab_{\layer-1}\|^2/\tdim+\sigma^2.
\end{equation}
Exploiting the circular symmetry of $u_{\layer}$, we have 
\begin{equation}\label{eqn.equal.dist.component.vm.um}
  \big|[\zb(\deltab)]_{M-\layer+1}\big| \equaldist |v_\layer + u_\layer|. 
\end{equation}
Thus,  
\begin{align} \label{eqn.equaldist.zb.vm.um}
\P\Big[\big|[\zb(\deltab)]_{M-\layer+1}\big| \leq C_\infty\Big] & = 
\P\big[|v_\layer + u_\layer| \leq C_\infty\big] \\[2mm]
& = \int_{0}^{\infty} \P\Big[|v_\layer + u_\layer| \leq C_\infty\big| v_\layer = v \Big]  g_{\layer}(v).
\label{eqn.int.represent}
\end{align}
For given $v_\layer = v$, the RV $\frac{2}{\sigma_\layer^2}|v + u_\layer|^2$ is non-central $\chi^2_{2}$-distributed with non-centrality parameter 
$\frac{2 v^2}{\sigma_\layer^2}$. Thus, (see \cite[Corollary 1.3.5]{muir05})
\begin{equation}\label{eqn.non.central.chi.distribution}
  \P\Big[|v_\layer + u_\layer| \leq C_\infty\big| v_\layer = v \Big] = \sum_{s = 0}^{\infty} e^{-\frac{v^2}{\sigma_m^2}} \Big(\frac{v}{\sigma_m}\Big)^{\! 2 s} \frac{1}{s!} 
  \,\, \gamma_{s+1}\! \bigg(\frac{C_\infty^2}{\sigma_m^2}\bigg). 
 \end{equation}
Inserting \eqn{eqn.dist.p} and \eqref{eqn.non.central.chi.distribution} into \eqn{eqn.int.represent}  yields 
\[
\P\big[|v_\layer + u_\layer| \! \leq \! C_\infty\big] \!= \!  \sum_{s = 0}^{\infty}
 \frac{2\, \gamma_{s+1} (C_\infty^2/\sigma_m^2) \tdim^{\layer+L}}{ s!\, \sigma_m^{2 s}\, \Gamma(\layer\!+\!L) |\deltanb_{\tdim-\layer+1}|^{2(\layer+L)} }\,  
  \!\!\int_{0}^{\infty}\! \!\! \!  v^{2(s+\layer+L)-1} e^{-v^2\Big(\frac{\tdim}{|\deltanb_{\tdim-\layer+1}|^2} + \frac{1}{\sigma_m^2}\Big)} dv.
\] 
Here, the integral can easily be rewritten such that the integrand is the pdf of a $\chi_{2(s+\layer+L)}$-distributed RV (cf.\ \eqn{eqn.chi}), 
 which then yields  
\[
 \int_{0}^{\infty}\! \!\!  v^{2(s+\layer+L)-1} e^{-v^2\Big(\frac{\tdim}{|\deltanb_{\tdim-\layer+1}|^2} + \frac{1}{\sigma_m^2}\Big)} dv = \frac{1}{2}\Gamma(s+\layer+L) 
 \bigg(\frac{\tdim}{|\deltanb_{\tdim-\layer+1}|^2} + \frac{1}{\sigma_m^2} \bigg)^{-(s+\layer+L)}.
\]
Finally, using $\Gamma(a) = (a-1)!$ for positive integers $a$, we get    
\begin{equation}\label{eqn.dist.intermediate}
\P\big[|v_\layer + u_\layer| \leq C_\infty\big] =   \sum_{s = 0}^{\infty} D_{s}(\deltab_\layer) \,
\gamma_{s+1}\! \bigg(\frac{C_\infty^2}{\sigma_m^2}\bigg)
\end{equation}
where 
\begin{equation}\label{eqn.coefficient.schlange}
D_{s}(\deltab_\layer) =  {s\!+\!\layer\!+\!L\!-\!1 \choose \layer\!+\!L\!-\!1} p(\deltab_\layer)^{\layer+L} \, (1-p(\deltab_\layer))^{s} \,
\end{equation}
and, as defined in \eqn{eqn.parameter}, 
\[	
	p(\deltab_\layer) =   \frac{\sigma_\layer^2}{\sigma_\layer^2 + |\deltanb_{\tdim-\layer+1}|^2/\tdim} =   \frac{\|\deltab_{\layer-1}\|_2^2+\tdim\sigma^2}{\|\deltab_{\layer}\|_2^2+\tdim\sigma^2}. 
\]
 In the remainder of this section, we show that the infinite summation in \eqref{eqn.dist.intermediate} can be avoided.  We use $p_\layer = p(\deltab_\layer)$ to simplify notation and 
we start by noting that \eqn{eqn.dist.intermediate} can be written as 
\begin{equation}\label{eqn.dist.intermediate2}
\P\big[|v_\layer + u_\layer| \leq C_\infty\big] =   \frac{p_\layer^{\layer+L}}{(\layer+L-1)!} \int_0^{\frac{C_\infty^2}{\sigma_m^2}} 
\left[
\sum_{s = 0}^{\infty} \Bigg(\prod_{i=1}^{\layer+L-1}\!\!(s+i)\Bigg) \, \frac{[(1\!-\!p_\layer)t]^{s}}{s!} \,
\right]
e^{-t}
dt
\end{equation}
where the identity \eqn{eqn.gamma} for the lower incomplete Gamma function was used. With $g(x) = e^x x^{\layer+L-1}$ and 
the series expansion $e^x = \sum_{s=0}^{\infty} \frac{x^s}{s!}$, we have that 
\[
	g^{(\layer+L-1)}(x)= \sum_{s = 0}^{\infty} \Bigg(\prod_{i=1}^{\layer+L-1}\!\!(s+i)\Bigg) \, \frac{x^{s}}{s!}.
\]
On the other hand, by Leibniz's law for the differentiation of products of functions, we also have
\[
	g^{(\layer+L-1)}(x)  = \sum_{l=0}^{\layer+L-1} {\layer\!+\!L\!-\!1\choose l} \frac{(\layer+L-1)!}{(\layer+L-1-l)!} \,e^x x^{\layer+L-1-l}.
\] 
 Thus, \eqn{eqn.dist.intermediate2} can equivalently be written as
 \[
\P\big[|v_\layer + u_\layer| \leq C_\infty\big] =   \sum_{l = 0}^{\layer+L-1} {\layer\!+\!L\!-\!1\choose l}\,  \frac{p_\layer^{\layer+L} 
(1\!-\!p_\layer)^{\layer+L-1-l}}{\Gamma(\layer+L-l)} \int_0^{\frac{C_\infty^2}{\sigma_m^2}} t^{\layer+L-l-1} e^{-p_\layer t} dt.
\]
By substituting $t' = p_\layer t$ and again using identity \eqn{eqn.gamma}, we finally get 
\begin{equation}\label{eqn.final.appendix.bin.mixture}
	\P\big[|v_\layer + u_\layer| \leq C_\infty\big] = \sum_{l = 0}^{\layer+L-1} {\layer\!+\!L\!-\!1\choose l}\,  p_\layer^{l} \,
	(1\!-\!p_\layer)^{\layer+L-1-l}\, \gamma_{\layer+L-l} \bigg(p_\layer \frac{C_\infty^2}{\sigma_m^2}\bigg)
\end{equation}	
which, noting that  
\[
\frac{p_\layer}{\sigma_\layer^2} = \frac{1}{\sigma^2 + \|\deltab_\layer\|^2/\tdim},
\]
concludes the derivation of \eqref{eqn.dist}.

\section{Sum Representation of $\big|[\zb(\deltab)]_{M-\layer+1}\big|^2$} \label{app.behboodian}

In the following, we prove \eqref{eqn.equaldist.tm} based on the following theorem. \\
{ {\bf Theorem} {\bf \cite{behboodian72}.}
{\em Consider the RVs  
\begin{equation}\label{eqn.zl}
	z^{(l)} =   g\big(y^{(l)}_{1}, y^{(l)}_2, \dots, y^{(l)}_{\tind}\big), \quad l = 0, \dots, \tind
\end{equation}
where $y^{(l)}_{i}$, $i = 1, \dots, \tind$, for every $l$, are statistically independent RVs with pdfs equal to $f_1(x)$ if 
$i \leq l$ and $f_2(x)$ otherwise. If $g(\cdot)$ is a symmetric function (i.e., $g(\cdot)$ is unchanged by any permutation of its arguments), then the pdf of 
\begin{equation}\label{eqn.z.g}
	z = g(y_1, y_2, \dots, y_{\tind})
\end{equation}
where the $y_i$, $i = 1, \dots, \tind$, are i.i.d.\ with mixture pdf 
 \begin{equation}\label{eqn.basic.mixture}
 f_{y_i}(x) = p f_1(x) + (1-\!p) f_2(x), \quad 0 \leq p \leq 1 
 \end{equation}
is given by 
\begin{equation}\label{eqn.gen.binomial.dist}
	f_z(x) = \sum_{l=0}^{\tind}B_{l}\,   f_{\!z^{(l)}}(x)
\end{equation}
with 
\[
B_{l} =  {\tind \choose l} p^{\,l} \,(1-p)^{\tdim-l}. 
\]
Here, $f_{z^{(l)}}(x)$, $l = 0, \dots, \tind$,  denotes the pdf of $z^{(l)}$ specified in \eqref{eqn.zl}.}

We apply this theorem to the case at hand by defining $f_{1}(x) = \delta(x)$ and $f_{2}(x) = f_{\!\chi^2_2}(\argument)$ and setting $a = \layer+L-1$. 
Furthermore, we take  $g(\cdot) $ as 
\begin{equation}\label{eqn.z.g.specific}
	g\big(x_{1}, x_2, \dots, x_{ \layer+L-1}\big) =   \frac{\|\deltab_\layer\|_2^2/\tdim+\sigma^2}{2}\Bigg( \!\sum_{i=1}^{\layer+L-1}\!\!\! x_{i} \Bigg) 
\end{equation}
which implies
\begin{equation}\label{eqn.f.zl}
	f_{z^{(l)}}(x) = \frac{2}{\|\deltab_\layer\|_2^2/\tdim+\sigma^2}\, f_{\chi^2_{2(\layer+L-1-l)}}\bigg( \frac{2 x}{\|\deltab_\layer\|_2^2/\tdim+\sigma^2} \bigg) 
\end{equation}
if $l < \layer+L-1$  and $f_{z^{(l)}}(x) = \delta(x)$ if $l = \layer+L-1$ for the pdfs of the RVs $z^{(l)}$ defined in \eqref{eqn.zl}. Using \eqref{eqn.gen.binomial.dist},  we thus get
\[
	f_{z}(x) = \frac{2}{\|\deltab_\layer\|_2^2/\tdim+\sigma^2} \sum_{l=0}^{\layer+L-1} \! \! B_{l} \, f_{\chi^2_{2(\layer+L-1-l)}}\!\bigg( \frac{2 x}{\|\deltab_\layer\|_2^2/\tdim+\sigma^2} \bigg) 
\]
with the corresponding  cdf essentially given by the RHS of  \eqref{eqn.dist} but with two missing degrees of freedom in the $\chi^2$-distributed RVs underlying the individual terms in the sum.
To compensate for these two missing degrees of freedom, we construct the RV
\begin{equation}\label{eqn.z.prime}
	t^2_{\layer} = z +  \frac{\|\deltab_\layer\|_2^2/\tdim+\sigma^2}{2} \, \gamma^2
\end{equation}
with $\gamma^2  \sim \chi^2_2$ being statistically independent of $z$. Noting that $(f_{\chi^2_{a}} \ast f_{\chi^2_{b}})(x) = f_{\chi^2_{a+b}}(x)$, we obtain 
\[
f_{t^2_{\layer}}(x) = \frac{2}{\|\deltab_\layer\|_2^2/\tdim+\sigma^2} \sum_{l=0}^{\layer+L-1} \! \! B_{l} \, f_{\chi^2_{2(\layer+L-l)}}\!\bigg( \frac{2 x}{\|\deltab_\layer\|_2^2/\tdim+\sigma^2} \bigg) 
\]
or, equivalently,  
\[
\text{P}\big[t^2_{\layer} \leq x \big] =  \sum_{l=0}^{\layer + L-1} \!\! B_l \, \gamma_{\layer+L-l}\bigg(  
\frac{x}{\|\deltab_\layer \|_2^2/\tdim + \sigma^{2}}\bigg)
\]
thus, by comparison with \eqref{eqn.dist},  establishing that $t^2_{\layer} \equaldist \big|[\zb(\deltab)]_{M-\layer+1}\big|^2$. Finally, \eqref{eqn.z.prime} together with 
\eqref{eqn.z.g} and \eqref{eqn.z.g.specific} shows \eqref{eqn.sum.represent}.

\section{Bounds on Lower Incomplete Gamma Function}\label{app.incomplete.gamma.function}

In this section, we summarize properties of the lower (regularized) incomplete Gamma function 
\begin{equation}\label{eqn.gamma}
\gamma_{a}(\garg) 
=  \frac{1}{\Gamma(a)}\int_0^{\garg} y^{a-1} e^{-y} dy, \quad \garg, a \in \Rnum, \,\, \garg, a \geq 0 
\end{equation}
needed in this paper. In the remainder of this section, we will furthermore assume that $a \in \Nnum$, which is the most relevant case for our results. We start by noting that $\gamma_{a}(\garg)$ can  equivalently  be written as \cite[Sec.\ 6.5]{Abr65}
\begin{align}
	\gamma_{a}(\garg)  	& = e^{-\garg} \sum_{i=a}^{\infty} \frac{\garg^i}{i!}   \label{eqn.gamma.series1} \\
						& = 1-e^{-\garg} \sum_{i=0}^{a-1} \frac{\garg^i}{i!} \label{eqn.gamma.series2}. 
\end{align}
An immediate consequence of  \eqref{eqn.gamma.series2} is $\gamma_1(\garg) = 1-e^{-\garg}$. From \eqref{eqn.gamma.series1}, we can directly infer that 
 \begin{equation}\label{eqn.lower.bound.gamma.higher.orders}
 \gamma_{a_{1}\!}(\garg) \geq \gamma_{a_{2}\!}(\garg), \quad a_{1} \leq a_{2}. 
 \end{equation}
Furthermore, we have \cite[Eq.\ (5.4)]{gautschi98} 
\begin{equation}      
	\Big(1-e^{-\frac{1}{\sqrt[a]{a!}}\garg}\Big)^a \leq \gamma_a(\garg) \leq \big(1-e^{-\garg}\big)^a.   \label{eqn.UB.on.gamma}
\end{equation}
We will also need the relation 
\begin{equation}\label{eqn.UB.root.lower.DOFs}
	\left[\gamma_{a_{1}}(\garg)\right]^{\frac{1}{a_{1}}} \geq \left[\gamma_{a_{2}}(\garg)\right]^{\frac{1}{a_{2}}}, \quad a_{1} \leq a_{2} 
\end{equation}
which will be proved by showing that $\left[\gamma_{a}(\garg)\right]^{\frac{1}{a}}$ is a nonincreasing function of $a\in \Nnum$, i.e., 
\begin{equation}\label{eqn.incomplete.gamma.root.dec.a}
\left[\gamma_{a}(\garg)\right]^{\frac{1}{a}} \geq \left[\gamma_{a+1}(\garg)\right]^{\frac{1}{a+1}}. 
\end{equation}
The proof is by induction. For $a = 1$, we have $\gamma_{1}(\garg) \geq \left[\gamma_{2}(\garg)\right]^{\frac{1}{2}}$, which follows from \eqref{eqn.UB.on.gamma}. 
It remains to show that 
\begin{equation}\label{eqn.induct.1}
\left[\gamma_{n}(\garg)\right]^{\frac{1}{n}} \geq \left[\gamma_{n+1}(\garg)\right]^{\frac{1}{n+1}}, \quad n\in \Nnum 
\end{equation}
implies 
\begin{equation}\label{eqn.induct.2}
\left[\gamma_{n+1}(\garg)\right]^{\frac{1}{n+1}} \geq \left[\gamma_{n+2}(\garg)\right]^{\frac{1}{n+2}}. 
\end{equation}
To this end, we use \cite[Lemma 3]{merkle_93} which states that 
\begin{equation}\label{eqn.induct.3}
\gamma_{n+1}(\garg) \geq   \left[\gamma_{n}(\garg)\right]^{\frac{1}{2}}  \left[\gamma_{n+2}(\garg)\right]^{\frac{1}{2}}. 
\end{equation}
Inserting \eqref{eqn.induct.1} into \eqref{eqn.induct.3}, we get 
\[
\gamma_{n+1}(\garg) \geq   \left[\gamma_{n+1}(\garg)\right]^{\frac{1}{2} \frac{n}{n+1}}  \left[\gamma_{n+2}(\garg)\right]^{\frac{1}{2}}
\]
which gives 
\[
\left[\gamma_{n+1}(\garg)\right]^{\frac{1}{2} \frac{n+2}{n+1}} \geq  \left[\gamma_{n+2}(\garg)\right]^{\frac{1}{2}}
\]
establishing \eqref{eqn.induct.2} thereby concluding the proof. 

We will finally show that 
\begin{equation}\label{eqn.gamma.bound}
	 \gamma_{a}\bigg(\frac{x_{1}}{1+x_{2}}\bigg) \geq \gamma_{a}(x_{1}) \, (1+x_{2})^{-a}
\end{equation}
for any $x_{1}, x_{2} \geq 0$. Inserting into the definition \eqref{eqn.gamma} yields 
\[
	 \gamma_{a}\bigg(\frac{x_{1}}{1+x_{2}}\bigg)  =  \frac{1}{\Gamma(a)}\int_0^{\frac{x_{1}}{1+x_{2}}}\! y^{a-1} e^{-y} dy
\]
which, upon substituting $\tilde{y} = (1+x_{2}) y$,  can be rewritten as 
\[
 \gamma_{a}\bigg(\frac{x_{1}}{1+x_{2}}\bigg) = (1+x_{2})^{-a}  \frac{1}{\Gamma(a)} \int_0^{x_{1}}  \tilde{y}^{a-1} e^{-\frac{\tilde{y}}{1+x_{2}}} d\tilde{y}.
\]
Since $e^{-\frac{\tilde{y}}{1+x_{2}}}\geq e^{-\tilde{y}}$  for $x_{2} \geq 0$, we arrive at  \eqref{eqn.gamma.bound}.

\section{Asymptotics of Radii}\label{app.calc.linearM.SD2}

\subsection{Asymptotics of $\rtwo^2$ in \eqref{eqn.ctwo.epsilon}} 
For fixed SNR (i.e., fixed $\sigma^2$) and fixed $\epsilon$, 
the asymptotic ($\rdim \rightarrow \infty$) behavior of $\rtwo^2  =\sigma^2\,\gamma_{\rdim}^{-1}\!\left(1-\epsilon\right)$ can be obtained as follows.  
According to  \cite[Eq.\ (2.13)]{gautschi98}
\[
	\gamma_{\rdim+1}\big(\rdim+\sqrt{2 \rdim}\,x\big) = 1-Q\big(\sqrt{2}\,x\big) + {\cal O}\big(1/\sqrt{N}\,\big), \quad\rdim \rightarrow \infty
\]
for  $x \in \Rnum$, $0 \leq x < \infty$. 
Therefore, we have 
\[
	\rtwo^2 = \sigma^2\Big(N-1+\sqrt{N-1}\,\, Q^{-1}\!\!\left(\epsilon + {\cal O}\big(1/\sqrt{N}\,\big)\right)\Big)
\]
where $Q^{-1}\!\big(\epsilon + {\cal O}\big(1/\sqrt{N}\,\big)\big) =  {\cal O}(1)$ showing 
that $\rtwo^2 \equalasympt \sigma^2 N$, $\rdim \rightarrow \infty$. 

\subsection{Asymptotics of $\rinf^2$ in \eqref{eqn.cinf.epsilon} }
For fixed  SNR (i.e., fixed $\sigma^2$) and fixed $\epsilon$, the asymptotic ($\rdim \rightarrow \infty$) behavior of $\rinf^2 = -\sigma^2\,\text{log}\!\left(1-\sqrt[\rdim]{1-\epsilon}\,\right)$ is obtained as follows.  
We have $\sqrt[\rdim]{1-\epsilon} = 1 + {\cal O}(1)/N$, $\rdim \rightarrow \infty$. Thus,  
\[
  \rinf^2 = \sigma^2\,\text{log}(\rdim)+{\cal O}(1), \quad\rdim \rightarrow \infty
  \]
which shows that $\rinf^2 \equalasympt \sigma^2\, \text{log}(\rdim)$, $\rdim \rightarrow \infty$. 

\section{Asymptotic Behavior of $\P\big[\|\zb_\layerk(\deltab_\layerk)\hspace{-0.04cm}\|_\infty \leq \rinf\big]$}\label{app.asympt.prop.b.l.inf}

In the following, we characterize the asymptotic (in SNR) behavior of $\P\big[\|\zb_\layerk(\deltab_\layerk)\hspace{-0.04cm}\|_\infty \leq \rinf\big]$. 
This is done by splitting the product on the RHS in  
\eqref{eqn.order.stat} into three parts, which are treated separately (recall the definition of $\Nzeros(\deltab_\layerk)$ in Section \ref{sec.TPB.asymptotic.analyis} as the index of the 
first erroneous tree level and the definition of $\rinfsigma$ in \eqref{eqn.rinfsigma.def}). 

\begin{itemize}
\item $\layer =1,\dots, \Nzeros(\deltab_\layerk)-1$:  We have $[\zb(\deltab)]_{\tdim-\layer+1} = [\nb]_{\tdim-\layer+1}$, which is $\CN(0,\sigma^2)$ so that 
$\P\Big[ \big|[\zb(\deltab)]_{\tdim-\layer+1} \big| \leq \rinf\Big] = \gamma_{1}\!\left(\rinfsigma\right)$. Hence, 
the first part is given by
\begin{equation}\label{eqn.asympt.part1.final}
	 \left[\gamma_{1}\!\left(\rinfsigma\right)\right]^{L}   \prod_{\layer=1}^{\Nzeros(\deltab_\layerk)-1} \P\Big[ \big|[\zb(\deltab)]_{\tdim-\layer+1} \big| \leq \rinf\Big] 
	 =  \left[\gamma_{1}\!\left(\rinfsigma\right)\right]^{\Nzeros(\deltab_\layerk)-1+L}.  
\end{equation}
\item $\layer = \Nzeros(\deltab_\layerk)$: 
The second part corresponds to the first erroneous tree level associated with $\deltab_\layerk$. Here, we start by noting that \eqref{eqn.parameter} yields
\[
 p(\deltab_\layer)  = \frac{\tdim \sigma^2}{\|\deltab_\layer\|_2^2+\tdim \sigma^2}
\]
where we used $\|\deltab_{\layer-1}\|_2^2 = 0$.
We thus have
\begin{equation}\label{eqn.asympt.part2.p}
 p(\deltab_\layer)  \equalasympt  \left(\SNR\, \|\deltab_{\layer}\|_2^2/\tdim \right)^{-1}, \quad  \SNR \rightarrow \infty
\end{equation}
and 
\begin{equation}\label{eqn.asympt.part2.p2}
	1-p(\deltab_\layer)  \equalasympt 1, \quad  \SNR \rightarrow \infty. 
\end{equation}
Furthermore, 
\[
\gamma_{\layer+L-l}\bigg(  
\frac{C_\infty^2}{\|\deltab_{\layer} \|_2^2/\tdim + \sigma^{2}}\bigg) = \gamma_{\layer+L-l}\bigg(  
\frac{\rinfsigma}{1+\SNR\,\|\deltab_{\layer} \|_2^2/\tdim}\bigg) 
\]
and \eqref{eqn.gamma.series1} implies that 
\begin{align}
 \gamma_{\layer+L-l}\bigg(  
\frac{\rinfsigma}{1+\SNR\,\|\deltab_{\layer} \|_2^2/\tdim}\bigg) & \equalasympt
 \nonumber \\ 
& \hspace{-2.0cm}   \frac{1}{(\layer+L-l)!} \,
\rinfsigma^{\layer+L-l}    \left(\SNR\,\|\deltab_{\layer}\|_2^2/\tdim \right)^{-(\layer+L-l)}, \quad \SNR \rightarrow \infty. 
\label{eqn.asympt.part2.gamma}
\end{align}
With \eqref{eqn.dist}  and \eqref{eqn.asympt.part2.p} -- \eqref{eqn.asympt.part2.gamma}, we finally arrive at 
\begin{equation}\label{eqn.asympt.part2.final}
\text{P}\Big[\big|[\zb(\deltab)]_{M-\layer+1}\big| \leq C_\infty \Big]  \equalasympt D\big(\layer \big)   
\left(\SNR\, \|\deltab_{\layer}\|_2^2/\tdim \right)^{-(\layer+L)}, \quad  \SNR \rightarrow \infty
\end{equation}
where  
\[
 D\big(\layer \big)  =   \sum_{l=0}^{\layer+L-1} { \layer\!+\!L\!-\!1 \choose l} \frac{1}{(\layer \! +\!L\!-\!l)!}
	\,\rinfsigma^{\layer + L -l}.
\]

\item $\layer = \Nzeros(\deltab_\layerk)+1, \dots, \layerk$: For these tree levels, we have $\|\deltab_{\layer-1}\|_2^2 \neq 0$, which yields 
\begin{equation}\label{eqn.asympt.part3.p}
 p(\deltab_{\layer})  \equalasympt  \frac{\|\deltab_{\layer-1}\|_2^2}{\|\deltab_{\layer}\|_2^2}, \quad  \SNR \rightarrow \infty. 
\end{equation}
Combining this result with \eqref{eqn.asympt.part2.gamma} and \eqref{eqn.dist}, we thus obtain 
\[
 \text{P}\Big[\big|[\zb(\deltab)]_{M-\layer+1}\big| \leq C_\infty \Big]  \equalasympt \rinfsigma \left(\frac{\|\deltab_{\layer-1}\|_2^2}{\|\deltab_{\layer}\|_2^2}\right)^{\!\!\! \layer+L-1} \!\!\!\!
 \left(\SNR\, \|\deltab_{\layer}\|_2^2/\tdim \right)^{-1},  \quad  \SNR \rightarrow \infty 
\]
so that
\begin{align}
	\prod_{\layer=\Nzeros(\deltab_\layerk)+1}^{\layerk}\!\!\! \P\Big[ \big|[\zb(\deltab)]_{\tdim-\layer+1} \big| \leq \rinf\Big] & \equalasympt \nonumber \\
	&\hspace{-3cm}
	 \rinfsigma^{\layerk-\Nzeros(\deltab_\layerk)} \,\SNR^{-(\layerk-\Nzeros(\deltab_\layerk))} \!\!\!\!\!\!\!
	\prod_{\layer=\Nzeros(\deltab_\layerk)+1}^{\layerk} \!\!\!\!\!\frac{ (\|\deltab_{\layer-1}\|_2^2/M)^{\layer+L-1}}{(\|\deltab_{\layer}\|_2^2/M)^{\layer+L}}, \quad \SNR \rightarrow \infty. \label{eqn.asympt.part3.final}
\end{align}
Next, note that 
\begin{equation}\label{eqn.asympt.simplification}
  \prod_{\layer=\Nzeros(\deltab_\layerk)+1}^{\layerk} \!\!\!\!\!\frac{ (\|\deltab_{\layer-1}\|_2^2/M)^{\layer+L-1}}{(\|\deltab_{\layer}\|_2^2/M)^{\layer+L}} 
   =  \frac{ (\|\deltab_{\Nzeros(\deltab_\layerk)}\|_2^2/M)^{\Nzeros(\deltab_\layerk)+L}}{(\|\deltab_{\layerk}\|_2^2/M)^{\layerk+L}}. 
\end{equation}
\end{itemize}
Combining \eqref{eqn.asympt.part1.final}, \eqref{eqn.asympt.part2.final}, \eqref{eqn.asympt.part3.final}, and \eqref{eqn.asympt.simplification} finally yields \eqref{eqn.asymptotic.behavior.Pbk.inf}. 

\section{Properties of $A\big(\Nzeros(\deltab_\layerk)\big)$}
\subsection{Limit of $A\big(\Nzeros(\deltab_\layerk)\big)$ for $\rinfsigma \rightarrow 0$}\label{app.Am0.limit}
We want to prove that 
\begin{equation}\label{eqn.Abo.lim.UB.app}
\underset{\rinfsigma \rightarrow 0} {\lim}\,\, 
A\big(\Nzeros(\deltab_\layerk)\big) = 1. 
\end{equation}
With  \eqref{eqn.gamma.series1}, we can write 
\[
	 \left[\gamma_{1}\!\left(\rinfsigma\right)\right]^{\Nzeros(\deltab_\layerk)+L-1}  = \rinfsigma^{\Nzeros(\deltab_\layerk)+L-1}(1+o(1))^{\Nzeros(\deltab_\layerk)+L-1}, \quad \rinfsigma \rightarrow 0 
\]
which gives
\[
	A\big(\Nzeros(\deltab_\layerk)\big) = \,(1+o(1))^{\Nzeros(\deltab_\layerk)+L-1}
 	\!\!\sum_{l=0}^{\Nzeros(\deltab_\layerk)+L-1} \!\!\! {\Nzeros(\deltab_\layerk)\!+\!L\!-\!1 \choose l} \frac{1}{(\Nzeros(\deltab_\layerk)\! +\!L\!-\!l)!}
	\rinfsigma^{\Nzeros(\deltab_\layerk)+L-1-l} 
\] 
for $\rinfsigma \rightarrow 0$ establishing \eqref{eqn.Abo.lim.UB.app}.
\subsection{Monotonicity of $A\big(\Nzeros(\deltab_\layerk)\big)$}\label{app.Am0.decreasing.function}

In the following, we show that $A\big(\Nzeros(\deltab_\layerk)\big)$ in \eqref{eqn.function.Am0} is a nonincreasing function of $\rinfsigma$ 
(or, equivalently, noting that $\rinfsigma = -\text{log}\!\left(1-\sqrt[\rdim]{1-\epsilon}\,\right)$, $A\big(\Nzeros(\deltab_\layerk)\big)$ is a nondecreasing function of $\epsilon$). 
This will be done by setting 
$x = \rinfsigma$, 
$\widehat{\layer} = \Nzeros(\deltab_\layerk) + L$, and by showing that 
\[
f(x) = \left[\gamma_{1}\!\left(x\right)\right]^{\widehat{\layer}-1} \sum_{l=0}^{\widehat{\layer}-1} {\widehat{\layer}\!-\!1 \choose l} \frac{1}{(\widehat{\layer}-\!l)!}
	\, x^{\!-l}
\]
is a nonincreasing function of $x \geq 0$, or equivalently, 
$f'(x) \leq 0$, for $x \geq 0$. For  $\widehat{\layer} = 1$ this holds trivially as $f(x) = 1$. We therefore consider the case $\widehat{\layer} \geq 2$ in what follows. 
The condition $f'(x) \leq 0$, for $x \geq 0$, is equivalent to 
\begin{equation}\label{eqn.monot.decreasing.Ab.first.sum}
\frac{e^{x}-1}{(\widehat{\layer}-1)}
\sum_{l=0}^{\widehat{\layer}-1} {\widehat{\layer}\!-\!1 \choose l} \frac{l}{(\widehat{\layer}-\!l)!}	\, x^{\!-l-1}
\geq 
\sum_{l=0}^{\widehat{\layer}-1} {\widehat{\layer}\!-\!1 \choose l} \frac{1}{(\widehat{\layer}-\!l)!}	\, x^{\!-l}, \quad x \geq 0.
\end{equation}
Multiplying both sides of \eqref{eqn.monot.decreasing.Ab.first.sum} by $x^{\widehat{\layer}} \geq 0$, and substituting $i = \widehat{\layer}-l$, it remains to show that   
\begin{equation}\label{eqn.monot.decreasing.Ab.p.q}
p(x) 
\geq 
q(x), \quad \text{for}\,\, x \geq 0
\end{equation}
 where
 \begin{equation}
p(x)   = (e^{x}\!-\!1) \sum_{i=1}^{\widehat{\layer}} \frac{\widehat{\layer}\!-\!i}{\widehat{\layer}\!-\!1} \,a_{i} \,x^{i-1} \label{eqn.poly.p.x}
\end{equation}
and
\begin{equation}
q(x)  = \sum_{i=1}^{\widehat{\layer}} a_{i}	\, x^{i} \label{eqn.poly.q.x}
\end{equation}
with 
\begin{equation}\label{eqn.def.coefficients.ai}
a_{i} =  {\widehat{\layer}\!-\!1 \choose i\!-\!1} \frac{1}{i!}\,. 
\end{equation}
Here, we used $ {\widehat{\layer}- 1 \choose \widehat{\layer} - i} =  {\widehat{\layer} -1 \choose i - 1}$.
Evidently, a sufficient condition for \eqref{eqn.monot.decreasing.Ab.p.q} to hold is that $p(0) \geq q(0)$ and $p'(x)  \geq q'(x)$, for $x\geq 0$.
 Successively applying this argument,  \eqref{eqn.monot.decreasing.Ab.p.q} can be shown by proving that  
\begin{equation}\label{eqn.monot.increasing.A.cond.1}
  p^{(n)}\!(x) \big|_{x \,= \,0}  \geq q^{(n)}\!(x) \big|_{x\, = \,0}, \quad \text{for}\,\, n = 0,\dots, \widehat{\layer}
\end{equation}
and
\begin{equation}\label{eqn.monot.increasing.A.cond.2}
p^{(\widehat{\layer}+1)}(x) \geq  q^{(\widehat{\layer}+1)}(x), \quad x\geq 0.
\end{equation}
Condition \eqref{eqn.monot.increasing.A.cond.2}  can be verified by noting that $p^{(\widehat{\layer}+1)}(x)  \geq 0$ for $x\geq 0$ (cf. \eqn{eqn.poly.p.x})  and  
$q^{(\widehat{\layer}+1)}(x) = 0$ since $q(x)$ in \eqref{eqn.poly.q.x}
 is a polynomial of degree $\widehat{\layer}$.  It thus remains to establish \eqref{eqn.monot.increasing.A.cond.1}. Since 
 we have $p(0) = 0$ and $q(0) = 0$, it follows that  \eqref{eqn.monot.increasing.A.cond.1} is trivially satisfied for $n=0$. It therefore remains to show \eqref{eqn.monot.increasing.A.cond.1} for 
 $n = 1,\dots, \widehat{\layer}$.  
By Leibniz's law for the differentiation of products of functions, we obtain 
\[
g^{(n)}(x)  \big|_{x \,= \,0} = \begin{cases} {n \choose i-1} (i-1)! \,, & i \leq n \\ 0\,, & i = n+1,\dots,  \widehat{\layer}
\end{cases} 
\]
for $g(x) =  (e^{x}\!-\!1) \,x^{i-1}$, 
which yields 
\[
p^{(n)}(x) \big|_{x \,= \,0}  = \sum_{i=1}^{n}   {n \choose i-1} \frac{\widehat{\layer}\!-\!i}{\widehat{\layer}\!-\!1} \,a_{i}\, (i-1)!\,.  
\]
For the RHS of \eqref{eqn.monot.increasing.A.cond.1} we get
\[
 q^{(n)}(x)\big|_{x\, = \,0}  = a_{n}\, n!\,.
\]
Using \eqref{eqn.def.coefficients.ai}, the condition \eqref{eqn.monot.increasing.A.cond.1} 
can thus be rewritten as  
\begin{equation}\label{eqn.monot.increasing.A.cond.final}
 \sum_{i=1}^{n}   {n \choose i-1} {\widehat{\layer}\!-\!2 \choose i-1}  \frac{1}{i}  \geq   {\widehat{\layer}\!-\!1 \choose n-1}, \quad n = 1,\dots, \widehat{\layer}.
\end{equation}
 Note that \eqref{eqn.monot.increasing.A.cond.final} is trivially satisfied for $n = 1$. It thus remains to consider 
 $n = 2,\dots, \widehat{\layer}$. The RHS of \eqref{eqn.monot.increasing.A.cond.final} can be written as 
 \begin{equation}\label{eqn.binomial.coefficient.split}
 {\widehat{\layer}\!-\!1 \choose n-1} = {\widehat{\layer}\!-\!2 \choose n-2}+{\widehat{\layer}\!-\!2 \choose n-1}. 
 \end{equation}
The proof is concluded by showing that the sum of the two terms on the left hand side of \eqref{eqn.monot.increasing.A.cond.final} corresponding to 
$i = n$ and $i = n-1$ is greater than or equal to the RHS in \eqref{eqn.binomial.coefficient.split}. A direct comparison shows that this is the case if 
\[
 {n \choose n-1}  \frac{1}{n}  \geq 1 \quad \text{and} \quad {n \choose n-2}   \frac{1}{n-1}  \geq 1
\]
for $n = 2,\dots, \widehat{\layer}$. This is now easily verified by noting  that ${n \choose n-1}/n = 1$ and  ${n \choose n-2}/(n-1) = n/2$. 
\section{Properties of $\rratio(\epsilon)$} \label{app.radii.ratio}
Using definition \eqref{eqn.radii.ratio.def} with \eqref{eqn.cinf.epsilon} and \eqref{eqn.ctwo.epsilon}, we have 
\begin{equation}\label{eqn.radii.ratio.epsilon}
\rratio(\epsilon) = \frac{\gamma_{\rdim}^{-1}\!\left(1-\epsilon\right)}{\gamma_{1}^{-1}\big(\left(1-\epsilon \right)^{1/\rdim}\!\big)}
\end{equation}
by noting that $\gamma_{1}(x) = 1-e^{-x}$ (see Appendix \ref{app.incomplete.gamma.function}). 

\subsection{Limits of $\rratio(\epsilon)$} \label{app.radii.limit}     
\subsubsection{Limit of $\rratio(\epsilon)$ for $\epsilon \rightarrow 1$}  \label{app.radii.limit.eps1}     
We want to prove that 
\[
\underset{\epsilon \rightarrow 1} {\lim}\,\, \rratio(\epsilon) = \sqrt[\rdim]{\rdim!}\,. 
\]
Setting $x = 1-\epsilon$, this amounts to showing that 
\begin{equation}\label{eqn.radii.ratio.equivalent}
\underset{x \rightarrow 0} {\lim}\,\,   \frac{\gamma_{\rdim}^{-1}(x)}{\gamma_{1}^{-1}\big(x^{1/\rdim}\big)}  = \sqrt[\rdim]{\rdim!}\,. 
\end{equation}
We start by considering the numerator in \eqref{eqn.radii.ratio.equivalent} and note that \eqref{eqn.gamma.series1} implies
\[
	\gamma_{\rdim}(y) = \frac{1}{\rdim!} y^{\rdim}(1+o(1)), \quad y \rightarrow 0 
\]
and thus
\begin{equation}\label{eqn.radii.ratio.equivalent.enumerator}
	\gamma_{\rdim}^{-1}(x) =  \sqrt[\rdim]{\rdim!} \, x^{1/\rdim}\, {(1+o(1))^{-1}}, \quad x \rightarrow 0. 
\end{equation}
Similarly, for  the denominator in \eqref{eqn.radii.ratio.equivalent}, we obtain 
\[
	\gamma_{1}^{-1}\big(x^{1/\rdim}\big) =   x^{1/\rdim}\, {(1+o(1))^{-1}}, \quad x \rightarrow 0 
\]
which, together with \eqn{eqn.radii.ratio.equivalent.enumerator}, establishes \eqref{eqn.radii.ratio.equivalent}. 

\subsubsection{Limit of $\rratio(\epsilon)$ for $\epsilon \rightarrow 0$}\label{app.radii.limit.eps0}     
We want to prove that 
\[
\underset{\epsilon \rightarrow 0} {\lim}\,\, \rratio(\epsilon) = 1. 
\]
Again, setting $x = 1-\epsilon$, this amounts to showing that 
\begin{equation}\label{eqn.radii.ratio.equivalent2}
\underset{x \rightarrow 1} {\lim}\,\,   \frac{\gamma_{\rdim}^{-1}(x)}{\gamma_{1}^{-1}\big(x^{1/\rdim}\big)}  = 1. 
\end{equation}
We therefore need to prove that $\gamma_{\rdim}^{-1}(x) \equalasympt  \gamma_{1}^{-1}\big(x^{1/\rdim}\big)$, $x \rightarrow 1$.  
Starting with the denominator in  \eqref{eqn.radii.ratio.equivalent2}, we  first note that 
\begin{equation}\label{eqn.radii.ratio.denominator2}
   \gamma_{1}^{-1}\big(x^{1/\rdim}\big)  = \text{log}\!\left(\frac{1}{1-x^{1/\rdim}}\right).   
\end{equation}
Next, we have 
\[
	x^{1/\rdim} = (1-(1-x))^{1/\rdim} = 1-\frac{1}{\rdim}(1-x)+{\cal O}((1-x)^{2}), \quad x \rightarrow 1
\] 
and hence 
\[
	\frac{1}{1-x^{1/\rdim}} = \frac{\rdim}{1-x} (1+{\cal O}(1-x)), \quad x \rightarrow 1
\] 
which finally yields 
\[
    \text{log}\!\left(\frac{1}{1-x^{1/\rdim}}\right) =  \text{log}\!\left(\frac{1}{1-x}\right) + \text{log}(\rdim) + {\cal O}(1-x) \equalasympt \text{log}\!\left(\frac{1}{1-x}\right), \quad x \rightarrow 1
\]
establishing that  
\begin{equation}\label{eqn.radii.ratio.denominator2}
 	\gamma_{1}^{-1}\big(x^{1/\rdim}\big)  \equalasympt \text{log}\!\left(\frac{1}{1-x}\right), \quad x \rightarrow 1.
\end{equation} 
For the numerator in \eqref{eqn.radii.ratio.equivalent2}, we first note that $\underset{x \rightarrow \infty} {\lim}\,\,  \gamma_{\rdim}(x) = 1$, which
implies that the  $x \rightarrow 1$  asymptote 
of the inverse function $\gamma_{\rdim}^{-1}(x)$
can be obtained by characterizing the $x \rightarrow \infty$ asymptote of 
$\gamma_{\rdim}(x)$. It follows from \eqref{eqn.gamma.series2} that 
\[
	\gamma_{\rdim}(x) = 1- \frac{1}{(\rdim-1)!} e^{-x} x^{\rdim-1} (1+o(1)), \quad x \rightarrow \infty
\]
which yields  
\[
	\text{log}\big((\rdim-1)! \,(1-\gamma_{\rdim}(x))\big) = -x + (\rdim-1) \text{log}(x) + o(1),  \quad x \rightarrow \infty
\] 
and hence 
\[
 \text{log}\big((\rdim-1)! \,(1-\gamma_{\rdim}(x))\big) \equalasympt -x, \quad x \rightarrow \infty.
\]
Now setting $x = \gamma_{\rdim}^{-1}(y)$, we finally get 
\begin{equation*}
   \gamma_{\rdim}^{-1}(y)  \equalasympt  -\text{log}\big((\rdim-1)! \,(1-y)\big) \equalasympt \text{log}\left(\frac{1}{1-y}\right), \quad y \rightarrow 1.
\end{equation*}
Together with  \eqref{eqn.radii.ratio.denominator2}, this implies \eqref{eqn.radii.ratio.equivalent2}.    

\subsection{Monotonicity of $\rratio(\epsilon)$} \label{app.radii.ratio.monotonicity}     
In the following, we show that  $\rratio(\epsilon)$ in \eqref{eqn.radii.ratio.epsilon} is a nondecreasing function of $\epsilon$ on the interval $[0, 1]$. This will be accomplished by setting 
$1-\epsilon = \gamma_{\rdim}(x)$, $x \in \Rnum$, $x\geq 0$, and showing that the function $f(x) = x/g(x)$ with 
\[   
  g(x) =  -\text{log}\Big(1- \left[\gamma_{\rdim}\!\left(x\right)\right]^{\frac{1}{\rdim}}\Big) 
\]
is nonincreasing in $x \geq 0$, or equivalently
\[
f'(x) = \frac{g(x)-g'(x)x}{g^2(x)} \leq 0, \quad \text{for}\,\,x \geq 0. 
 \]   
It thus remains to show that 
\begin{equation}\label{eqn.monotonicity.conv.cond}
	g(x)-g'(x)x \leq 0, \quad \text{for}\,\,x \geq 0. 
\end{equation}
Next, we note that $g(x)$ is convex for $x \geq 0$ if and only if the first-order convexity condition $g(x)+g'(x)(y-x) \leq g(y)$ holds for all $x,y \geq 0$  \cite[Eq.\ (3.2)]{boyd_conv_opt01}.  
This first-order convexity condition evaluated at $y=0$  becomes \eqref{eqn.monotonicity.conv.cond} by noting that  $g(0) = 0$.
Consequently, it is sufficient to show that $g(x)$ is a convex function for $x \geq 0$ or, equivalently, that $1-\left[\gamma_{\rdim}\!\left(x\right)\right]^{\frac{1}{\rdim}}$ is {\em log-concave} for $x \geq 0$.  
The function $1-\left[\gamma_{\rdim}\!\left(x\right)\right]^{\frac{1}{\rdim}}$ is a complementary cdf, which can be written as 
\begin{align*}	
1-\left[\gamma_{\rdim}\!\left(x\right)\right]^{\frac{1}{\rdim}}& = \int_{x}^{\infty} \! \left(\left[\gamma_{\rdim}\!\left(t\right)\right]^{\frac{1}{\rdim}} \right)' dt 
\end{align*}
where  $\left(\left[\gamma_{\rdim}\!\left(x\right)\right]^{\frac{1}{\rdim}} \right)'$ denotes the corresponding pdf. Using the fact that log-concavity of a pdf implies that the corresponding complementary cdf is also log-concave 
\cite[Theorem 3]{Bagnoli05}, it is sufficient to show that  
\begin{align}
  \left(\left[\gamma_{\rdim}\!\left(x\right)\right]^{\frac{1}{\rdim}} \right)' & =  \frac{1}{\rdim} \left[\gamma_{\rdim}\!\left(x\right)\right]^{\frac{1}{\rdim}-1}\, \gamma_{N}'(x) \nonumber \\
  \label{eqn.sqrt.gamma.derivative}
 & =  \frac{e^{-x}}{\rdim\, \Gamma(\rdim)} \left( \frac{x}{\left[\gamma_{\rdim}\!\left(x\right)\right]^{\frac{1}{\rdim}}} \right)^{\rdim-1}  
 \end{align}
is log-concave for $x \geq 0$. Here, we used $\gamma_{N}'(x) = e^{-x} x^{N-1} / \Gamma(N)$ (cf.\ \eqref{eqn.gamma}). 
The log-concavity (or log-convexity) of functions is preserved by the multiplication with exponentials (which themselves are log-convex and log-concave), 
by positive scaling, and by taking positive powers \cite{boyd_conv_opt01}, i.e., $e^{a x} v(x)$,  
$b v(x)$, $[v(x)]^{b}$, $a,b \in \Rnum$, $b > 0$,  is log-concave (log-convex) if $v(x)$ is log-concave (log-convex). 
Therefore, \eqn{eqn.sqrt.gamma.derivative} is log-concave if  $x^{\rdim} e^{-x}/\gamma_{\rdim}(x)$ (obtained by multiplying the RHS of \eqref{eqn.sqrt.gamma.derivative} by $\rdim\, \Gamma(\rdim) e^{x}$, taking the corresponding result to the power of $\rdim/(\rdim-1)$ followed by multiplication by $e^{-x}$) is log-concave. Equivalently,  \eqn{eqn.sqrt.gamma.derivative} is log-concave for $x \geq 0$ if 
\[
	h(x) =  \gamma_{\rdim}(x) x^{-\rdim} e^{x}	
\]
is log-convex for $x \geq 0$.  Next, with the series expansion \eqref{eqn.gamma.series1} for $\gamma_{\rdim}(x)$, we obtain 
\[
	h(x) =  \sum_{i=0}^{\infty} \frac{x^{i}}{(i+\rdim)!}.
\]
Using the series representation of the confluent hypergeometric function 
\[
	F(a,b,x) = \sum_{i=0}^{\infty} \frac{(a)_{i}}{(b)_{i}} \frac{x^{i}}{i!}
\]
where $(\cdot)_{i}$ denotes the Pochhammer symbol, i.e., $(a)_{i} = a(a+1)\cdots(a+i-1)$,  $(b)_{i} = b(b+1)\cdots(b+i-1)$ with $(a)_{0} = (b)_{0} = 1$, we can write 
\[
	h(x) = \frac{1}{\rdim!}\,F(1,\rdim+1,x). 
\]
With the integral representation of $F(a,b,x)$ \cite{Abr65}, we finally get 
\begin{equation}\label{eqn.montonicity.integral.confluent.hyper.fun}
	h(x) =  \frac{1}{\Gamma(\rdim)} \int_{0}^{1} e^{x t} \,(1-t)^{\rdim-1} dt.
\end{equation}
Applying the integration property of log-convex functions \cite[p.\ 106]{boyd_conv_opt01}, which states that 
log-convexity of $v(x,y)$ in $x$ for each $y$ in some set ${\cal C}$ implies log-convexity of 
$u(x) = \int_{y \in {\cal C}} v(x,y) dy$, we can conclude that $h(x)$ is log-convex for $x \geq 0$ if the integrand in \eqref{eqn.montonicity.integral.confluent.hyper.fun} is log-convex in $x$ for each $t \in [0,1]$.  
The proof is concluded by noting that this is trivially the case as the integrand, for each $t \in [0,1]$, is proportional to an exponential function (which is log-convex) for all $t$. 
\section{Calculation of  $\P\Big[ \big\|[\zb(\deltab)]_{\tdim-\layer+1} \big\|_{\tinfty} \leq \rtilde\Big]$}\label{app.distribution.schlange.independent}
In the following, we derive an analytic expression for $\P\Big[ \big\|[\zb(\deltab)]_{\tdim-\layer+1} \big\|_{\tinfty} \leq \rtilde\Big]$ under the assumption that 
$b_{\tdim-\layer+1}$ is {\em purely real, purely imaginary,} or {\em equal to zero}. The real and imaginary parts of $[\zb(\deltab)]_{\tdim-\layer+1}$ are given by  
\begin{align*}
	[\zb(\deltab)]_{\real,\tdim-\layer+1} & = R_{\tdim-\layer+1,\tdim-\layer+1}\deltanb_{\real, \tdim-\layer+1} + u_{\real, \layer}\\
	[\zb(\deltab)]_{\imag,\tdim-\layer+1} & = R_{\tdim-\layer+1,\tdim-\layer+1}\deltanb_{\imag, \tdim-\layer+1} + u_{\imag, \layer}. 
\end{align*}
Here,  $u_{\layer} \sim \CN(0,\sigma_\layer^2)$ is specified in \eqref{def.vm.um} ($\sigma_{\layer}^2$ is specified in \eqref{eqn.def.sigmam})
and $R_{\tdim-\layer+1,\tdim-\layer+1} \in \Rnum$.  Under the assumption that 
$b_{\tdim-\layer+1}$ is purely real, purely imaginary, or equal to zero, $[\zb(\deltab)]_{\real,\tdim-\layer+1}$ and $[\zb(\deltab)]_{\imag,\tdim-\layer+1}$ are statistically independent, which yields 
\begin{equation}\label{eqn.dist.intermediate.tilde.product}
\P\Big[ \big\|[\zb(\deltab)]_{\tdim-\layer+1} \big\|_{\tinfty} \leq \rtilde\Big] = \P\Big[ | [\zb(\deltab)]_{\text{R},\tdim-\layer+1}| \leq \rtilde\Big]
 \P\Big[ |[\zb(\deltab)]_{\text{I},\tdim-\layer+1}| \leq \rtilde\Big]. 
\end{equation}
Let us first assume that $\deltanb_{\tdim-\layer+1}$ is purely real, i.e., $\deltanb_{\tdim-\layer+1} = \deltanb_{\real, \tdim-\layer+1} \neq 0$.  
Similar to \eqref{eqn.equal.dist.component.vm.um}, we can write $\big|[\zb(\deltab)]_{\real, M-\layer+1}\big| \equaldist |v_{\layer} + u_{\real,\layer}|$ and 
$\big|[\zb(\deltab)]_{\imag, M-\layer+1}\big| = |u_{\imag,\layer}|$, where $v_{\layer} 
= R_{\tdim-\layer+1,\tdim-\layer+1}|\deltanb_{\tdim-\layer+1}|$ 
is a scaled $\chi_{2(\layer+L)}$-distributed RV with pdf \eqref{eqn.dist.p} and $u_{\real,\layer}$ and $u_{\imag,\layer}$ are i.i.d.\ $\N(0,\sigma^2_\layer/2)$. 
The RV $\frac{\sqrt{2}}{\sigma_{\layer}}|u_{\imag,\layer}|$ is thus $\chi_{1}$-distributed, which gives 
\begin{equation}\label{eqn.dist.intermediate.tilde.first.term}
\P\Big[ |u_{\imag,\layer}| \leq \rtilde\Big]  =  \gamma_{\frac{1}{2}}\!\bigg(\frac{C_\tinfty^2}{\sigma^2_\layer}\bigg).
\end{equation}
For given $v_\layer = v$, the RV $\frac{2}{\sigma_\layer^2}|v + u_{\real,\layer}|^2$ is non-central $\chi^2_{1}$-distributed with non-centrality parameter 
$\frac{2 v^2}{\sigma_\layer^2}$. Thus, following the steps \eqref{eqn.non.central.chi.distribution} -- \eqref{eqn.dist.intermediate}, we obtain 
\begin{equation} \label{eqn.dist.intermediate.tilde} 
\P\big[|v_\layer + u_{\real,\layer}| \leq \rtilde\big] =   \sum_{s = 0}^{\infty} D_{s}(\deltab_\layer) \,
\gamma_{s+\frac{1}{2}}\! \bigg(\frac{\rtilde^2}{\sigma_m^2}\bigg)
\end{equation}
where $D_{s}(\deltab_\layer) $ was defined in \eqref{eqn.coefficient.schlange}. Note that the only difference between \eqref{eqn.dist.intermediate.tilde}  and \eqref{eqn.dist.intermediate} is the occurrence of the factor $1/2$ instead of the factor $1$ in the index of the incomplete Gamma function. 
As a result, however, it seems that \eqref{eqn.dist.intermediate.tilde} cannot be expressed as a finite sum as was done for \eqref{eqn.dist.intermediate} to arrive at \eqref{eqn.final.appendix.bin.mixture}. 
The final expression \eqref{eqn.dist.schlange} now follows by combining \eqref{eqn.dist.intermediate.tilde.product}, \eqref{eqn.dist.intermediate.tilde.first.term},  and \eqn{eqn.dist.intermediate.tilde}. 
The cases $\deltanb_{\tdim-\layer+1} = \deltanb_{\text{I}, \tdim-\layer+1} \neq 0$ and $\deltanb_{\tdim-\layer+1} = 0$ can be seen to result in  \eqref{eqn.dist.schlange} by following the steps 
\eqref{eqn.dist.intermediate.tilde.product} -- \eqref{eqn.dist.intermediate.tilde} properly modified.

\bibliographystyle{ieeetr}      %
\bibliography{tf-zentral_loc,bib_see} 
        
\end{document}